\documentclass[aps,prd,twocolumn,showpacs,notitlepage,eqsecnum,
superscriptaddress,nofootinbib,longbibliography]{revtex4-2}

\usepackage[centertags]{amsmath} \usepackage{amssymb} \usepackage{latexsym}
\usepackage{enumerate} \usepackage{graphicx} \usepackage{mathrsfs}
\usepackage[colorlinks]{hyperref} \usepackage{stmaryrd} \usepackage{import}
\usepackage{tensor} \usepackage[usenames,dvipsnames]{xcolor} \usepackage{bm}
\usepackage{multirow} \usepackage{breakurl} \usepackage{float}
\usepackage{verbatim} \usepackage{mathrsfs}
\usepackage{booktabs}
\usepackage{makecell}
\usepackage[normalem]{ulem} 
\usepackage{longtable}
\allowdisplaybreaks[1]

\definecolor{CiteColor}{rgb}{0,0.5,0} \hypersetup{citecolor=CiteColor}
\definecolor{RefColor}{rgb}{0.55,0,0} \hypersetup{linkcolor=RefColor}
\definecolor{darkgreen}{rgb}{0.2,0.7,0.2}

\begin{document}

\title{Advancing the Effective-One-Body Framework in the Test-Mass Limit}

\author{Nami Nishimura}
\affiliation{Max Planck Institute for Gravitaional Physics (Albert Einstein Institute), Am Mühlenberg 1, Potsdam 14476, Germany}
\affiliation{Department of Physics, University of Maryland, College Park, MD 20742, USA}

\author{Alessandra Buonanno}
\affiliation{Max Planck Institute for Gravitaional Physics (Albert Einstein Institute), Am Mühlenberg 1, Potsdam 14476, Germany}
\affiliation{Department of Physics, University of Maryland, College Park, MD 20742, USA}

\author{Guglielmo Faggioli}
\affiliation{Max Planck Institute for Gravitaional Physics (Albert Einstein Institute), Am Mühlenberg 1, Potsdam 14476, Germany}

\author{Maarten van de Meent}
\affiliation{Max Planck Institute for Gravitaional Physics (Albert Einstein Institute), Am Mühlenberg 1, Potsdam 14476, Germany}
\affiliation{Center of Gravity, Niels Bohr Institute, Blegdamsvej 17, 2100 Copenhagen, Denmark}

\author{Gaurav Khanna}
\affiliation{Department of Physics and Center for Computational Research, University of Rhode Island, Kingston, RI 02881, USA}
\affiliation{Department of Physics and Center for Scientific Computing and Data
Science Research, University of Massachusetts, Dartmouth, MA 02747, USA}

\begin{abstract}
We present \texttt{SEOB-TML}, an enhanced effective-one-body (EOB) framework for the test-mass limit, optimized for quasi-circular, spin-aligned binary black holes.
On the dynamical side, we introduce a quadrupole-factorized ($Q$-factorized) prescription that maps the total energy flux—including horizon absorption—onto a single $(2,2)$ mode baseline.
This approach effectively captures higher-order multipole contributions without explicit mode summation, while simultaneously leading to a dramatic reduction in fractional flux errors.
To ensure a smooth transition to the post-merger stage, we replace traditional next-to-quasicircular corrections with a phenomenological ansatz, enabling a flexible, mode-dependent attachment prescription.
For the merger–ringdown stage, we utilize quasi-normal mode coefficients extracted from numerical waveforms via \texttt{qnmfinder} to explicitly model mode-mixing effects.
These enhancements lead to a substantial reduction in residuals, capturing the complex physical modulations prominent in retrograde configurations.
Additionally, we implement the $(2,0)$ mode across the full waveform, further extending the model's physical coverage and accuracy.
Overall, our framework generates highly accurate late inspiral–merger–ringdown waveforms for extreme-mass-ratio systems, 
significantly reducing dephasing and improving the near-merger reconstruction. 
We demonstrate the performance of \texttt{SEOB-TML} against the current state-of-the-art \texttt{SEOBNRv5HM} model, highlighting how our specialized developments extend the reliability of the EOB framework into the test-mass limit.
\end{abstract}

\maketitle
\twocolumngrid

\section{Introduction}
The first detection of gravitational waves (GWs) by LIGO-Virgo~\cite{LIGOScientific:2018mvr, LIGOScientific:2020ibl, LIGOScientific:2021usb, LIGOScientific:2019lzm} in 2015 marked a milestone in gravitational physics. Since then, the LIGO-Virgo-KAGRA (LVK)~\cite{LIGOScientific:2016aoc, LIGOScientific:2021djp, KAGRA:2013rdx, LIGOScientific:2025slb, LIGOInstrumentWhitePaper,VirgoInstrumentWhitePaper} Collaboration has reported more than two hundred signals from the coalescence of binary compact objects, mainly stellar-mass black holes (BHs) and neutron stars.
Matched filtering, which is used to infer source properties such as masses and spins, demands highly accurate waveform models, and modeling errors will become increasingly limiting as detector sensitivity improves.
Next generation observatories such as the Einstein Telescope~\cite{Punturo_2010,ET:2019dnz,ET:2025xjr}, Cosmic Explorer~\cite{Evans:2021gyd} and LISA~\cite{amaro2017laser, LISA:2024hlh} will be characterized by higher sensitivity, leading to increased detection rates by several orders of magnitude and extending GW observations to lower frequency regions. 
With upcoming detectors, signals from binaries at increasingly asymmetric mass ratios will become accessible, ranging from $10^{-4}$–$10^{-2}$ for intermediate-mass-ratio inspirals (IMRIs) to $10^{-7}$–$10^{-4}$ for extreme-mass-ratio inspirals (EMRIs).
In this context, it becomes essential to develop precise waveform models in the test-mass limit (TML), both to capture the dynamics of EMRIs and to strengthen the connection between waveform modeling across different mass-ratio regimes. 

A central ingredient in the success of GW observations has been the development of accurate and computationally efficient waveform models for binary BH systems. 
Numerical relativity (NR) simulations~\cite{Pretorius:2005gq, Campanelli:2005dd,Baker:2005vv} provide the most faithful description of the full coalescence and are particularly valuable for comparable-mass binaries.
However, their high computational cost prevents dense coverage of the parameter space, making it essential to construct models that combine analytical methods with NR input. 
Several complementary strategies have been developed to produce complete inspiral–merger–ringdown (IMR) waveforms. 
The NR surrogate models~\cite{Blackman:2015pia,Blackman:2017dfb,Blackman:2017pcm, Varma:2018mmi, Varma:2019csw, Williams:2019vub, Rifat:2019ltp,Islam:2021mha,Islam:2022laz,Yoo:2022erv, Islam:2024zqo, Rink:2024swg,Maurya:2025shc,Nee:2025nmh} interpolate NR data in a reduced-order basis and thus achieve very high accuracy, but they are restricted to the region of parameter space covered by NR simulations. 
Another widely used class is the phenomenological models known as \texttt{IMRPhenom}, built in both frequency-domain (FD)~\cite{Pan:2007nw,Ajith:2007qp, Ajith:2009bn, Santamaria:2010yb, Hannam:2013oca,Husa:2015iqa,Khan:2015jqa, London:2017bcn,Khan:2018fmp,Khan:2019kot,Pratten:2020fqn,Pratten:2020ceb,Garcia-Quiros:2020qpx, Hamilton:2021pkf, Thompson:2023ase, Colleoni:2024knd, Hamilton:2025xru, Ramos-Buades:2026kbq} and time-domain (TD)~\cite{Estelles:2020twz,Estelles:2020osj, Estelles:2021gvs, Rossello-Sastre:2024zlr, Planas:2025feq}. 
Finally, the EOB formalism~\cite{Buonanno:1998gg, Buonanno:2000ef,Damour:2000we, Damour:2001tu, Buonanno:2005xu} provides a powerful analytical framework that combines various analytical approximation methods while maintaining the flexibility to facilitate systematic calibration against NR results.

In the EOB approach, the binary dynamics are mapped to those of an effective particle moving in a deformed Schwarzschild or Kerr spacetime, with the deformation parameter determined by the symmetric mass ratio. 
This mapping enables the construction of accurate and efficient waveforms through a close interplay between analytical approximations and NR calibration. 
Among waveform families, EOB-based models, including the SEOBNR models~\cite{Buonanno:2006ui,Buonanno:2007pf,Buonanno:2009qa,Pan:2011gk,Damour:2012ky,Damour:2015isa,Pan:2009wj,Taracchini:2012ig,Taracchini:2013rva,Bohe:2016gbl,Cotesta:2018fcv,Ossokine:2020kjp,Pompiliv5,RamosBuadesv5,Gamboa:2024imd,Gamboa:2024hli,Ramos-Buades:2021adz,Estelles:2025zah,VandeMeentv5} 
and the \texttt{TEOBResumS} models~\cite{Damour:2014sva,Nagar:2015xqa,Gamba:2024cvy,Albanesi:2025txj,Nagar:2018zoe,Nagar:2019wds,Nagar:2020pcj,Riemenschneider:2021ppj,Chiaramello:2020ehz,Gamba:2021ydi,Nagar:2018gnk,Rettegno:2019tzh,Akcay:2020qrj,Nagar:2018plt, Nagar:2024oyk, Albanesi:2024xus}
are distinguished by their robustness across mass ratios and spins, and by their ability to provide complete IMR waveforms at relatively low computational cost. 
By construction, EOB models incorporate the TML, since the formalism is built as a deformation of Kerr or Schwarzschild spacetime.
Historically, the systems composed of a small BH of mass $\mu$ orbiting a large BH with mass M such that the symmetric mass-ratio $\nu = \mu/ M \ll 1$ served as a laboratory for developing physically motivated ansatze to describe the inspiral, plunge, merger and ringdown (RD) within the EOB framework~\cite{Damour:2008gu}. 
Indeed, several studies~\cite{Nagar:2006xv, Damour:2007xr,  Damour:2008gu, Barausse:2011kb, Taracchini:2013wfa,Taracchini:2014zpa, Albanesi:2021rby, Albanesi:2022ywx, Albertini:2022rfe, Albertini:2022dmc, vandeMeent:2023ols, Albertini:2023aol, Albanesi:2023bgi, Albertini:2024rrs, Albanesi:2024fts, Faggioli:2024ugn, Leather:2025nhu, Faggioli:2025hff, Nagni:2025cdw} have shown that the insights from BH perturbation theory can significantly improve EOB waveform modeling.
Results from the TML have already been successfully incorporated to inform and refine waveform models across the full mass-ratio spectrum.

From the wide range of existing EOB formulations, we select the state-of-the-art \texttt{SEOBNRv5HM} model~\cite{Pompiliv5} as our baseline.
This multipolar IMR model, developed for quasi-circular, spinning, nonprecessing BBHs, combines post-Newtonian (PN) results for the Hamiltonian~\cite{Balmelli:2015zsa, Khalil:2020mmr}, the radiation reaction (RR) force, and waveform modes~\cite{Henry:2022dzx}.
It includes the GW spin-weighted spherical harmonic modes $(2,2), (2,1), (3,3), (3,2), (4,4), (4,3), (5,5)$ and is calibrated to 442 NR waveforms, all produced with the pseudo-Spectral Einstein code (SpEC) of the Simulating eXtreme Spacetime (SXS) collaboration~\cite{Blackman:2015pia,Blackman:2017dfb,Varma:2018mmi,Varma:2019csw, Yoo:2022erv, Bohe:2016gbl,SXScatalog1,Boyle:2019kee, Chu:2015kft, Hemberger:2013hsa, Scheel:2014ina, Lovelace:2014twa, LIGOScientific:2016kms,Lovelace:2016uwp,LIGOScientific:2016sjg,Kumar:2015tha,Mroue:2013xna, Scheel:2025jct}. 
Additional inputs from a  set of 13 Teukolsky waveforms with $q = 1000$ supplement the calibration in the small-mass-ratio regime~\cite{Barausse:2011kb, Taracchini:2014zpa}. 
For comparable-mass binaries, \texttt{SEOBNRv5HM} achieves good accuracy, validated by unfaithfulness studies against NR simulations and is routinely employed in GW parameter estimation thanks to its efficiency. 
However, in the TML its performance is more limited: PN approximations degrade in the strong-field regime, and calibration to the TML waveform data remains sparse compared to NR coverage. 

In this article, we present \texttt{SEOB-TML}, an enhanced EOB framework optimized for the TML. This model introduces a new factorized flux prescription that maps the total energy flux—including horizon absorption—onto a single $(2,2)$ mode baseline. 
To ensure a smooth transition to the post-merger stage, we utilize a flexible, mode-dependent phenomenological ansatz that enables individualized attachment prescriptions. For the merger–RD stage, we explicitly model mode-mixing effects by incorporating quasi-normal mode (QNM) coefficients extracted via \texttt{qnmfinder}~\cite{qnmfinder_code, Mitman:2025hgy}.
Additionally, we extend the model’s physical coverage by including the $(2,0)$ mode throughout the entire waveform.
We demonstrate the performance of \texttt{SEOB-TML} by validating it against \texttt{SEOBNRv5HM} within the TML.
This study serves as a preliminary exploration of integrating EMRI waveforms into the EOB framework; while these results are exploratory, they aim to inform future refinements of the SEOBNR family and ultimately extend its reliability across a wider mass-ratio spectrum.

The article is structured as follows.
In Sec.~\ref{sec:perturbation_theory}, we describe the numerical methods used to compute the fluxes and TD waveforms.
Sec.~\ref{sec:RRForce} provides a brief overview of the EOB dynamics and details our specific implementation of the analytical fluxes, which are then compared against numerical data to assess the accuracy of the RR force.
In Sec.~\ref{sec:ModelingInspPlunge}, we outline the construction of the multipolar inspiral-plunge waveform modes, highlighting improvements and differences with respect to \texttt{SEOBNRv5HM}. 
The modeling of the merger–RD is detailed in Sec.~\ref{sec:ModelingMergerRingdown}, where we construct new ansatze using QNM coefficients extracted from numerical waveforms to capture the mode mixing.  
Sec.~\ref{sec:TotalIMR} highlights the improvement in the total IMR waveform in the TML compared to \texttt{SEOBNRv5HM}, while Sec.~\ref{sec:h20mode} reports the implementation of the $(2,0)$ mode across the full IMR.
Finally, Sec.~\ref{sec:Conclusion} summarizes our main conclusions and outlines directions for future work.

\section*{Notation}
We adopt natural units where $c = G = 1$ and consider a non-spinning test-mass (TM) $\mu = \nu M$ orbiting a Kerr BH with mass $M$ and dimensionless spin $a = J/M^2$. 
The Kerr spacetime is described using Boyer-Lindquist coordinates $(T,R,\theta, \varphi)$ and the TM motion is characterized by the canonical phase-space variables $(R, \varphi, P_R, P_{\varphi})$.
Throughout this work, we employ dimensionless, scaled quantities defined as 
\begin{equation}
t = \frac{T}{M}, \; \; \; \;   r = \frac{R}{M}\; \; \; \;  p_r = \frac{P_R}{\mu}\; \; \; \; p_{\varphi} = \frac{P_{\varphi}}{M \mu}
\end{equation}
Correspondingly, the Hamiltonian and the RR force are normalized by $\mu$.  
The orbital frequency is denoted $\Omega$, from which we define a dimensionless velocity parameter $x^{1/2} = v_{\Omega} = (M \Omega)^{1/3}$. 
For ease of notation, we set $M=1$ throughout the remainder of the paper.

\section{Numerical approach using perturbation theory}
\label{sec:perturbation_theory}
In this section, we describe the numerical framework used to compute equatorial TM trajectories from inspiral through merger. We then outline the generation of the corresponding TD Teukolsky waveforms, which serve as the benchmark for our EOB prescriptions.
We restrict our system of interest to a non-spinning TM orbiting a Kerr BH on a quasi-circular orbit in the equatorial plane. 
Such a system can be described using BH perturbation theory, which approximates the dynamics of the binary by expanding the Einstein field equations in powers of $\nu$. 
At zeroth-order, the motion is governed by a geodesic in the background geometry of the primary mass, while higher-order perturbative corrections account for the gravitational field of the smaller body and its backreaction on the spacetime.
In this work, we compute the waveforms at leading order in $\nu$ by solving the Teukolsky master equation~\cite{Teuk_original}.

In Boyer–Lindquist coordinates, the Teukolsky equation takes the form
\begin{align}
&-\left[ \frac{(r^2 + a^2)^2}{\Delta} - a^2 \sin^2\theta \right] \partial_{tt} \Psi 
- \frac{4ar}{\Delta} \partial_{t\varphi} \Psi \nonumber \\
&- 2s \left[ r - \frac{r^2 - a^2}{\Delta} + i a \cos\theta \right] \partial_t \Psi 
+ \Delta^{-s} \partial_r \left( \Delta^{s+1} \partial_r \Psi \right) \nonumber \\
&+ \frac{1}{\sin\theta} \partial_\theta \left( \sin\theta \, \partial_\theta \Psi \right) 
+ \left[ \frac{1}{\sin^2\theta} - \frac{a^2}{\Delta} \right] \partial_{\varphi\varphi} \Psi \nonumber \\
&+ 2s \left[ \frac{a(r-1)}{\Delta} + \frac{i \cos\theta}{\sin^2\theta} \right] \partial_\varphi \Psi 
- \left( s^2 \cot^2\theta - s \right) \Psi \nonumber \\
&= -4\pi (r^2 + a^2 \cos^2\theta) \mathcal{T},
\label{eq:teukolsky_master_eq}
\end{align}
with $\Delta = r^2 - 2r + a^2$. 
The parameter $s$ denotes the spin weight of the field; for our purposes, we set $s = -2$ to describe the evolution of the curvature perturbations.
In this case, the Teukolsky master field $\Psi$ is directly related to the Weyl scalar $\psi_4$, which represents outgoing gravitational radiation, via the relation $\Psi = (r - i a \cos\theta)^4 \psi_4$.
The driving term of this evolution is the source term $\mathcal{T}$, which is constructed from the energy-momentum tensor of the TM. 
Since $\mathcal{T}$ contains Dirac delta functions localized at the particle’s coordinate position, the resulting field depends explicitly on the prescribed trajectory. 
Consequently, the numerical solution of Eq.~\eqref{eq:teukolsky_master_eq} requires a high-fidelity description of the orbital dynamics, which we detail below.

\subsection{Numerical dynamics computation}
To describe the dynamics entering the source term, we integrate Hamilton’s equations (Eq.~\eqref{eq:HamiltonEq}) augmented by the RR force that accounts for the dissipation of energy and angular momentum due to gravitational radiation.
The generation of trajectories that faithfully reproduce the equatorial quasi-circular dynamics of a TM in Kerr spacetime at leading-order in $\nu$ follows the methodology established in Refs.~\cite{Taracchini:2013wfa, Taracchini:2014zpa}.
This approach utilizes numerical fluxes computed for equatorial circular geodesics as a function of the orbital frequency.
To fully implement the procedure introduced in Ref.~\cite{Taracchini:2014zpa}, numerical fluxes are required down to orbital radii approaching the inner light-ring.
To address this requirement, we compute the FD Teukolsky fluxes using \texttt{ModGEMS}~\cite{vandeMeent:2014raa, vandeMeent:2015lxa, vandeMeent:2016pee, vandeMeent:2017bcc}.
This code solves the Teukolsky equation in FD with a geodesic source using a semi-analytical implementation of the Mano-Suzuki-Takasugi formalism \cite{Mano:1996vt, Mano:1996gn, Fujita:2004rb, Fujita:2009us}.
These fluxes are computed on the same radial grid used in Ref.~\cite{Taracchini:2014zpa}.
Once the total numerical fluxes—comprising both infinity and horizon contributions—are obtained, they are interpolated over the orbital frequency to evolve the equations of motion toward the event horizon, located at $r_{+} = 1 + \sqrt{1 - a^2}$. 
Since stable circular orbits do not exist beyond the photon orbit radius $r_{\mathrm{LR}}$, we define a transition radius $r_{\text{min}} = r_{\mathrm{LR}} + 0.01$, such that for $r > r_{\text{min}}$, the RR force is computed directly from the interpolated FD Teukolsky fluxes as explicitly shown in Eq.~(2) of Ref.~\cite{Taracchini:2013wfa}.
In the subsequent plunge region $r_{+} < r \leq r_{\text{min}}$, we turn off the RR force following the strategy of Ref.~\cite{Taracchini:2014zpa}.
This is justified because the motion of the TM, in this regime, is well approximated by a geodesic, and the influence of GW fluxes becomes negligible compared to the conservative dynamics as the TM approaches the horizon.
Finally, we verify the consistency of our \texttt{ModGEMS} implementation by comparing it against the publicly available fluxes from the BH Perturbation Toolkit~\cite{BHPToolkit}. This comparison covers the full extent of the Toolkit’s available data—up to the innermost stable circular orbit (ISCO)—which was produced in Refs.~\cite{Taracchini:2014zpa, Hughes:1999bq, Drasco:2005kz}.

\subsection{Numerical waveform computation}
Having determined the TM trajectory, we compute the emitted GWs by feeding the dynamics of the TM into the source term $\mathcal{T}$ and numerically solving the Teukolsky equation (Eq.~\eqref{eq:teukolsky_master_eq}).
The specific implementation and numerical infrastructure used to compute the TD Teukolsky waveforms are detailed in Refs.~\cite{Sundararajan:2007jg, Sundararajan:2008zm, Sundararajan:2010sr, Field:2020rjr, Zenginoglu:2011zz}. 
This framework employs hyperboloidal time slicing, a technique that enables the direct extraction of GW modes at future null infinity without the need for extrapolation.
To maintain high accuracy over long-duration simulations, the code leverages GPU-based parallel computing.
Although this study does not focus on the late-time tails of the radiation, the code is capable of evolving the system well beyond the QNM phase, accurately capturing the subsequent power-law decay regime once the QNMs have sufficiently subsided.
The complex GW strain $h(t; \iota, \varphi) = h_{+}(t; \iota, \varphi) - i h_{\times}(t; \iota, \varphi)$ is recovered at future null infinity by performing a double time integration of the Weyl scalar $\psi_4$, which is related to the strain via:
\begin{equation}
\psi_4 = \frac{1}{2} \frac{d^2}{dt^2} \left( h_{+} - i h_{\times} \right).
\end{equation}
In general, the total strain $h(t; \iota, \varphi)$, is expanded in the basis of $-2$ spin-weighted spherical harmonics
\begin{equation}
h(t; \iota, \varphi) = \sum_{\ell=2}^{\infty} \sum_{m=-\ell}^{\ell} {}_{-2}Y_{\ell m}(\iota, \varphi) h_{\ell m}(t),
\label{eq:EOBGWmodes}
\end{equation}
where $\iota$ denotes the binary's inclination angle with respect to the orbital plane, and $\varphi$ is the azimuthal direction to the observer.

\section{Computation of Radiation Reaction Force}
\label{sec:RRForce}
In this section, we compare analytical and numerical GW fluxes. 
We propose a refined factorization and resummation of the EOB flux at infinity that offers improved accuracy and a more streamlined structure in the TML compared to that of \texttt{SEOBNRv5HM}.
After illustrating the effectiveness and precision of our proposed analytical flux, we further enhance it by including the absorption flux at the BH horizon, implemented using a consistent factorized prescription.

\subsection{Overview of the factorized-mode flux}
\label{sec:EOBdynamics}
The EOB approach~\cite{Buonanno:1998gg, Buonanno:2000ef, Damour:2000we, Damour:2001tu, Buonanno:2005xu} maps the two-body problem onto the effective motion of a single body of mass $\mu$ in a deformed Schwarzschild or Kerr spacetime. 
This mapping facilitates the construction of an EOB Hamiltonian that encapsulates the conservative dynamics of the system.
Dissipative effects are subsequently incorporated via the RR force, which accounts for the energy and angular momentum flux carried by gravitational waves to both null infinity and the BH horizon.
Together, these yield Hamilton's equations, which can be integrated to obtain the effective dynamics from which the multipolar gravitational waveform at null infinity can be analytically computed.

In order to obtain a first-order description of extreme-mass-ratio binaries, we set the $\nu$ corrections to zero in the conservative sector and retain the leading-order contributions to the RR force.
The Hamiltonian of interest reduces to the Kerr Hamiltonian restricted to equatorial orbits $(\theta = \pi/2, p_{\theta} =0)$
\begin{equation} \label{eq:Kerr Hamiltonian}
H = \Lambda^{-1} \left( 2 a p_{\phi} + \sqrt{\Delta p_{\phi}^{2} r^{2} + \Delta^{2} \Lambda \frac{p_{r}^{2}}{r} + \Delta \Lambda r }\right)
\end{equation}
where $\Lambda = r^3 + 2a^2 + a^2 r$.
Instead of the radial momentum $p_r$, we use $p_{r_*}$, the momentum conjugate to the tortoise radial coordinate $r_*$. The tortoise coordinate is related to the Boyer-Lindquist coordinate $r$ by 
\begin{equation}
\begin{aligned}
dr_{*} &= \frac{r^2 + a^2}{\Delta} \, dr = \frac{1}{\xi(r)} \, dr, \\
p_{r_*} &= \xi(r) \, p_r.
\end{aligned}
\end{equation}
This is a common practice~\cite{Damour:2007xr, Pan:2009wj} to avoid spurious numerical singularities near the event horizon.  

The evolution of the TM dynamics is provided by Hamilton's equations of motion
\begin{equation}
\begin{aligned}
\dot{r} &= \xi \frac{\partial H}{\partial p_{r_*}}, & 
\dot{p_{r_*}} &= -\xi \frac{\partial H}{\partial r} + \frac{p_{r_*}}{p_{\phi}} \mathcal{F}_{\phi}, \\
\dot{\phi} &= \frac{\partial H}{\partial p_{\phi}}, & 
\dot{p_{\phi}} &= \mathcal{F}_{\phi}.
\end{aligned}
\label{eq:HamiltonEq}
\end{equation}
where the RR force $\mathcal{F}_{\phi}$ is obtained by summing GW modes in the factorized form $h_{\ell m}^{F}$~\cite{Damour:2007yf, Damour:2008gu, Damour:2007xr, Pan:2010hz}, 
\begin{equation}
\mathcal{F}_{\phi} = - \frac{\Omega}{8 \pi}  \underset{\ell=2}{\overset{\ell_{\text{max}}}{\sum}} \underset{m=1}{\overset{\ell}{\sum}} m^{2} |d_{L}h^{F}_{\ell m}|^{2}
\label{eq:EOB_summedflux}
\end{equation}
with $d_{L}$ being the luminosity distance of the binary to the observer. In \texttt{SEOBNRv5HM}, Eq.~\eqref{eq:EOB_summedflux} is truncated at $\ell_{\text{max}} = 8$ since the contribution of the higher modes is negligible in the comparable-mass case. 
The factorized inspiral modes~\cite{Damour:2008gu,Damour:2007xr,Pan:2010hz} are written as
\begin{equation}
h_{\ell m}^{F} = h_{\ell m}^{N} \hat{S}_{\text{eff}} T_{\ell m} f_{\ell m} e^{i \delta_{\ell m}} \, ,
\label{eq:hlmF}
\end{equation}
where the first factor $h_{\ell m}^{N}$ is the leading (Newtonian) order waveform and its explicit expression is~\cite{Damour:2008gu, Pan:2010hz}
\begin{equation}
h_{\ell m}^{N} = \frac{\nu}{d_L} n_{\ell m} c_{\ell+\epsilon_{\ell m}}\left( \nu \right) v_{\Omega}^{\ell + \epsilon}Y_{\ell - \epsilon_{\ell m}, -m} \left( \frac{\pi}{2}, \phi \right)
\label{eq:hNewt}
\end{equation}
Here, $v_{\Omega} \equiv x^{1/2} = \Omega^{1/3}$ while $Y_{\ell m}$ denotes the scalar spherical harmonics.
The parity of the mode is defined as
\begin{equation} 
    \epsilon_{\ell m} =
    \begin{cases}
        0, & \ell + m \; \text{is even} \\
        1, & \ell + m \; \text{is odd} 
    \end{cases}
\end{equation}
with the function $n_{\ell m}$ given by
\begin{equation} 
    n_{\ell m} =
    \begin{cases}
        \frac{8 \pi \left( i m \right)^{\ell}}{\left( 2 \ell + 1\right)!!} \sqrt{\frac{\left( l + 1\right)\left( l + 2\right)}{\ell \left( l - 1\right)}}, & \ell + m \; \text{is even} \\
        \frac{-16 i \pi \left( i m\right)^{\ell}}{\left( 2 \ell + 1\right)!!} \sqrt{\frac{\left( 2\ell + 1\right)\left(\ell + 2 \right)\left( \ell^2 - m^2\right)}{\left( 2\ell -1\right)\left( \ell + 1\right)\ell \left( \ell - 1\right)}}, & \ell + m \; \text{is odd} 
    \end{cases}
\end{equation}
where the coefficient $c_{k}(\nu)$ is
\begin{equation}
c_{k}\left( \nu \right) = \left( \frac{1 - \sqrt{1 - 4 \nu}}{2}\right)^{k-1} + \left( -1\right)^{k} \left( \frac{1 + \sqrt{1 - 4 \nu}}{2}\right)^{k-1}
\end{equation}
Note that the \texttt{SEOBNRv5HM} model employs the variable
\begin{equation}
v_{\varphi} \equiv \Omega \left( \frac{\partial H_{\text{EOB}}}{\partial p_{\phi}}\right) ^{-2/3} |_{p_r = 0}
\end{equation}
instead of $v_{\Omega}$ in Eq.~\eqref{eq:hNewt}, which simplifies to $v_{\varphi} = \Omega \left( r^{3/2} + a\right)^{2/3}$ in the TML. 
Since choosing $v_{\Omega}$ instead of $v_{\varphi}$ is more applicable to generic TML orbits~\cite{Faggioli:2024ugn, Gamboa:2024hli}—as the requirement of $p_r = 0$ can be problematic for eccentric motion—we make this change which does not significantly impact the accuracy of the quasi-circular model.
The leading contribution of the tail factor $T_{\ell m}$~\cite{Blanchet:1997jj}, generated by the backscattering of the GWs off the Kerr background is 
\begin{equation}
T_{\ell m} = \frac{\Gamma \left( \ell + 1 - 2 i k \right)}{\Gamma \left( \ell + 1\right)} e^{\pi k} e^{2 i k \text{ln}\left( 2 k r_0\right)}
\end{equation}
where $\Gamma \left( z\right)$ is the Euler Gamma function, $k = m \Omega$ and $r_0 = 2/ \sqrt{e}$. 
The effective source term $\hat{S}_{\text{eff}}$ is given by either the effective energy or the orbital angular momentum, such that 
\begin{equation} 
    \hat{S}_{\text{eff}} =
    \begin{cases}
        H, & \ell + m \; \text{is even} \\
        p_{\phi} v_{\Omega} & \ell + m \; \text{is odd} 
    \end{cases}
\end{equation}
The remaining components of the factorized modes are expressed as an amplitude $f_{\ell m}$ and a phase $\delta_{\ell m}$, which are computed such that the expression of $h_{\ell m}^{F}$ agree with the PN-expanded modes. 
Following the prescription in~\cite{Damour:2008gu}, $f_{\ell m}$ is further resummed as $f_{\ell m} = (\rho_{\ell m})^{\ell}$. This transformation mitigates the growth of the 1PN coefficient, which otherwise increases linearly with $\ell$ and could degrade the accuracy of the resummation at higher multipoles.
\footnote{In the original \texttt{SEOBNRv5HM} model, the factor $f_{\ell m}$ is decomposed into non-spinning and spinning contributions for odd-$m$ modes, such that $f_{\ell m} = (\rho_{\ell m}^{\mathrm{NS}})^{\ell} + f_{\ell m}^{S}$ (See Eq.(34) in Ref~\cite{Pompiliv5}). 
However, we find that maintaining the form $f_{\ell m} = \rho_{\ell m}^{\ell}$ for all $m$ leads to better agreement in both the energy flux and the waveform amplitudes when compared against Teukolsky data. Consequently, the \texttt{SEOBNRv5HM} results reported throughout this paper incorporate this modification.}
We remark that we are considering the waveform at leading order in $\nu$, therefore all the subleading $\nu$-dependencies are switched off in the factors of Eqs.~\eqref{eq:EOB_summedflux} and \eqref{eq:hlmF}, except for the leading Newtonian contribution in Eq.~\eqref{eq:hNewt}.

The mode-sum factorized flux in Eq.~\eqref{eq:EOB_summedflux} has been widely explored and its effectiveness and accuracy have been validated for comparable mass case. 
However, in the TML, PN-expanded formulae begin to perform poorly before reaching the ISCO, particularly for systems with large prograde spins.
This behavior is expected: as the BH spin increases, the ISCO shifts toward smaller radii and higher orbital frequencies, leading to a progressive degradation of the PN expansion in this higher relativistic regime.
The accuracy of the flux is a crucial bottleneck for EMRI modeling, as the accumulated dephasing can lead to significant systematic biases in parameter estimation. 
As demonstrated in Ref~\cite{Khalvati:2025znb}, the relative flux error must remain below the mass-ratio of the system to preserve waveform reliability for LISA data analysis.
In principle, one could improve the fidelity of the EOB flux by leveraging the high-order PN terms available in the TML—up to $11$PN for the spinning case and $22$PN for the non-spinning case.
However, increasing the PN order does not necessarily reduce the relative error near the ISCO for prograde orbits~\cite{Fujita:2014eta}.
Furthermore, reaching the required accuracy for rapidly rotating primaries ($a \gtrsim 0.9$) would necessitate truncating the multipolar sum at $\ell_{\text{max}} \gtrsim 30$~\cite{Khalvati:2025znb}. 
Such high-order expansions and large mode-sums introduce a prohibitive computational cost, making them unsuitable for fast waveform generation.
To overcome these challenges, we propose an alternative approach that avoids the multipole sum entirely.
This method models the total energy emission by mapping the full flux onto a single $(2, 2)$ mode baseline—a strategy designed to reduce complexity while maintaining high accuracy even in the strong-field regime.
\subsection{$Q$-factorized flux at infinity}
\label{sec:Qfactorized_inf}
\begin{figure*}[t]
\centering
\includegraphics[width=\textwidth]{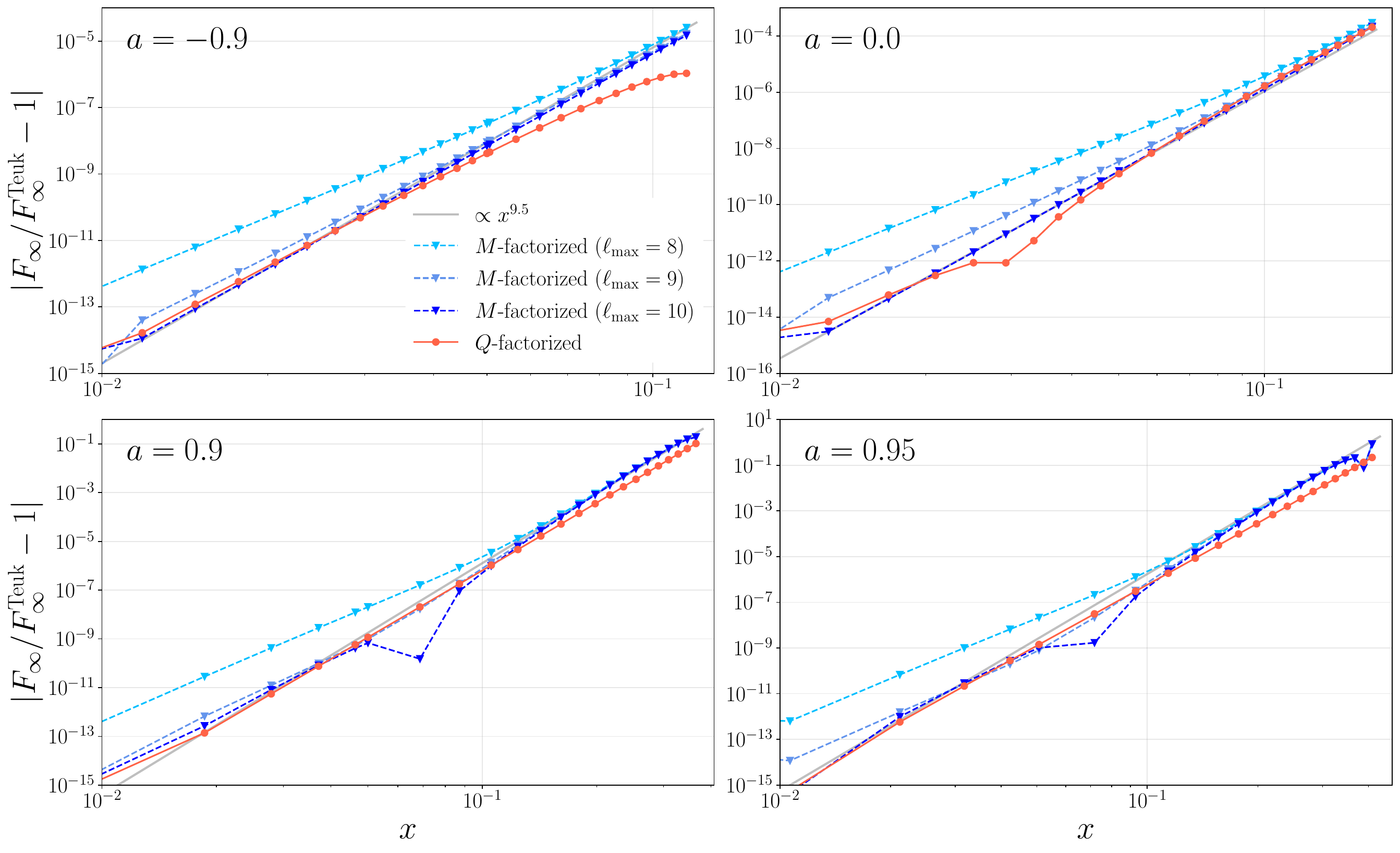}
\caption{Absolute value of the fractional difference between the analytical EOB fluxes and FD Teukolsky data for spins $a \in \{-0.9, 0, 0.9, 0.95\}$.
The 9PN $M$-factorized flux is shown in shades of blue, corresponding to truncation at $\ell_{\text{max}} = 8, 9, 10$, 
while the 9PN $Q$-factorized flux is shown in red. The $x$-axis corresponds to a dimensionless velocity parameter $x = \Omega^{2/3}$ and all fluxes are extended up to the ISCO.
The dashed lines indicate the expected $x^{9.5}$ scaling.
The $Q$-factorized flux effectively captures higher-multipole contributions, as evidenced by its low-frequency alignment with the $\ell_{\text{max}}=10$ $M$-factorized flux. Furthermore, the $Q$-factorized prescription demonstrates superior accuracy in the strong-field regime across all considered spins.}
\label{fig:SingleTermFluxComp}
\end{figure*}
In this section, we formalize the alternative factorization method hinted at above. 
We focus specifically on the flux at infinity, as the horizon absorption contributions will be incorporated later in our analysis.
Recall that, in \texttt{SEOBNRv5HM}, the normalized energy flux at infinity is computed by summing the factorized modes as follows: 
\begin{align}
\frac{1}{\nu^2} \dot{E} &= \frac{1}{\nu^2} \frac{x^3}{8 \pi} \sum_{\ell = 2}^{\ell_{\text{max}}} \sum_{m=1}^{\ell} m^2 |d_{L} h_{\ell m}^{F}|^2
\label{eq:EOBfluxSum}
\end{align}
which relates to Eq.~\eqref{eq:EOB_summedflux} via $\mathcal{F}_{\phi} = - x^{-3/2} dE/dt$. 
For the remainder of this work, we refer to Eq.~\eqref{eq:EOBfluxSum} as the mode-sum factorized ($M$-factorized) flux.
In contrast, our proposed analytical flux is constructed solely from the factorized $(2,2)$ mode.
We denote this expression as the quadrupole-factorized ($Q$-factorized) flux, defined as:
\begin{align}
\frac{1}{\nu^2} \dot{E} &= \frac{1}{\nu^2}\dot{E}_{22} \; \beta^{4} \nonumber \\
        &=\frac{1}{\nu^2}  \frac{x^3}{2 \pi} d_{L}^2 \; \left| h_{22}^{N} \, \hat{S}_{\text{eff}} \, T_{22} \, \rho_{22}^{2} \right|^{2} \; \beta^{4}(x)
\label{eq:Steff}
\end{align}
where the multiplicative parameter $\beta (x)$ is expressed as a PN series:
\begin{equation}
\beta(x)= 1 + \sqrt{x} \, \beta_{\frac{1}{2}} + x \, \beta_{1} + x^{\frac{3}{2}} \, \beta_{\frac{3}{2}} + \dots +  x^n \, \beta_{n} .
\label{eq:PNbeta}
\end{equation}
The exponent of $4$ in $\beta^4(x)$ is chosen to remain consistent with the structural form of the dominant $(2,2)$ mode contribution in the standard EOB framework. 
Specifically, as seen in Eq.~\eqref{eq:hlmF} and its substitution into Eq.~\eqref{eq:EOBfluxSum}, the flux contribution of the $\ell=2$ mode results in an overall factor of $\rho_{22}^4$. 
By adopting the same scaling for our calibration factor $\beta(x)$, we ensure that our proposed flux preserves the established resummation features already implemented within the \texttt{SEOBNRv5HM} model. 
Furthermore, while we also tested exponents of $1, 2,$ and $3$, we found that a fourth-power scaling yields the best numerical performance.

In order to determine the coefficients $\beta_{1/2}, \beta_{1}, \dots, \beta_{n}$ in Eq.~\eqref{eq:PNbeta}, we match the RHS of Eq.~\eqref{eq:Steff}, evaluated in the circular limit, with the PN-expanded flux for circular orbits, which is known up to 11th-PN order in the spinning case~\cite{Fujita:2014eta} and 22th-PN order in the non-spinning case~\cite{Fujita:2012cm}. 
In the circular-orbit limit, the individual components entering the factorized $\dot{E}_{22}$ on the RHS reduce to  
\begin{align}
h_{22}^{N}(x) &= \frac{\nu}{d_{L}} \, 8 \sqrt{\frac{\pi}{5}} \, x , \\
\hat{S}_{\text{eff}}^{(\epsilon_{22})}(x,a) &= \frac{1 - \frac{2}{r} + a \left(\frac{1}{r}\right)^{3/2}}
{\sqrt{1 - \frac{3}{r} + 2a \left(\frac{1}{r}\right)^{3/2}}} \label{eq:Seff} , \\
T_{22}(x) &= \sqrt{ \frac{ 2\pi \, x^{3/2} \left( 1 + 16  x^{3} \right) \left( 4 + 16  x^{3} \right) }
{ 1 - e^{-8 \pi  x^{3/2}} } } .
\end{align}
Here, the radius of circular orbit is determined as a function of the velocity $x$ and spin $a$.
\begin{equation}
r = (x^{-3/2} - a)^{2/3}
\end{equation}
For the $\rho_{22}$ factor in the analytical flux, we adopt the expression expanded to 5PN order in the TML.
This choice ensures the most accurate behavior in the strong-field regime for high prograde spins without requiring additional resummation within our framework.
The highest PN order of the $\beta$-coefficients, $\beta_{n}$, is determined by the PN order to which the circular-orbit flux on the LHS is expanded.  
In this work, we carry out the matching up to 9PN order, the motivation for this choice, balancing accuracy across the entire frequency range, is detailed at the end of this subsection.

In the following, we compare two analytical energy fluxes; (i) the $M$-factorized flux truncated at $\ell_{\text{max}} = 8, 9, 10$, and (ii) the $Q$-factorized flux, both implemented at 9PN order.
Figure~\ref{fig:SingleTermFluxComp} shows the fractional difference between analytical EOB fluxes with respect to the FD Teukolsky flux, both at infinity. 
We consider circular orbits with spins $a = -0.9, 0.0, 0.9,$ and $0.95$, and all data are extended up to the ISCO, where the radius $r_{\rm ISCO}$ is given by~\cite{Bardeen:1972fi}
\begin{equation}
\begin{aligned}
r_{\text{ISCO}} &= 3 + Z_2 \mp \sqrt{(3-Z_1)(3+Z_1+2Z_2)}, \\
Z_1 &= 1 + (1-a^2)^{1/3} \left[ (1+a)^{1/3} + (1-a)^{1/3} \right], \\
Z_2 &= \sqrt{3a^2 + Z_1^2}.
\end{aligned}
\end{equation} 
In the low-frequency region, for all spin values considered, the $Q$-factorized flux achieves an accuracy comparable to the factorized-mode flux that requires summation of modes up to $\ell = 10$.  
The slopes of the fractional residuals are compatible with the expected scaling, $x^{(n+1/2)}$ at low $x$, where $n=9$ corresponds to the PN order of the analytical fluxes we considered.  
This demonstrates that the $Q$-factorized flux effectively encodes higher-mode contributions without explicitly summing over each mode, leading to substantial computational savings.  
In the strong-field regime, the $Q$-factorized flux also outperforms the $M$-factorized flux. 
For $a = -0.9$, it is approximately an order of magnitude more accurate at the ISCO, while for $a = 0.9$, it is roughly half an order of magnitude more accurate. 
\begin{figure}
\centering
\includegraphics[width=\columnwidth]{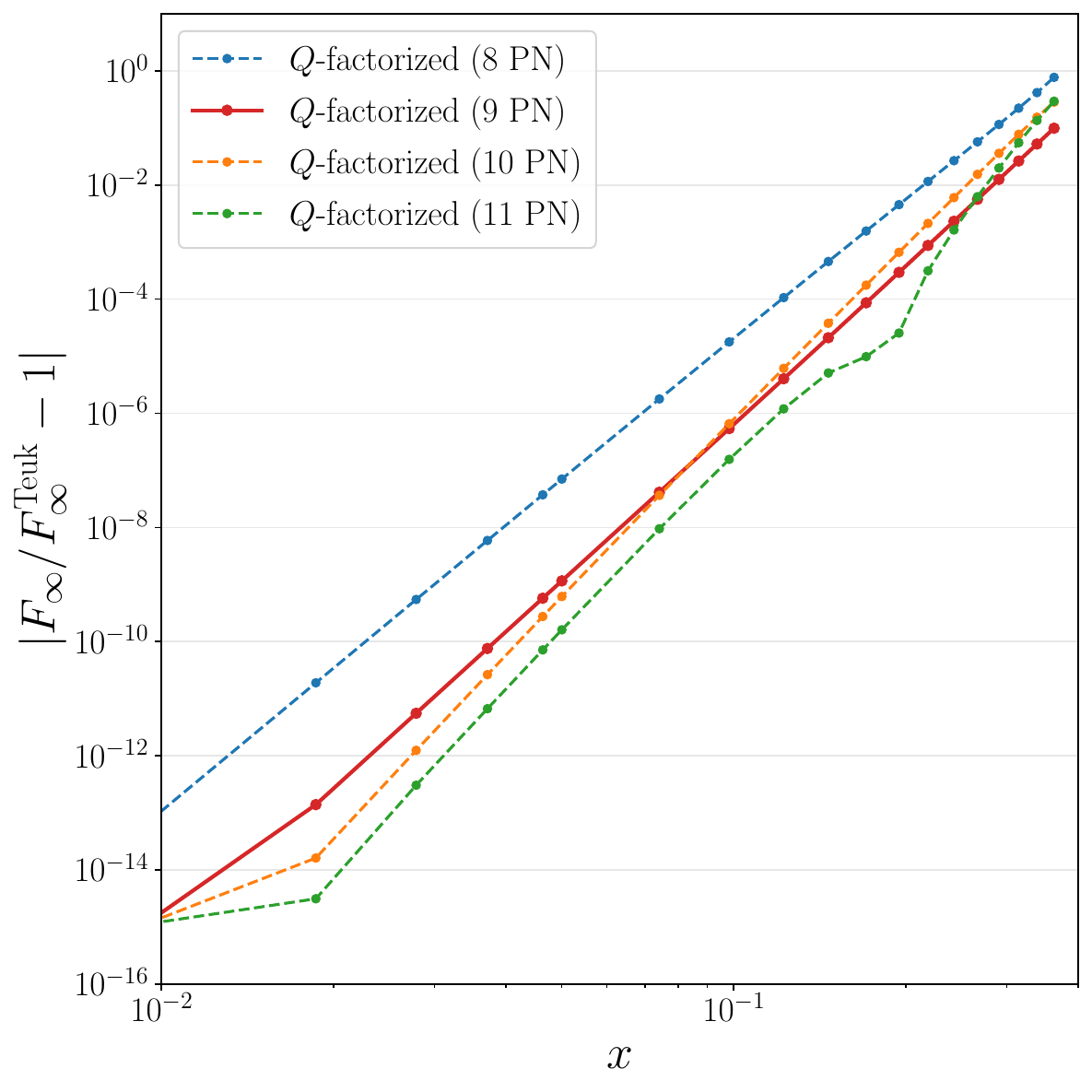} 
\caption{Absolute value of the fractional difference between $Q$-factorized fluxes at various PN orders and the FD Teukolsky flux for $a = 0.9$, all extended up to the ISCO.
While increasing the PN order beyond 9PN yields only marginal improvement at low frequencies, it degrades the accuracy in the strong-field regime; for this reason, the 9PN flux (red) is adopted in this work.}
\label{fig:steff_spin090}
\end{figure}

As previously noted, the series for $\beta(x)$ in Eq.~\eqref{eq:PNbeta} are truncated at $n=9$. 
We justify this choice by observing the behavior of the PN expansion at high orders.
Figure~\ref{fig:steff_spin090} shows the fractional difference between the $Q$-factorized flux at different PN orders and the numerical Teukolsky data for $a = 0.9$.
From 8PN to 9PN, the fractional error in the low-frequency regime decreases significantly.
However, increasing the order to 10PN or 11PN yields only minor additional improvement at low frequencies but noticeably worsens the accuracy in the strong-field regime.
We observe these features for other $a \geq 0.9$ cases as well. 
Since modeling fluxes for high prograde spins is particularly challenging, we adopt the 9PN order in this work, as it offers the best balance between low- and high-frequency accuracy.

\subsection{$Q$-factorized flux at the horizon} 
\label{sec:Q_horizon}
\begin{figure}[t]
\centering
\includegraphics[width=\columnwidth]{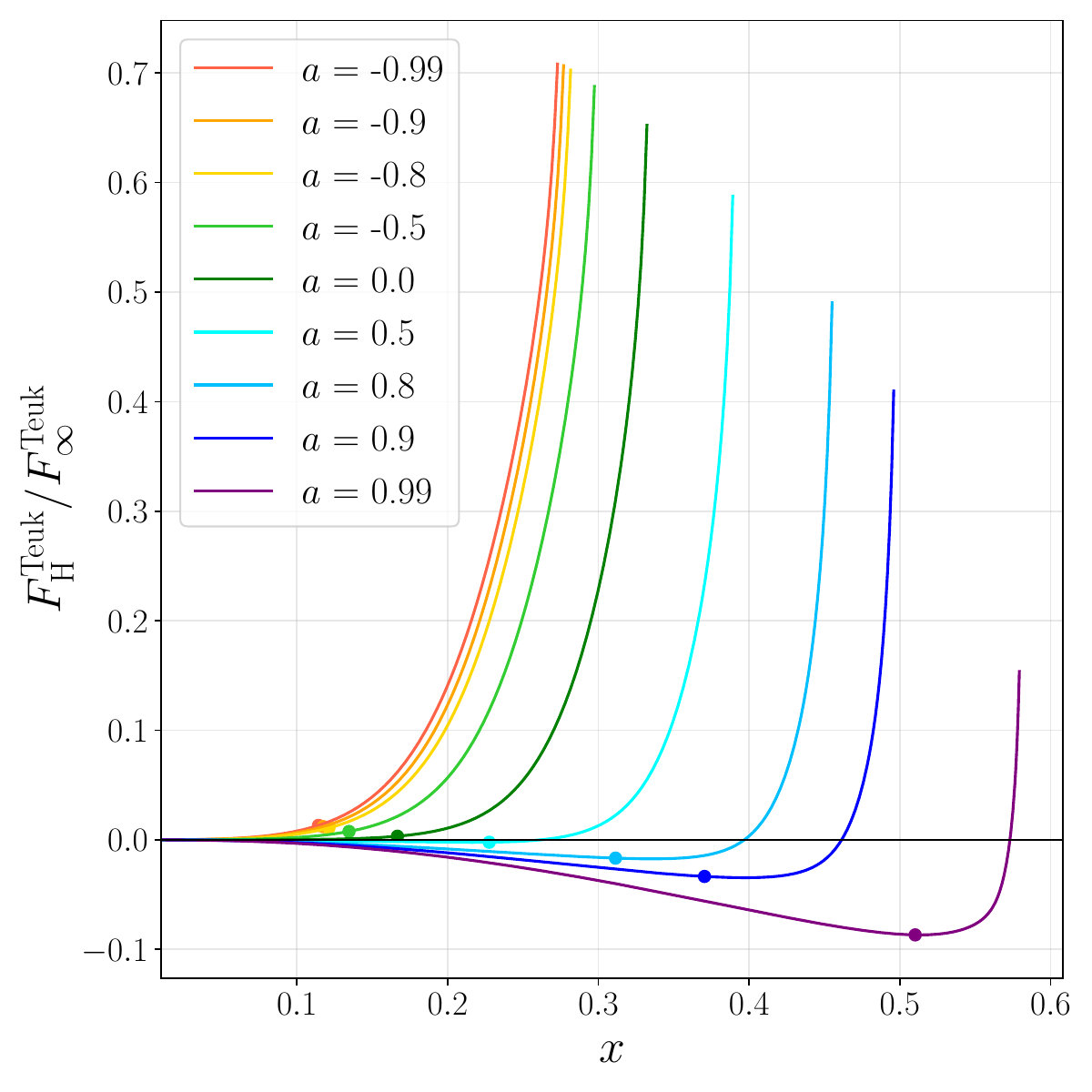} 
\caption{Ratio of the energy flux absorbed by the horizon, $F_H$, to the energy flux radiated to infinity, $F_{\infty}$, for different spin values, computed from the FD Teukolsky fluxes. 
The fluxes are extended up to $r = r_{\mathrm{LR}} + 0.01$, and the locations of the respective ISCOs are indicated by dots.}
\label{fig:FhorFinf_LR}
\end{figure}

While the majority of gravitational radiation typically escapes to future null infinity, a non-negligible fraction of the energy and angular momentum is absorbed by the BH horizon. 
This effect has been extensively studied in the literature~\cite{Tagoshi:1997jy,Taylor:2008xy,Poisson:2014gka,Chatziioannou:2012gq,Chatziioannou:2016kem,Saketh:2022xjb}. 
For comparable-mass, quasi-circular binaries, this contribution is typically negligible. 
However, in systems with extreme mass ratios~\cite{Datta:2024vll,Datta:2019epe,Maselli:2017cmm,Hughes:2001jr} or significant eccentricity~\cite{Datta:2023wsn}, the cumulative effect over long inspirals can lead to non-negligible phase shifts. 
In BH perturbation theory, horizon absorption enters at $2.5$PN order relative to the leading flux to infinity for spinning BHs, and at $4$PN order for non-spinning cases~\cite{Tagoshi:1997jy, Poisson:1994yf}.
Figure~\ref{fig:FhorFinf_LR} shows the ratio of the flux at the horizon to the flux at infinity, $F_{H}^{\text{Teuk}}/F_{\infty}^{\text{Teuk}}$, as a function of $x$ for different spin parameters, extended up to $r_{\text{min}} = r_{\mathrm{LR}} - 0.01$. 
The location of the ISCO is indicated as a dot for each spin. 
At the ISCO, the horizon flux typically contributes only a few percent of the flux to infinity for spins $a \lesssim 0.7$, but the effect grows substantially for rapidly spinning BHs, reaching almost $14\%$ in the extremal case~\cite{Colleoni:2015ena}.
A distinctive feature is that the horizon flux becomes negative for prograde spins. 
This behavior arises from the factor $(\Omega - \Omega_H)$ in the horizon flux, predicted by BH perturbation theory, where $\Omega_H$ is the horizon frequency:  
\begin{equation}
\Omega_H = \frac{a}{2 r_+}, \qquad r_+ = 1 + \sqrt{1 - a^2}.
\label{eq:OmegaH}
\end{equation}
Hence, for $a>0$ with $\Omega < \Omega_H$, the horizon flux is negative. This reflects the extraction of rotational energy from the BH through superradiance, known as a Penrose-like process in which energy is transferred from the ergosphere to the orbital motion. 
Beyond the ISCO, the orbital frequency exceeds the horizon frequency, which drives the flux back to positive values~\cite{Taracchini:2013wfa}.  

Neglecting horizon absorption can lead to significant dephasing in the waveforms. 
As Figure~\ref{fig:FhorFinf_LR} illustrates, for $a \leq 0$, the horizon flux increases the rate of dissipation, leading to a shorter inspiral. By contrast, for $a > 0$, the inversion of energy flow due to superradiance causes the inspiral to last longer. 
For instance, Ref.~\cite{Taracchini:2014zpa} compared two trajectories, one including and one excluding the horizon flux, and constructed the corresponding $(2,2)$ waveforms spanning $100$ GW cycles before the ISCO, aligned at low frequencies. 
They found that the phase difference at ringdown was approximately $-2$ rad for the non-spinning case, whereas it grew to nearly $+23$ rad for $a=0.99$.
For long inspirals relevant to space-based detectors, lasting $\sim 10^6$ cycles, the accumulated impact of horizon absorption is even more significant. 

Several EOB models have integrated horizon absorption effects into their analytical flux prescriptions.
For the Schwarzschild case, it was first implemented into the EOB framework via a resummation $F^{H}(x; \nu) = \sum_{\ell = 2}^{\ell_{\text{max}}} \sum_{m=1}^{\ell} F_{\ell m}^{H_{LO}} (x; \nu) \; \hat{S}_{\text{eff}}^2(x) \; \left( \rho_{\ell m}^H (x; \nu)\right)^{2 \ell} $ where $F_{\ell m}^{H_{LO}}$ indicates the leading order contribution to the absorbed fluxes and $\rho_{\ell m}^H$ is the residual amplitude correction given at $1^{+3}$ PN (effective 4PN approximation using the polynomial obtained Taylor-expanding up to formal 4PN) for $\ell = m = 2$ and $\ell = 2, m = 1$ cases~\cite{Nagar:2011aa}.
Later, higher modes $\rho_{\ell m}$ were included, up to $\ell = 4$~\cite{Bernuzzi:2012ku}, resulting in a reduction of the phasing error for the non-spinning case~\cite{Albertini:2022dmc}. 
In the present work, we introduce a new factorization for the horizon flux that mirrors the structural form of the $Q$-factorized flux at infinity. 

For the factorized resummation of the horizon flux, the modal energy flux is decomposed as~\cite{Taracchini:2013wfa}
\begin{equation}
\dot{E}_{\ell m}^{H} = \frac{32}{5} \nu^2 \left( 1 - \frac{\Omega}{\Omega_H} \right) 
x^{15/2} \eta_{\ell m}^{N,H} (\hat{S}_{\text{eff}})^2 (\rho_{\ell m}^H)^{2\ell}
\end{equation} 
Recall that when $0<\Omega < \Omega_H$, $\dot{E}_{H}$ becomes negative. The factor $\left( 1 - \Omega /\Omega_H\right)$ is included to account for the sign of the modal energy flux at the horizon due to superradiance.
The factor $\eta_{\ell m}^{H}$ represents the leading term and takes the form 
\begin{equation}
\eta_{\ell m}^{N}  = x^{2\left(\ell -2 \right) + \epsilon} n_{\ell m}^{H} c_{\ell m}^{H}
\end{equation}
where 
\begin{equation} 
    n_{\ell m}^{H} =
    \begin{cases}
      -\frac{5}{32} \frac{(\ell +1) (\ell +2)}{\ell (\ell - 1)} \frac{2\ell + 1}{\left( (2\ell + 1)!!\right)^2}
      \frac{(\ell - m)!}{\left( (\ell - m)!!\right)^2} \frac{(\ell + m)!}{\left( (\ell + m)!!\right)^2}, {\scriptstyle \epsilon_{\ell m}=0}\\
      -\frac{5}{8 \ell^2} \frac{(\ell +1) (\ell +2)}{\ell (\ell - 1)} \frac{2\ell + 1}{\left( (2\ell + 1)!!\right)^2}
      \frac{\left( (\ell - m)!!\right)^2}{(\ell - m)!} \frac{ \left( (\ell + m)!!\right)^2}{(\ell + m)!}, {\scriptstyle \epsilon_{\ell m}=1}, 
    \end{cases}
\end{equation}

\begin{equation}
c_{\ell m}^{H}= \frac{1}{a} \prod_{k=0}^{\ell} \left[ k^2 + \left(m^2 - k^2\right) a^2 \right],
\end{equation}
The definition for the effective source $\hat{S}_{\text{eff}}$ is the same as Eq.~\eqref{eq:Seff} and $\epsilon$ is the parity. 
The remaining factor $\rho_{\ell m}^H$ is the $2\ell$-th root of the residual amplitude of the modal energy flux. 

\begin{figure*}[t]
\centering
\includegraphics[width=\textwidth]{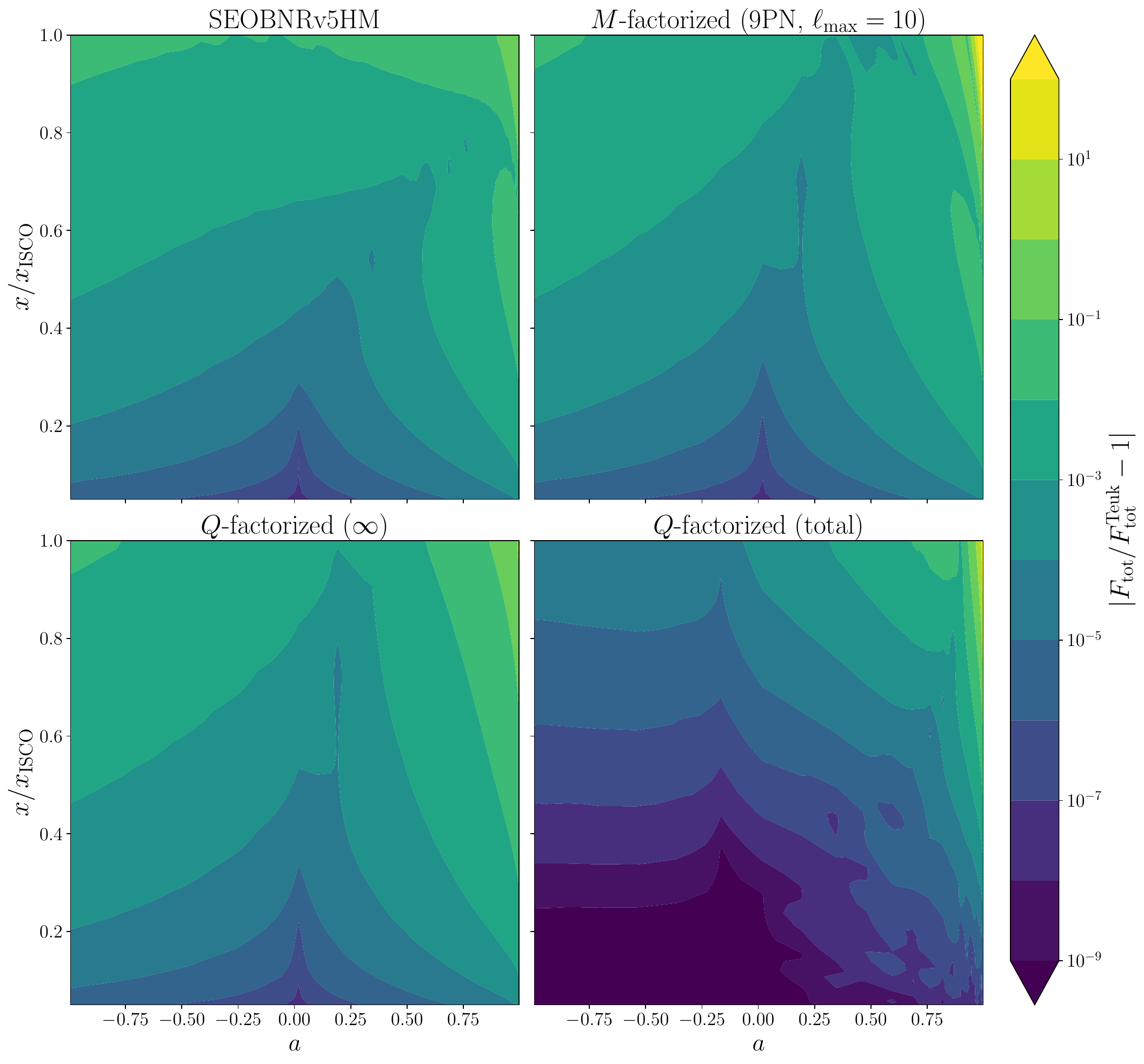}
\caption{
Fractional error in the energy flux across the spin-frequency parameter space for four analytical prescriptions.
The color map represents the absolute fractional difference $|F_{\text{tot}} / F_{\text{tot}}^{\text{Teuk}} - 1|$ relative to the numerical FD Teukolsky flux. The $x$-axis spans the dimensionless spin range $a \in [-0.993, 0.998]$, and the $y$-axis denotes the orbital frequency parameter $x$ rescaled by its value at the ISCO.
\textbf{Top left:} The \texttt{SEOBNRv5HM} flux evaluated in the TML.
\textbf{Top right:} 9PN $M$-factorized flux truncated at $\ell_{\max} = 10$. 
\textbf{Bottom left:} The proposed $Q$-factorized flux at infinity (Eq.~\eqref{eq:Steff}).
\textbf{Bottom right:} The full $Q$-factorized flux including horizon absorption (Eq.~\eqref{eq:SteffH}).
The inclusion of horizon absorption in the $Q$-factorized prescription dramatically reduces errors in the low-frequency regime, an improvement that extends to the ISCO for retrograde spins. While performance degrades for high prograde spins near the ISCO, the $Q$-factorized approach generally outperforms the $M$-factorized flux across the majority of the parameter space using significantly fewer multipolar modes.}
\label{fig:flux_comparison_contour}
\end{figure*}

\begin{table*}[t]
\caption{Relative fractional errors (in percent) of various analytical flux prescriptions compared to the numerical FD Teukolsky results. All values are evaluated at the ISCO for selected dimensionless spin parameters $a$.
For high retrograde spins ($a \approx -0.94$), the total $Q$-factorized flux provides a 2.5-order-of-magnitude improvement over the three alternative prescriptions. While the model remains high-fidelity up to relatively high prograde spins, performance degrades in the near-extremal regime ($a \gtrsim 0.95$).}
\label{tab:FluxISCOComp}
\begin{ruledtabular}
\begin{tabular}{ l c c c c }
Spin ($a$) & \multicolumn{1}{c}{\texttt{SEOBNRv5HM} (TML)} & \multicolumn{1}{c}{$M$-factorized (9PN)} & \multicolumn{1}{c}{$Q$-factorized ($\infty$)} & \multicolumn{1}{c}{$Q$-factorized (total)} \\ 
\hline
$-0.942$ & 1.48\%  & 1.24\%  & 1.24\%  & 0.00395\% \\
$0.019$  & 1.15\%  & 0.332\%  & 0.287\%  & 0.0259\% \\
$0.900$  & 10.8\% & 16.0\% & 15.7\% & 0.685\% \\
$0.959$  & 30.2\% & 955\% & 43.9\% & 70.3\% \\
\end{tabular}
\end{ruledtabular}
\end{table*}

To ensure consistency with our $Q$-factorized treatment of the flux at infinity in Eq.~\eqref{eq:Steff}, we model the horizon flux using only the dominant (2,2) mode weighted by a correction factor $\alpha(x)$:
\begin{align}
\frac{1}{\nu^2} \dot{E}_H &= \frac{1}{\nu^2}\dot{E}_{22H} \; \alpha (x)\nonumber \\
        &= \frac{32}{5} x^{15/2} \left( 1 - \frac{\Omega}{\Omega_H} \right) 
\eta_{22}^{N,H} \hat{S}_{\text{eff}}^2 (\rho_{22}^H)^{4} \alpha(x) \, ,
\label{eq:SteffH}
\end{align}
Following the form of $\beta(x)$ in Eq.~\eqref{eq:PNbeta}, this multiplicative factor is defined as:
\begin{equation}
\alpha(x) = 1 + \sqrt{x} \, \alpha_{\frac{1}{2}} + x \, \alpha_{1} + x^{\frac{3}{2}} \, \alpha_{\frac{3}{2}} + \dots +  x^n \, \alpha_{n} .
\label{eq:PNalpha}
\end{equation}
In order to obtain the $\alpha$ coefficients, we expand the RHS of Eq.~\eqref{eq:SteffH} evaluated in the circular limit, in terms of $x$ and match it with the PN-expanded horizon flux at each PN order.
This PN-expanded horizon flux is known up to 11PN for Kerr BHs and 22.5PN for Schwarzschild BHs~\cite{Fujita:2014eta}. 
For $\rho_{22}^H$, we adopt the expression extended up to 5PN order in the TML, and match the LHS and RHS up to 6.5PN order relative to $F_{\infty}^{N}$ (4PN order relative to $F_{H}^N$), which results in $\alpha (x)$ including up to $\alpha_{4}$.
This choice of PN order provides the best agreement with numerical fluxes in the strong-field region. Since higher PN terms offer diminishing improvements at an increased computational cost, this represents a practical compromise between accuracy and efficiency.

\subsection{Total flux comparison}
\label{sec:total_flux_comparison}
Finally, we assess the performance of the proposed $Q$-factorized prescriptions by comparing them against the FD Teukolsky fluxes generated using ModGEMS~\cite{Honet:2025lmk}.
We assess four different analytical fluxes: 
\begin{itemize}
    \item \textbf{SEOBNRv5HM (TML):} The $M$-factorized flux prescription of the \texttt{SEOBNRv5HM} model, evaluated at the TML ($\nu = 0$).
    \item \textbf{$M$-factorized (9PN):} The traditional mode-sum factorized flux calculated to $9$PN order and truncated at a high multipolar order of $\ell_{\max} = 10$.
    \item \textbf{$Q$-factorized ($\infty$):} The prescription for the flux at infinity defined in Eq.~\eqref{eq:Steff}. The multiplicative parameter $\beta(x)$ is determined by matching up to $9$PN order.
    \item \textbf{$Q$-factorized (total):} The full prescription which combines the $9$PN $Q$-factorized flux at infinity with the horizon absorption flux described in Eq.~\eqref{eq:SteffH}.
\end{itemize}

Figure~\ref{fig:flux_comparison_contour} illustrates the absolute fractional difference $|F_{\text{tot}} / F_{\text{tot}}^{\text{Teuk}} - 1|$ between these analytical models and the FD Teukolsky fluxes across the spin-frequency parameter space.
To facilitate comparison across all spins, the orbital frequency $x$ is rescaled by its value at the ISCO. A standout feature of these results is that the inclusion of horizon absorption in the $Q$-factorized prescription yields a dramatic reduction in error for the low-frequency and retrograde-spin regimes.
Specifically, in the weak-field regime, the absorption contribution is essential for achieving high-fidelity agreement, as seen in the bottom-right panel. For retrograde spins, this improvement persists up to the ISCO; at $a  \approx -0.95$, the total $Q$-factorized flux provides a 2.5-order-of-magnitude improvement over the other three prescriptions (see Table~\ref{tab:FluxISCOComp}).
Naively, one might expect the improvement to be on the order of the horizon flux contribution itself—approximately $0.3\%$ at $a =0$ and $1.2\%$ for $a =-0.9$, as shown in Figure~\ref{fig:FhorFinf_LR}. However, the $Q$-factorized flux at infinity already possesses small residual errors that happen to be of similar magnitude but opposite sign of the horizon contribution. As a result, the two effects partially cancel near the ISCO, leading to a total error that is significantly lower than that of either component alone.

While the relative gains are less pronounced for intermediate prograde spins ($0 \lesssim a  \lesssim 0.8$), the total $Q$-factorized flux still exhibits a marked improvement in the low-frequency regime compared to the traditional $M$-factorized flux.
As the orbital frequency increases toward the ISCO, the performance of the $Q$-factorized flux becomes comparable to the $9$PN $M$-factorized flux, maintaining a similar error profile in the strong-field region.
Crucially, our prescription achieves this level of accuracy while relying solely on a dominant $(2,2)$ mode baseline, whereas the traditional approach requires a computationally expensive summation up to $\ell_{\max} = 10$. 
This demonstrates that the $Q$-factorized approach offers a significantly more efficient computational path without sacrificing accuracy across the majority of the parameter space.
We note, however, that in the extremal prograde regime ($a  \gtrsim 0.95$), the model's performance degrades, with errors reaching $\sim 70\%$ at the ISCO, as summarized in Table~\ref{tab:FluxISCOComp}. This extremal regime is discussed further in Appendix~\ref{app:extremal}.

Despite these limitations in the high-prograde regime, the $Q$-factorized prescription provides a substantial improvement in accuracy in the TML.
By effectively capturing the interplay between radiation to infinity and to horizon through a single quadrupole-based framework, the model eliminates the need for expensive multipolar summations while reaching sub-percent accuracy across the vast majority of the inspiral parameter space. 
These results demonstrate that $Q$-factorization is a robust and computationally efficient candidate for enhancing analytical flux formulations within the EOB framework.

\section{Modeling the inspiral-plunge waveform}
\label{sec:ModelingInspPlunge}
Up to this point, we have focused on improving the EOB dynamics through the implementation of new analytical fluxes.
In this section, we turn our attention to the construction of the multipolar waveform modes $h_{\ell m}$ for spinning, non-precessing quasi-circular binaries. 
Our approach builds upon the framework established in \texttt{SEOBNRv5HM}~\cite{Pompiliv5}, while highlighting the modifications and improvements introduced in our proposed model, \texttt{SEOB-TML}.

\subsection{Attachment time}
\label{sec:AttachmentTime}
In the EOB framework, the GW modes are further decomposed into inspiral-plunge and merger–RD sectors. 
\begin{equation} 
    h_{\ell m}(t) =
    \begin{cases}
        h_{\ell m}^{\text{insp-plunge}}(t), & t < t_{\text{match}}^{\ell m} \\
        h_{\ell m}^{\text{MR}}(t), & t > t_{\text{match}}^{\ell m}
    \end{cases}
\end{equation}
The attachment time $t_{\text{match}}^{\ell m}$ represents a conventional choice aimed at ensuring a smooth transition between the two modeling sectors.
For example, in \texttt{SEOBNRv5HM}, the merger–RD part is attached at the peak of the $(2,2)$ mode amplitude for all modes except $(5,5)$. 
In this work, we adopt a more flexible, mode-dependent prescription to account for the physical complexities of the merger. 

During the merger–RD phase, the radiation emitted by the remnant BH is described by a superposition of QNMs~\cite{Kokkotas:1999bd}.
These modes are characterized by discrete complex frequencies, labeled by angular numbers $(\ell, m)$ and the overtone index $n=0, 1, \dots$, where $n=0$ denotes the fundamental mode.
At the transition to the ringdown, two distinct mechanisms cause a given $(\ell, m)$ spherical harmonic mode to be composed of a superposition of QNMs beyond the primary $(\ell, m, n)$ state.
First, a basis mismatch arises because the waveform strain is usually decomposed in terms of $-2$ spin-weighted spherical harmonics, whereas the Teukolsky equation separates naturally into $-2$ spin-weighted spheroidal harmonics. This mismatch introduces mixing between modes with the same $m$ but different $\ell$.
Second, mixing is induced by the so-called retrograde modes, whose frequency equals the (negative complex conjugate of) the $(\ell,-m,n)$ state. 
These are excited when the orbital frequency changes sign during the last stage of plunge~\cite{Taracchini:2014zpa}. 
A detailed analysis of both effects will be presented in Sec.~\ref{sec:ModeMixing}. 
Here we note only that, for $\ell \neq m$ cases with negative spins, the onset of mixing occurs promptly, to the extent that even the amplitude peak is affected. 
\begin{figure}[t]
\centering
\includegraphics[width=\columnwidth]{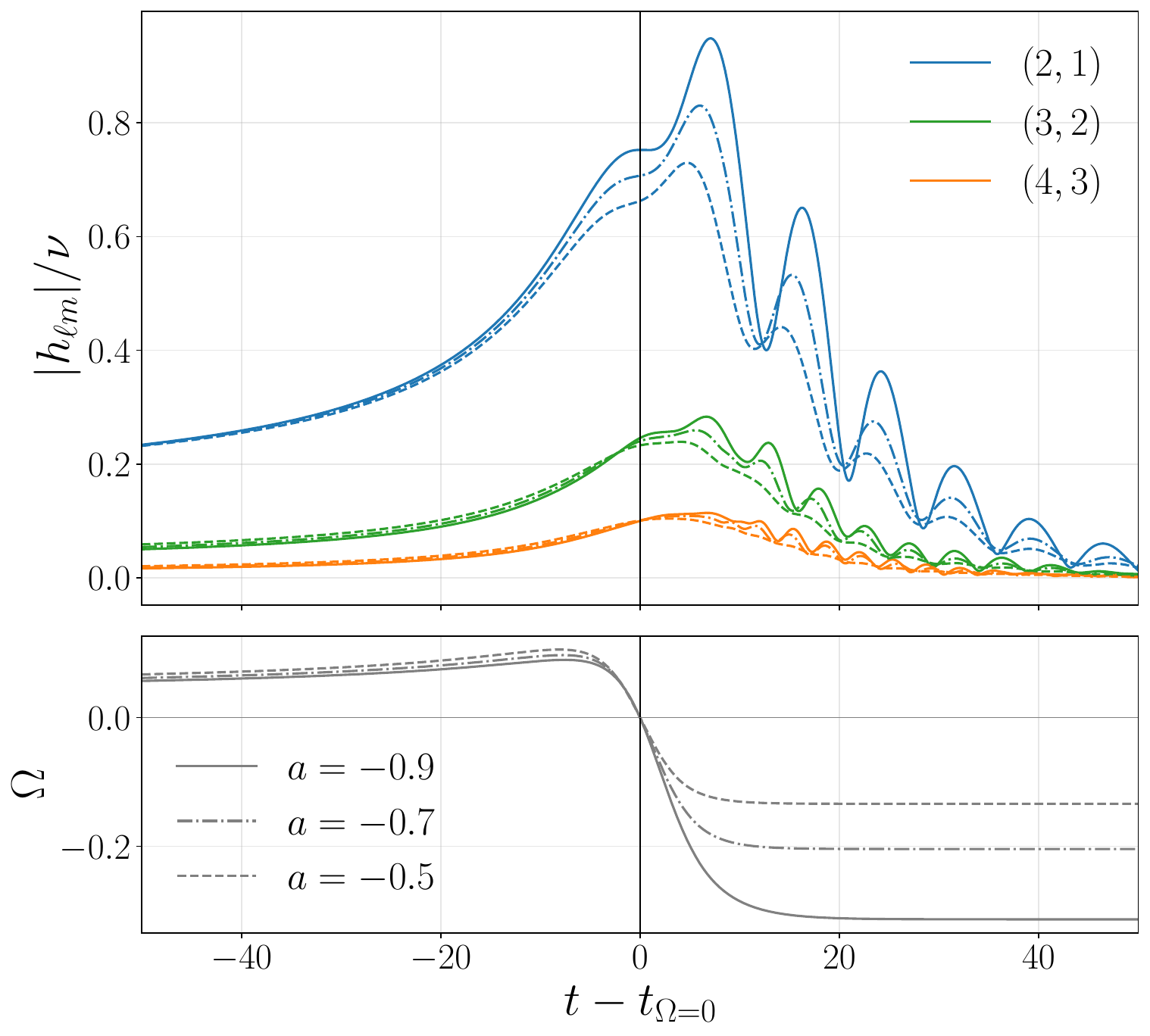} 
\caption{Evolution of the $\ell \neq m$ mode amplitudes (top panel) and the orbital frequency $\Omega$ (bottom panel) for various negative-spin configurations ($a = -0.5, -0.7, -0.9$). Solid, dash-dotted, and dashed lines correspond to decreasing spin magnitudes, as indicated in the legend. The time axis is centered at $t=0$, corresponding to the zero-crossing of $\Omega$.} 
\label{fig:h21attach}
\end{figure}

Figure~\ref{fig:h21attach} displays the amplitudes of the $\ell \neq m$ modes in the top panel and the evolution of the orbital frequency $\Omega$ in the bottom panel. Across both panels, different linestyles are used to distinguish specific spin magnitudes: $a = -0.9$ (solid), $a = -0.7$ (dash-dotted), and $a = -0.5$ (dashed). The time axis is shifted such that $t=0$ denotes the instant when the orbital frequency crosses zero.
Notably, the amplitude modulation caused by mode mixing begins almost simultaneously with $\Omega=0$ for the $(2,1)$ mode. For the $(3,2)$ and $(4,3)$ modes, the onset of mixing occurs with a delay of approximately $2.5$ and $3.5$, respectively. 
While the specific offset varies slightly with the BH spin, this dependence is negligible; consequently, we employ spin-independent offsets throughout this work.
Since the last stage of the inspiral–plunge is modeled using phenomenological hyperbolic functions, it is unable to reproduce the complex amplitude modulations introduced by mode mixing. We therefore defer the description of these features to the merger–RD sector. 

Based on these considerations, we define the attachment time in this work as 
\begin{align}
t_{\text{match}}^{\ell m} (a \geq 0) &= t^{\ell m}_{\text{peak}}, \quad \text{for all modes}, \nonumber \\[10pt]
t_{\text{match}}^{\ell m} (a < 0) &=
    \begin{cases}
        t^{\ell m}_{\text{peak}}, & (2,2), (3,3), (4,4), (5,5) \\[6pt]
        t^{\ell m}_{\Omega=0}, & (2,1), \\[6pt]
        t^{\ell m}_{\Omega=0} + 2.5, & (3,2), \\[6pt]
        t^{\ell m}_{\Omega=0} + 3.5, & (4,3).
    \end{cases}
\label{eq:AttachmentTime}
\end{align}
where $t^{\ell m}_{\text{peak}}$ denotes the time of the $(\ell,m)$ mode amplitude peak, and $t^{\ell m}_{\Omega=0}$ is the time when the orbital frequency crosses zero.  
Within the EOB framework, the attachment time is typically expressed in terms of the time interval $\Delta t^{\ell m}$ with respect to a reference time associated with a particular configuration of the dynamics of the binary~\cite{Cotesta:2018fcv, Riemenschneider:2021ppj, Pompiliv5}.
For example, \texttt{SEOBNRv5HM} defines the attachment time via the calibration parameter $\Delta t^{\ell m}_{\text{ISCO}}$: 
\begin{equation}
t^{\ell m}_{{\text{match}}}= t_{\text{ISCO}} + \Delta t^{\ell m}_{\text{ISCO}}
\end{equation}
where $t_{\text{ISCO}}$ is the time at which $r = r_{\text{ISCO}}$.
This ISCO-based prescription is not well suited for the TML. 
When the spin is large and negative, the particle continues to plunge for a long time after passing the ISCO. By contrast, for large positive spins, the plunge is rapid and, in some cases, the mode amplitude even peaks before the ISCO.  
For example, when $a=-0.95$ ($a=0.95$), the orbital frequency at the ISCO is $\Omega_{\text{ISCO}}=0.039$ ($0.27$), and the radial separation between the ISCO and the light ring is $4.90$ ($0.55$), yielding $\Delta t^{\ell m}_{\text{ISCO}} \simeq 455$ ($-25$).  
Thus, the low orbital frequency at the ISCO together with a large ISCO–horizon separation lead to significant delays in the peak amplitude for large negative spins, while the opposite occurs at large positive spins.
To avoid this wide variation across spins, we adopt the orbital frequency peak as a more robust reference time~\cite{Barausse:2011kb}.  
\begin{equation}
t^{\ell m}_{{\text{match}}}= t^{\Omega}_{\text{peak}} + \Delta t^{\ell m}_{\Omega \text{peak}}
\label{eq:attachOmegapeak}
\end{equation}
To determine $\Delta t^{\ell m}_{\Omega\text{peak}}$, we extract the mode-dependent matching times from TD Teukolsky waveforms and the peak orbital frequency times from trajectories computed employing FD Teukolsky fluxes. These values are then interpolated as a function of spin.

A subtlety arises when the spin approaches the nearly-extremal regime ($a \gtrsim 0.95$).  
In this case, the amplitude of several modes becomes almost flat, which makes it extremely difficult to robustly determine the time of the peak~\cite{Taracchini:2014zpa, Faggioli:2025hff}.
Physically, this is explained by the fact that the ISCO approaches the horizon as $a \to 1$, leading to a very short plunge following an extended quasi-circular inspiral, causing the amplitude peak to occur many cycles before the plunge and merger stage~\cite{Gralla:2016qfw}.
This indicates that the mode amplitude peak does not universally correspond to the merger stage across the entire parameter space~\cite{Faggioli:2025hff}.
To prevent this unreliable peak location at $a>0.95$ from contaminating the spin fits, we freeze the calibration at $a=0.95$ and use the same value of $\Delta t^{\ell m}_{\Omega\text{peak}}$ for all larger spins.

\subsection{Inspiral-plunge state}
\label{sec:intermediate}
\begin{figure}[t]
\centering
\includegraphics[width=\columnwidth]{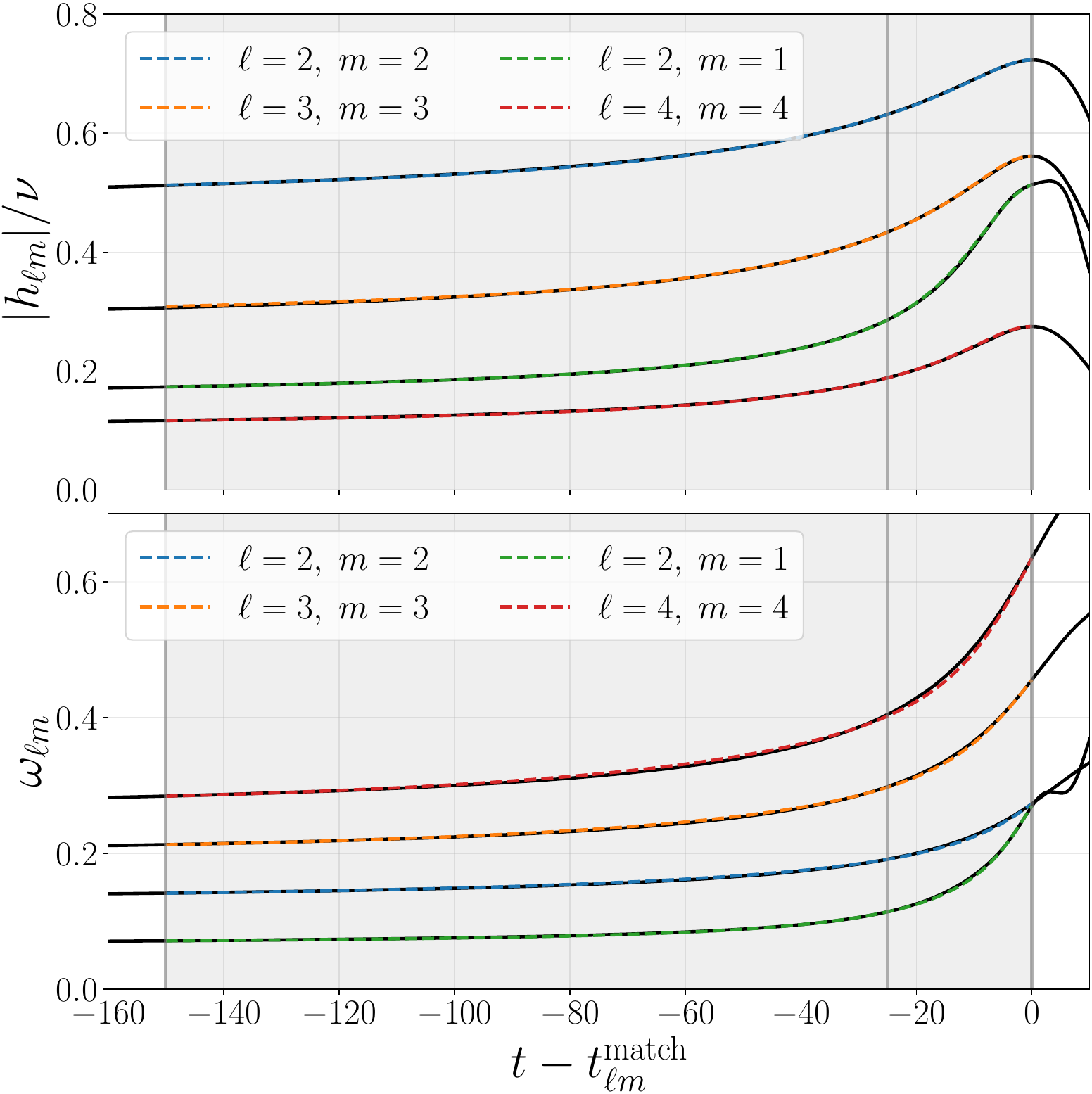}  
\caption{Amplitude (top) and frequency (bottom) of the four dominant modes for the non-spinning case. 
Colored dashed lines show the waveform modeled with the hyperbolic ansatz, while black solid lines show the TD Teukolsky waveforms. 
The shaded gray region indicates the time interval over which the hyperbolic ansatz is applied, with vertical grey lines marking $t_{\text{cut}}$, $t_{\text{cp}}$, and $t_{\text{match}}$.}
\label{fig:hypa0comp}
\end{figure}
As discussed in Sec.~\ref{sec:EOBdynamics}, improving the strong-field behavior of the PN-expanded waveform within the EOB formalism requires each waveform mode to be factorized and resummed, as prescribed in Eq.~\eqref{eq:hlmF}.
The residual amplitude corrections $\rho_{\ell m}$ are central building blocks in any EOB waveform model and require careful analytical treatment. 
As they are originally defined as PN series, their accuracy often degrades in the strong-field regime. 
To improve their behavior in the TML, various resummation strategies for $\rho_{\ell m}$ have been employed in the literature, such as Padé or inverse-Taylor constructions, notably in the \texttt{TEOBResum} family~\cite{Damour:2014sva,Nagar:2015xqa,Gamba:2024cvy,Albanesi:2025txj,Nagar:2018zoe,Nagar:2019wds,Nagar:2020pcj,Riemenschneider:2021ppj,Chiaramello:2020ehz,Gamba:2021ydi,Nagar:2018gnk,Rettegno:2019tzh,Akcay:2020qrj,Nagar:2018plt, Nagar:2024oyk, Albanesi:2024xus}.
However, such resummation prescriptions typically require mode-dependent choices of resummation order and structure. 
Therefore, in this work, for the modes $(2,2)$, $(2,1)$, $(3,2)$, $(3,3)$, $(4,3)$, $(4,4)$, and $(5,5)$, we truncate the PN expansion of $\rho_{\ell m}$ at 5PN order and introduce calibration parameters at 5.5PN and 6PN, obtained by fitting to numerical $\rho_{\ell m}$ data extracted from TD Teukolsky waveforms.
In particular, as discussed in Sec.~\ref{sec:RRForce}, since the analytical EOB flux does not depend explicitly on $\rho_{\ell m}$ for each mode, their primary role is to provide an accurate description of the early-inspiral waveform amplitudes, for which this calibrated truncation is sufficient.
For the phase $\delta_{\ell m}$, we use the same expression as \texttt{SEOBNRv5HM}, while setting $\nu=0$. 

Even with these treatments, the factorized waveform in Eq.~\eqref{eq:hlmF} becomes increasingly inaccurate as the system approaches the plunge.
To capture deviations from quasi-circular motion during the late inspiral and plunge, \texttt{SEOBNRv5HM} employs non–quasi-circular (NQC) corrections such that the inspiral–plunge waveform is modified as
\begin{equation}
    h_{\ell m}^{\text{insp-plunge}} = h_{\ell m}^{F} \, N_{\ell m}.
\end{equation}
where the NQC factor $N_{\ell m}$ is given by
\begin{align}
N_{\ell m} &= \left[ 1 + \frac{p_{r_*}^2}{(r \Omega)^2}
\left( a_1^{h_{\ell m}} + \frac{a_2^{h_{\ell m}}}{r} + \frac{a_3^{h_{\ell m}}}{r^{3/2}} \right) \right] \nonumber \\
&\quad \times \exp \left[ i \left( b_1^{h_{\ell m}} \frac{p_{r_*}}{r \Omega} + b_2^{h_{\ell m}} \frac{p_{r_*}^3}{r \Omega} \right) \right], \tag{36}
\end{align}
The NQC corrections are calibrated so that the waveform’s amplitude and frequency agree with numerical waveform values at the attachment point $t_{\text{match}}^{\ell m}$. 
In practice, the five constants 
($a_1^{h_{\ell m}}, a_2^{h_{\ell m}}, a_3^{h_{\ell m}}, b_1^{h_{\ell m}}, b_2^{h_{\ell m}}$) are fixed by imposing continuity conditions at this matching point.  
During the early inspiral, where $|p_{r_*}| \ll r\,\Omega$, the NQC factor naturally approaches unity.  
As the system evolves toward merger, the NQC correction is gradually activated, ensuring that it only modifies the waveform in the strong-field regime and avoids introducing spurious effects during the early inspiral. 
In \texttt{SEOBNRv5HM}, this construction is appropriate because the merger–RD attachment is performed near the peak of the $(2,2)$ mode amplitude for most modes. 
In our model, however, directly attaching the inspiral-plunge waveform using the NQC is less suitable. 
For positive spins, the attachment is performed at the peak of each individual mode, resulting in $\Delta t_{\ell m} = t^{\ell m}_{\text{peak}} - t^{22}_{\text{peak}} > 0$ in most cases. 
At such late times, some EOB dynamical quantities become unreliable.
Furthermore, as explained in Sec.~\ref{sec:AttachmentTime}, for certain modes with negative spins, the attachment occurs after the orbital frequency $\Omega$ has already crossed zero (e.g. $t_{43}^{\text{match}} = t_{\Omega = 0} + 3.5$), 
rendering the NQC basis ill-defined and causing unphysical oscillations. 

\begin{figure}[t]
\centering
\includegraphics[width=\columnwidth]{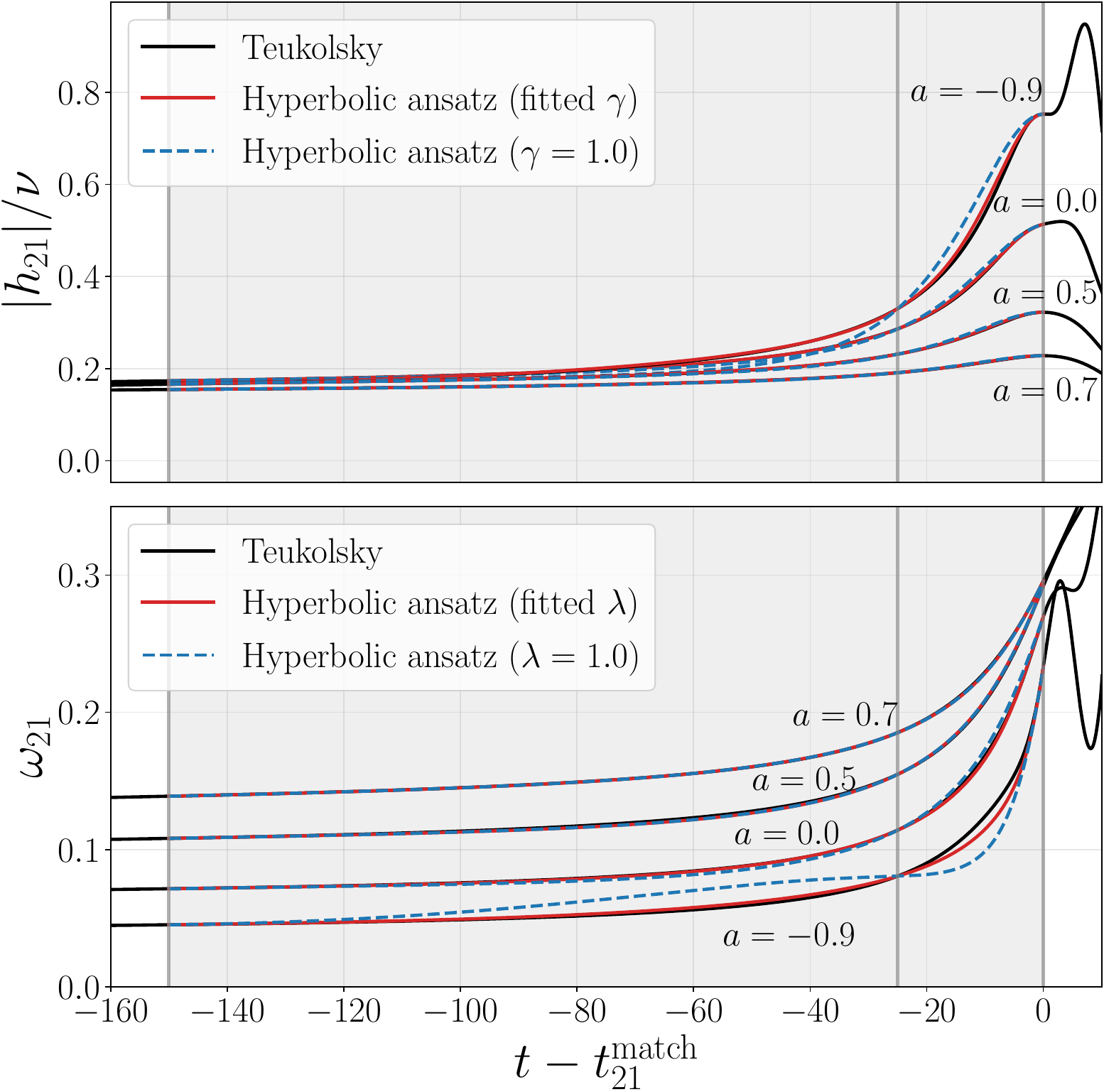} 
\caption{Amplitude (top) and gravitational frequency (bottom) of the $(2,1)$ mode during the inspiral–plunge for different spin values. 
The black solid line shows the TD Teukolsky waveform, the dashed line shows the hyperbolic ansatz (Eq.~\eqref{eq:hyp_amp} and Eq.~\eqref{eq:hyp_omega}) with original parameters $\gamma = \lambda = 1$, and the colored solid line shows the hyperbolic ansatz with fitted $\gamma$ and $\lambda$. 
The shaded gray region indicates the time interval over which the hyperbolic ansatz is applied, with vertical lines marking $t_{\text{cut}}$, $t_{\text{cp}}$, and $t_{\text{match}}$. 
Fitting $\gamma$ and $\lambda$ improves agreement with the Teukolsky waveform, particularly for the negative-spin cases.
}
\label{fig:hyp21mode}
\end{figure}

To circumvent these issues, we model the final stage of the inspiral-plunge without the NQC corrections.  
Following the approach in Ref.~\cite{Estelles:2020twz}, we instead employ phenomenological expressions based on the hyperbolic functions used in \texttt{IMRPhenomT}.
The waveform amplitude $H_{\ell m}^{\text{inter}}$ and gravitational frequency $\omega_{\ell m}^{\text{inter}}$ for the intermediate region is expressed as
\begin{equation}
\begin{split}
H_{\ell m}^{\text{inter}}(t) =\; & b_{0} + b_{1} \tau_{\ell m}^{2} + b_{2} \,\text{sech}^{1/7} \left( 2 \sigma_{\ell m 0}^I \;\tau_{\ell m} \right) \\
& + b_{3} \,\text{sech}\left(\gamma\; \sigma_{\ell m 0}^I \;\tau_{\ell m}  \right)+ b_4 e^{\tau_{\ell m}/4},
\label{eq:hyp_amp}
\end{split}
\end{equation}
\begin{equation}
\omega_{\ell m}^{\text{inter}}(t) = \sigma_{\ell m 0}^R \left[ 1 - \underset{k=0}{\overset{4}{\sum}} a_{k} \,\text{arcsinh}^{k} \left( \lambda \, \sigma_{\ell m 0}^I \, \tau_{\ell m}\right) \right],
\label{eq:hyp_omega}
\end{equation}
where $\tau_{\ell m} = t - t_{\ell m}^{\text{match}}$, $\sigma_{\ell m n} = \sigma_{\ell m n}^R - i \sigma_{\ell m n}^I$ with $\sigma_{\ell m n}$ denoting the QNM frequencies for the $(\ell,m)$ mode with overtone $n$.
The complex strain is then reconstructed as $h_{\ell m}^{\text{inter}}(t) = H_{\ell m}^{\text{inter}}(t)\; e^{i \phi_{\ell m}^{\text{inter}}(t)}$, with the phase defined by
\begin{equation}
\phi_{\ell m}^{\text{inter}}(t) = \phi_{\ell m}^{\text{insp-plunge}}|_{t_{\text{cut}}} + \int_{t' = t_{\text{cut}}}^{t} dt' \omega_{\ell m}^{\text{inter}}(t')
\label{eq:hypPhaseOffset}
\end{equation}
The phenomenological coefficients $a_i$ and $b_i$ are fixed by enforcing continuity and differentiability with the EOB inspiral waveform and with numerical input values, specifically the amplitude and frequency data extracted from Teukolsky waveforms.
Matching is performed at three points: the cutoff time of the factorized waveform, $t_{\text{cut}} = t_{\ell m}^{\text{match}} - 150$; an intermediate collocation point, $t_{\text{cp}} = t_{\ell m}^{\text{match}} - 25$; and the merger–RD attachment time, $t_{\ell m}^{\text{match}}$. 
The specific choices of $t_{\text{cut}}$ and $t_{\text{cp}}$ are empirical, providing a sufficiently wide window for the phenomenological ansatz while incorporating numerical waveform information for accurate late inspiral–plunge modeling.

Eqs.~\eqref{eq:hyp_amp} and~\eqref{eq:hyp_omega} define a linear system that uniquely determines all phenomenological coefficients. 
At $t_{\text{cut}}$, the ansatz is constrained to agree with the factorized EOB waveform: 
\begin{subequations}\label{eq:H_omega_defs}
\begin{align}
    H_{\ell m}^{\text{inter}} &= H_{\ell m}^{F}\big|_{t = t_{\text{cut}}}, \quad 
    \dot{H}_{\ell m}^{\text{inter}} = \dot{H}_{\ell m}^{F}\big|_{t = t_{\text{cut}}},\\
    \omega_{\ell m}^{\text{inter}} &= \omega_{\ell m}^{F}\big|_{t = t_{\text{cut}}}, \quad 
    \dot{\omega}_{\ell m}^{\text{inter}} = \dot{\omega}_{\ell m}^{F}\big|_{t = t_{\text{cut}}},
\end{align}
\end{subequations}
At $t_{\text{cp}}$ and $t_{\ell m}^{\text{peak}}$, the ansatz is matched to the corresponding Teukolsky waveform values: 
\begin{subequations}\label{eq:H_omega_defs2}
\begin{align}
    H_{\ell m}^{\text{inter}} &= H_{\ell m}^{\text{Teuk}}|_{t = t_{\text{cp}}}, \quad 
    H_{\ell m}^{\text{inter}}  = H_{\ell m}^{\text{Teuk}}|_{t = t_{\text{match}}},\\
    \dot{H}_{\ell m}^{\text{inter}} &= \dot{H}_{\ell m}^{\text{Teuk}}|_{t = t_{\text{match}}}, \\
    \omega_{\ell m}^{\text{inter}} &= \omega_{\ell m}^{\text{Teuk}}|_{t = t_{\text{cp}}}, \quad 
    \omega_{\ell m}^{\text{inter}} = \omega_{\ell m}^{\text{Teuk}}|_{t = t_{\text{match}}}, \\
    \dot{\omega}_{\ell m}^{\text{inter}} &= \dot{\omega}_{\ell m}^{\text{Teuk}}|_{t = t_{\text{match}}}.
\end{align}
\end{subequations}
Figure~\ref{fig:hypa0comp} shows a comparison of the amplitude and frequency of the four dominant modes—$(2,2)$, $(3,3)$, $(2,1)$, and $(4,4)$—constructed using the hyperbolic ansatz (colored dashed curves) against the corresponding Teukolsky waveforms (black solid curves) for the non-spinning case. 
The shaded region marks the time interval where Eq.~\eqref{eq:hyp_amp} and Eq.~\eqref{eq:hyp_omega} are applied, and the vertical lines mark $t = t_{\text{cut}}, t = t_{\text{cp}}$, and $t = t_{\text{match}}$.
Within this window, the phenomenological hyperbolic ansatz reproduces both the amplitude and frequency of the Teukolsky waveforms with good accuracy, demonstrating its effectiveness in modeling the late inspiral–plunge regime prior to merger. 

There are two main differences between our implementation and the original phenomenological ansatz of Ref.~\cite{Estelles:2020twz}.  
First, we have implemented an additional parameter $b_4$ in the amplitude ansatz.  
In their work, the attachment was always performed at the peak of each mode, yielding a vanishing amplitude derivative at $\tau_{\ell m} = 0$. 
In our model, however, the attachment time is modified for certain modes, particularly $\ell \neq m$ modes for the negative spin cases, where $t_{\ell m}^{\text{match}}$ does not necessarily coincide with the amplitude peak. 
To accommodate a nonzero derivative at the attachment, we introduce an additional term $b_4 e^{\tau_{\ell m}/4}$. 
This modification affects only the very late inspiral region, producing a nonzero derivative at $\tau_{\ell m}=0$ without degrading the behavior of the original ansatz. 
In cases where the attachment time is not shifted from the amplitude peak, we recover the original ansatz by setting $b_4=0$.
Second, in the original ansatz, the parameters $\gamma$ in Eq.~\eqref{eq:hyp_amp} and $\lambda$ in Eq.~\eqref{eq:hyp_omega} were fixed to unity. 
We find that, in TML, this choice is insufficient to reproduce the amplitude and gravitational frequency accurately for modes characterized by a steep growth in the late plunge, most notably the $(2,1)$ mode at large negative spins.

Figure~\ref{fig:hyp21mode} shows the $(2,1)$ mode during the inspiral–plunge, comparing three cases: the numerical Teukolsky waveform (black solid), the hyperbolic ansatz with the original parameters $\gamma = \lambda = 1$ (dashed), and the hyperbolic ansatz with $\gamma$ and $\lambda$ fitted to improve agreement (colored solid).
The comparison demonstrates that the original ansatz underestimates both the amplitude and the frequency evolution.
By allowing $\gamma$ and $\lambda$ to vary, the model captures the $(2,1)$ mode more accurately, particularly for retrograde spins where the signal evolves more steeply near the attachment time. While the default parameters remain sufficient for prograde cases, large negative spins require an increase in both $\gamma$ and $\lambda$. The optimized coefficients are subsequently interpolated and fitted as a function of spin for implementation in the model.

Overall, implementing the hyperbolic ansatz in the late inspiral–plunge instead of the traditional NQC corrections offers several key advantages. First, it allows the attachment of the merger–RD sector to occur after the peak of the $(2,2)$ mode. This flexibility is crucial, as a mode-dependent attachment point improves the modeling of the merger–RD transition for higher-order modes (see Sec.~\ref{sec:ModelingMergerRingdown} for further details).
Furthermore, while the NQC approach matches numerical inputs only at the attachment time, the hyperbolic ansatz includes an additional collocation point, $t_{\text{cp}}$, prior to attachment. By utilizing more numerical data points, this approach better constrains the late-plunge behavior and ensures a smoother transition to the merger–RD sector.

\section{Modeling merger–RD waveform}
\label{sec:ModelingMergerRingdown}
We now turn to the final stage of the waveform: the merger–RD.
\texttt{SEOBNRv5HM} is calibrated to a large set of spin-aligned, quasi-circular NR simulations covering comparable-mass binaries. However, it incorporates limited information from the TML, relying primarily on attachment-time values from 13 TD Teukolsky waveforms~\cite{Barausse:2011kb,Taracchini:2014zpa}.
This limited reliance explains why \texttt{SEOBNRv5HM} performs poorly in the TML, particularly for large positive or negative spins.
In this work, we follow the same fundamental merger–RD construction as in \texttt{SEOBNRv5HM}, but introduce several modifications to improve its performance in the TML.
In particular, we focus on modeling mode-mixing effects, which are significantly enhanced in the TML. By utilizing QNM coefficients extracted from numerical Teukolsky waveforms, we construct an ansatz that accurately captures this mixing behavior.
In the following, we first present the phenomenological merger–RD ansatz, highlighting its key departures from the \texttt{SEOBNRv5HM} framework. Subsequently, we detail our treatment of mode mixing and its implementation using the extracted QNM data.

\subsection{Phenomenological ansatz}
\label{sec:MRphenomAnsatz}
The damped gravitational emission from the remnant BH can be accurately described as linear perturbations of the Kerr solution, typically expressed as a linear combination of damped QNMs~\cite{Kokkotas:1999bd}.
Within this perturbative framework, the merger–RD modes, $h_{\ell m}^{\text{MR}}$, are constructed using a phenomenological ansatz informed by the TD Teukolsky waveforms. 
Following the approach in Refs.~\cite{Bohe:2016gbl, Cotesta:2018fcv, Pompiliv5}, we factorize the contribution of the fundamental QNM from the total waveform.
The resulting rescaled waveform is expressed as
\begin{equation}
	h_{\ell m}^{\text{MR}}(t) =  \tilde{A}_{\ell m0}(t) e^{i\tilde{\phi}_{\ell m0}(t)} e^{-i\sigma_{\ell m0}\left(t - t^{\ell m}_{\text{match}}\right)},
    \label{eq:hlm_MR}
\end{equation}
where the amplitude $\tilde{A}_{\ell m0}(t)$ is parameterized by 
\begin{equation}
\tilde{A}_{\ell m 0}(t) = 
\left( 
  c_{1,c}^{\ell m}  
  \tanh\left[ 
    c_{1,f}^{\ell m} \left(t - t^{\ell m}_{\text{match}}\right) + c_{2,f}^{\ell m} 
  \right] 
  + 
  c_{2,c}^{\ell m} 
\right)^{
  1/c_{3,c}^{\ell m} 
},
\label{eq:MRansatz_amp}
\end{equation}
The exponent $1/c_{3,c}^{\ell m}$ is motivated by the approach in Ref.~\cite{Albanesi:2021rby}, where such a term is introduced to enforce continuity of the second-order time derivative of the amplitude at the attachment time. 
This parameter is essential for suppressing unphysical amplitude peaks that otherwise tend to appear in the high-spin regime.
The phase $\tilde{\phi}_{\ell m0}(t)$ is given by
\begin{equation}
\tilde{\phi}_{\ell m0}(t) = \phi_{\text{match}}^{\ell m} - d^{\ell m}_{1,c} \; \log \left[ \frac{1 + d^{\ell m}_{2,f} e^{-d^{\ell m}_{1,f} \left(t - t^{\ell m}_{\text{match}} \right) } }{1+ d^{\ell m}_{2,f}}\right],
\label{eq:MRansatz_phase}
\end{equation}
where $\phi_{\text{match}}^{\ell m}$ is the phase of the inspiral-plunge mode $(\ell,m)$ at the attachment time.  
\begin{figure}[t]
\centering
\includegraphics[width=\columnwidth]{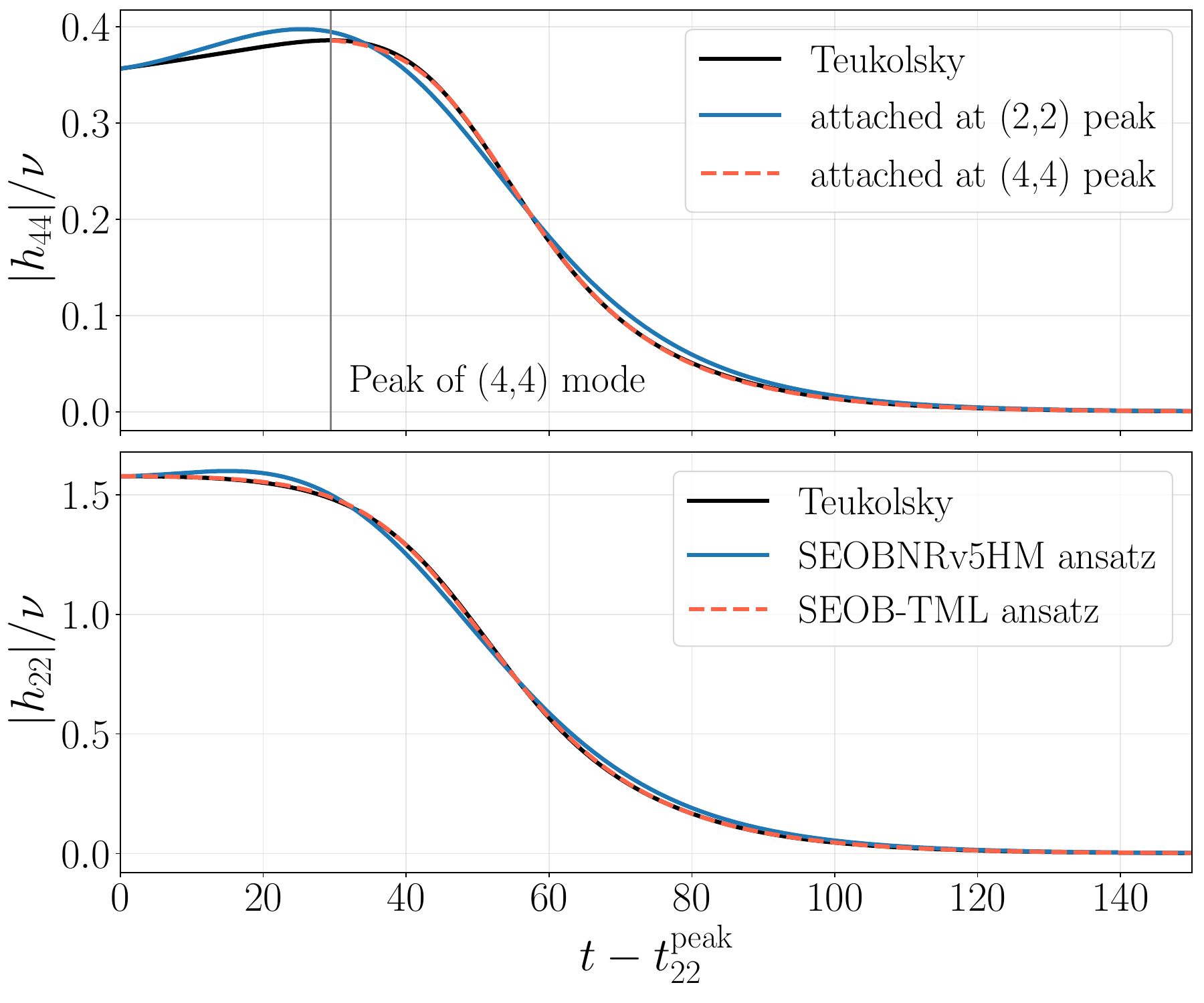} 
\caption{Mode amplitude comparison for prograde spin $a = 0.9$, where the waveforms in both panels are compared against the corresponding Teukolsky waveform (black).
The top panel displays the $(4,4)$ mode constructed by attaching the merger–RD ansatz at $t_{22}^{\text{peak}}$ (blue) versus its own peak at $t_{44}^{\text{peak}}$ (red), demonstrating that the latter improves the agreement. 
The bottom panel shows the $(2,2)$ mode constructed with the \texttt{SEOBNRv5HM} ansatz (blue) and our new model (red), illustrating how the inclusion of the $1/c_{3,c}^{\ell m}$ parameter provides a more faithful representation.}
\label{fig:AmpImprovement}
\end{figure}

The coefficients $d_{1,c}^{\ell m}$ and $c_{i,c}^{\ell m}$ ($i = 1,2,3$) are determined by enforcing that the amplitude and phase of the merger–RD modes in Eq.~\eqref{eq:hlm_MR} are continuously differentiable at $t = t_{\ell m}^{\text{match}}$. 
The amplitude ansatz depends on free parameters $c_{i,f}^{\ell m}$ ($i = 1,2$), the imaginary part of the QNM frequency, and input values extracted from the numerical waveforms. Moreover, 
\begin{subequations}
\begin{align}
c_{1,c}^{\ell m} &=  
\frac{\bigl( \dot{H}^{\text{\scalebox{0.85}{match}}}_{\ell m} + H^{\text{\scalebox{0.85}{match}}}_{\ell m} \; \sigma_{\ell m 0}^{I} \bigr) 
       \cosh^{2}\! c_{2,f}^{\ell m}}
     {c_{1,f}^{\ell m}\, c_{3,c}^{\ell m}}
\, (H^{\text{\scalebox{0.85}{match}}}_{\ell m})^{c_{3,c}^{\ell m} - 1} \label{eq:c1c}, \\[1em]
c_{2,c}^{\ell m} &= 
(H^{\text{\scalebox{0.85}{match}}}_{\ell m})^{c_{3,c}^{\ell m}}
- c_{1,c}^{\ell m} \, \tanh\!\; c_{2,f}^{\ell m} \label{eq:c2c}, \\[1em]
c_{3,c}^{\ell m}  &= \frac{1}{\bigl(\dot{H}^{\text{\scalebox{0.85}{match}}}_{\ell m} + H^{\text{\scalebox{0.85}{match}}}_{\ell m} \sigma_{\ell m 0}^{I}  \bigr)^{2}}\Bigl[ \left(\dot{H}^{\text{\scalebox{0.85}{match}}}_{\ell m}\right)^2
 - H^{\text{\scalebox{0.85}{match}}}_{\ell m} \ddot{H}^{\text{\scalebox{0.85}{match}}}_{\ell m}  \notag \\
  &\quad - 2\, c_{1,f}^{\ell m}\, H^{\text{\scalebox{0.85}{match}}}_{\ell m} \bigl(\dot{H}^{\text{\scalebox{0.85}{match}}}_{\ell m} + H^{\text{\scalebox{0.85}{match}}}_{\ell m} \sigma_{\ell m 0}^{I}  \bigr)\tanh\!\; c_{2,f}^{\ell m} \Bigr]\label{eq:c3c}
\end{align}
\end{subequations}
where $H^{\text{match}}_{\ell m}, \dot{H}^{\text{match}}_{\ell m},$ and $\ddot{H}^{\text{match}}_{\ell m}$ represent the $(\ell, m)$ mode amplitude and its first and second time derivatives, evaluated at the attachment time. 
Similarly, the phase ansatz depends on free parameters $d_{i,f}^{\ell m}$ ($i = 1,2$), the real part of the QNM frequency\footnote{Our nomenclature for the QNM frequencies follows the convention $\sigma_{\ell m 0} = \sigma_{\ell m 0}^R - i \sigma_{\ell m 0}^I$. This differs from \texttt{SEOBNRv5HM}, where the real and imaginary components are swapped in the notation (i.e., their $\sigma_{\ell m 0}^I$ corresponds to our $-\sigma_{\ell m 0}^R$). 
Consequently, while the underlying ansatz is identical to Eq.~(53) of Ref~\cite{Pompiliv5}, the resulting expression for $d_{1,c}^{\ell m}$ appears with a different index.}, and the mode frequency at $t = t_{\ell m}^{\text{match}}$:
\begin{align}
d_{1,c}^{\ell m} &= \bigl( \omega^{\text{match}}_{\ell m}  + \sigma_{\ell m 0}^{R} \bigr) \frac{1 + d_{2,f}^{\ell m}}{d_{1,f}^{\ell m} \; d_{2,f}^{\ell m}}.
\label{eq:d1c_equation}
\end{align} 
By applying the hyperbolic ansatz at the end of the inspiral-plunge (Sec.~\ref{sec:ModelingInspPlunge}), the waveform amplitude and phase are ensured to match the value of Teukolsky waveforms at the attachment time. 
This decouples the merger–RD modes from the EOB inspiral, allowing for an independent calibration of the two sectors.
The free parameters $c_{i,f}^{\ell m}$ and $d_{i,f}^{\ell m}$ are first obtained from least-squares fits to individual Teukolsky waveforms and then interpolated across spin $a$ using rational function fits. 
Potential divergences could occur if $c_{1,f}^{\ell m}$ or $c_{3,c}^{\ell m}$ vanished; however, in the explored parameter space $a \in (-1,1)$, these coefficients remain nonzero, preventing any pathological behavior.

There are two key differences between the present merger–RD model and that used in \texttt{SEOBNRv5HM}.  
The first difference concerns the attachment time. 
In \texttt{SEOBNRv5HM}, the merger–RD portion is attached at the peak of the $(2,2)$ mode for all modes except the $(5,5)$ mode. 
In contrast, as shown in Eq.~\eqref{eq:AttachmentTime}, for positive-spin cases, we attach the merger–RD portion at the peak of each individual mode.
The top panel of Fig.~\ref{fig:AmpImprovement} illustrates the necessity of this approach for spin $a = 0.9$.
Here, the $(4,4)$ Teukolsky waveform (black) is shown with the time axis shifted such that $t=0$ corresponds to the peak of $(2,2)$ mode. 
The $(4,4)$ peak, marked by the vertical line, occurs significantly later at $t \approx 30$.
Attaching the merger–RD model at the $(2,2)$ peak (blue) forces a premature transition and leads to poor agreement,
whereas attaching at the mode's natural peak (red) accurately captures the amplitude evolution.

The second difference lies in the amplitude ansatz.
Unlike \texttt{SEOBNRv5HM}, our model explicitly enforces second-derivative continuity through the parameter $1/c_{3,c}^{\ell m}$ in Eq.~\eqref{eq:MRansatz_amp}, which prevents unphysical amplitude peaks in large-spin cases.
As illustrated in the bottom panel of Figure~\ref{fig:AmpImprovement}, the new ansatz (red) reproduces the waveform amplitude more accurately than the QNM-rescaled ansatz of \texttt{SEOBNRv5HM} (blue), eliminating the spurious early-time peak and improving the late-time agreement.
This consistency is essential for refining the merger–RD model detailed in the following section, which incorporates both the fundamental $(\ell,m,0)$ mode and mode-mixing contributions.

\subsection{Mode Mixing}
\label{sec:ModeMixing}
As discussed in the previous section, the standard EOB framework models each spin-weighted spherical harmonic $(\ell,m)$ mode by factoring out the contribution of the fundamental $(\ell, m, 0)$ QNM. 
The remaining time-dependent amplitude $\tilde{A}_{\ell m0}(t)$ and phase $\tilde{\phi}_{\ell m0}(t)$ are then modeled using phenomenological ansatze, as shown in Eq.~\eqref{eq:hlm_MR}.
This approach, however, breaks down when mode mixing is significant, since the amplitude and phase of the waveform no longer evolve monotonically. 
In this subsection, we outline the two main physical origins of enhanced mode mixing observed in the TML and illustrate how it appears in the waveform, before introducing our modeling prescription. 

\subsubsection{Spheroidal–Spherical basis mismatch.}
As shown by the post-merger oscillations in the $(3,2)$ and $(4,3)$ modes~\cite{Buonanno:2006ui, Kelly:2012nd}, one major source of mixing arises from the projection of the gravitational strain onto different harmonic functions bases.
Specifically, while the Teukolsky waveforms are naturally decomposed in $-2$ spin-weighted spheroidal harmonics, our EOB model utilizes a $-2$ spin-weighted spherical basis. 
The two bases are related by an expansion of the form
\begin{equation}
_{-2} S_{\ell' m n} = \sum_{\ell \geq \max(|m|, 2)} \mu^{*}_{m \ell \ell' n}\, {}_{-2} Y_{\ell m},
\end{equation}
where $\mu^{*}_{m \ell \ell' n}$ are complex mode-mixing coefficients~\cite{London:2018nxs, Berti:2014fga}.  
Substituting this expansion into the waveform gives the relation
\begin{equation}
h_{\ell m}(t) = \sum_{\ell' \geq \max(|m|, 2)} \sum_{n \geq 0} {}^S h_{\ell' m n}(t)\, \mu^{*}_{m \ell \ell' n}.
\label{eq:SphericalSpheroidalMode}
\end{equation}
which shows that the spherical-harmonic mode $h_{\ell m}$ receives contributions from all spheroidal modes with the same $m$, but different $\ell$. 
We account for this effect in our model following the same strategy as in \texttt{SEOBNRv5HM}; details will be provided in Sec.~\ref{sec:ProgradeMix3243}. 

\subsubsection{Change in the sign of orbital frequency.}
A second source of mode mixing arises from the orbital motion of the particle during the plunge, in particular when the orbital frequency changes its sign in the case of negative BH spin.  
For $a <0$, the BH's spin angular momentum is anti-aligned with the orbital angular momentum. 
During the final plunge, frame dragging causes the particle's angular velocity to reverse sign and eventually lock onto the horizon frequency. 
This orbital reversal excites retrograde $(\ell,m)$ modes, whose complex frequencies are related to the prograde frequencies by $\sigma_{\text{retro}} = -\sigma_{\ell, -m, n}^*$.
This effect intensifies as the spin becomes more negative, driven by the widening gap between the light-ring and horizon frequencies.
For instance, when $a=-0.5$ ($a=-0.95$), we find $\Omega_{\mathrm{LR}} = 0.0495$ ($0.0393$) and $\Omega_{\text{H}} = -0.134$ ($-0.362$), yielding a frequency difference of $\Omega_{\text{H}} - \Omega_{\mathrm{LR}} = -0.183$ ($-0.401$).

\bigskip
In Ref.~\cite{Taracchini:2014zpa}, the merger–RD waveform $h_{\ell m}^{\text{MR}}$ was modeled as a linear superposition of $(\ell,m,n)$ modes in order to capture mode mixing. 
Schematically, their ansatz can be written as 
\begingroup
\footnotesize
\begin{equation}
\begin{aligned}
h_{\ell m}^{\text{MR}}(t) &=
\sum_{n=0}^{N-1} A_{{\scriptscriptstyle \ell\!,\!m\!,\!n}} 
e^{-i\sigma_{{\scriptscriptstyle \ell\!,\!m\!,\!n}}\!(t - t_{\rm match}^{\ell m})} \\[0.5mm]
&+ \mathscr{S}(t)\bigl[
A_{{\scriptscriptstyle \ell'\!,\!m\!,\!0}}
e^{-i\sigma_{{\scriptscriptstyle \ell'\!,\!m\!,\!0}}\!(t - t_{\rm match}^{\ell m})}\! +\! 
A_{{\scriptscriptstyle \ell\!,\!-m\!,\!0}}
e^{i\sigma_{{\scriptscriptstyle \ell\!,\!-m\!,\!0}}^{*}\!(t - t_{\rm match}^{\ell m})}
\bigr]
\end{aligned}
\end{equation}
\endgroup
where $N$ denotes the number of included overtones.
The term $\mathscr{S}(t) = \left[1 + \tanh[(t-t_s)/\tau_s]\right]/2$ is a sigmoid-like window function used to gradually activate the mixing terms after merger; the parameters $t_s$ and $\tau_s$ characterize the timing and duration of this transition, respectively.
The $A_{\ell, m, n}$ are the coefficients of the fundamental mode and its overtones, while $A_{\ell', m,0}$ and $A_{\ell,-m,0}$ quantify the strength of the interfering QNMs. 
Here, the index $\ell' \neq \ell$ denotes the spheroidal-mode contribution arising from the basis mismatch, while the $-m$ index accounts for the excitation of retrograde modes due to orbital frequency reversal.
These constants were determined phenomenologically: whenever mode mixing was clearly visible, $A_{\ell', m,0}$ and $A_{\ell,-m,0}$ were fixed by fitting the model GW frequency to the corresponding Teukolsky frequency over a suitable fitting window. 
Once the mixing amplitudes were fixed, the remaining $A_{\ell, m, n}$ were obtained through a hybrid matching procedure detailed in Ref~\cite{Pan:2011gk}.
In this way, all mixing-mode coefficients were calibrated numerically against the Teukolsky waveforms.  

A key distinction in our approach is that while Ref.~\cite{Taracchini:2014zpa} explicitly sums over overtones to model the $(\ell,m)$ contribution, our phenomenological ansatz—Eq.~\eqref{eq:hlm_MR}—factors out the fundamental $(\ell,m,0)$ mode. Consequently, the residual signal (which includes overtone contributions) is represented through the time-dependent amplitude and phase defined in Eqs.~\eqref{eq:MRansatz_amp} and~\eqref{eq:MRansatz_phase}.
Furthermore, while our strategy for incorporating additional mixing modes—detailed in Sec.~\ref{sec:ModelingMixingMode}—follows a similar logic to Ref.~\cite{Taracchini:2014zpa}, we have refined the numerical extraction of the QNM coefficients.
To determine the QNM content of the TD Teukolsky waveforms, we utilize \texttt{qnmfinder}~\cite{qnmfinder_code}, a reverse-search algorithm designed for robust QNM identification as presented in Ref.~\cite{Mitman:2025hgy}.

\subsection{Extracting QNM coefficients}
\label{sec:ExtractingQNMcoeffs}
As noted above, we utilize \texttt{qnmfinder} to extract the QNM coefficients from the TD Teukolsky data.
The method combines variable projection for free-frequency fits with a stability criterion on the complex amplitudes, ensuring that only physically meaningful QNMs are retained. 
A detailed discussion of the algorithm can be found in Ref.~\cite{Mitman:2025hgy}; here we summarize its main features. 
\begin{figure}[t]
\centering
\includegraphics[width=\columnwidth]{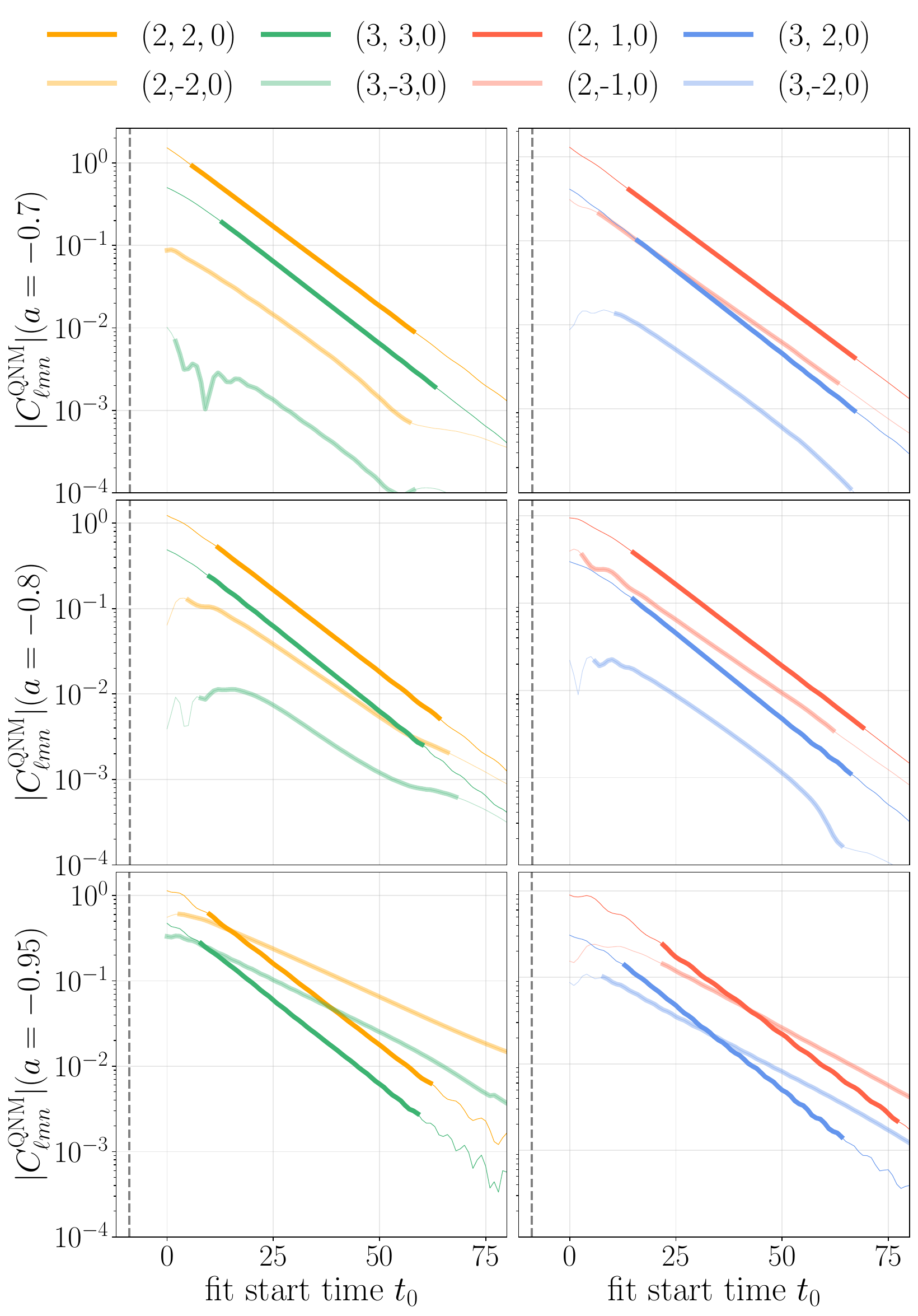}
\caption{The amplitude of stable QNM coefficients, $|C^{\rm QNM}_{\ell m n}|$, for negative-spin Teukolsky waveforms. 
Thin lines show fits starting from all candidate times, while thick lines indicate the most stable interval for each amplitude. 
Dark colors correspond to prograde modes $(\ell,m,0)$, and lighter colors to retrograde modes $(\ell,-m,0)$. 
The figure is split into two panels: left for $\ell=m$ modes and right for $\ell\neq m$ modes. 
The vertical dashed line marks the peak of the $(2,2)$ mode.
}
\label{fig:stableAqnm}
\end{figure}
\begin{figure}[t]
\centering
\includegraphics[width=\columnwidth]{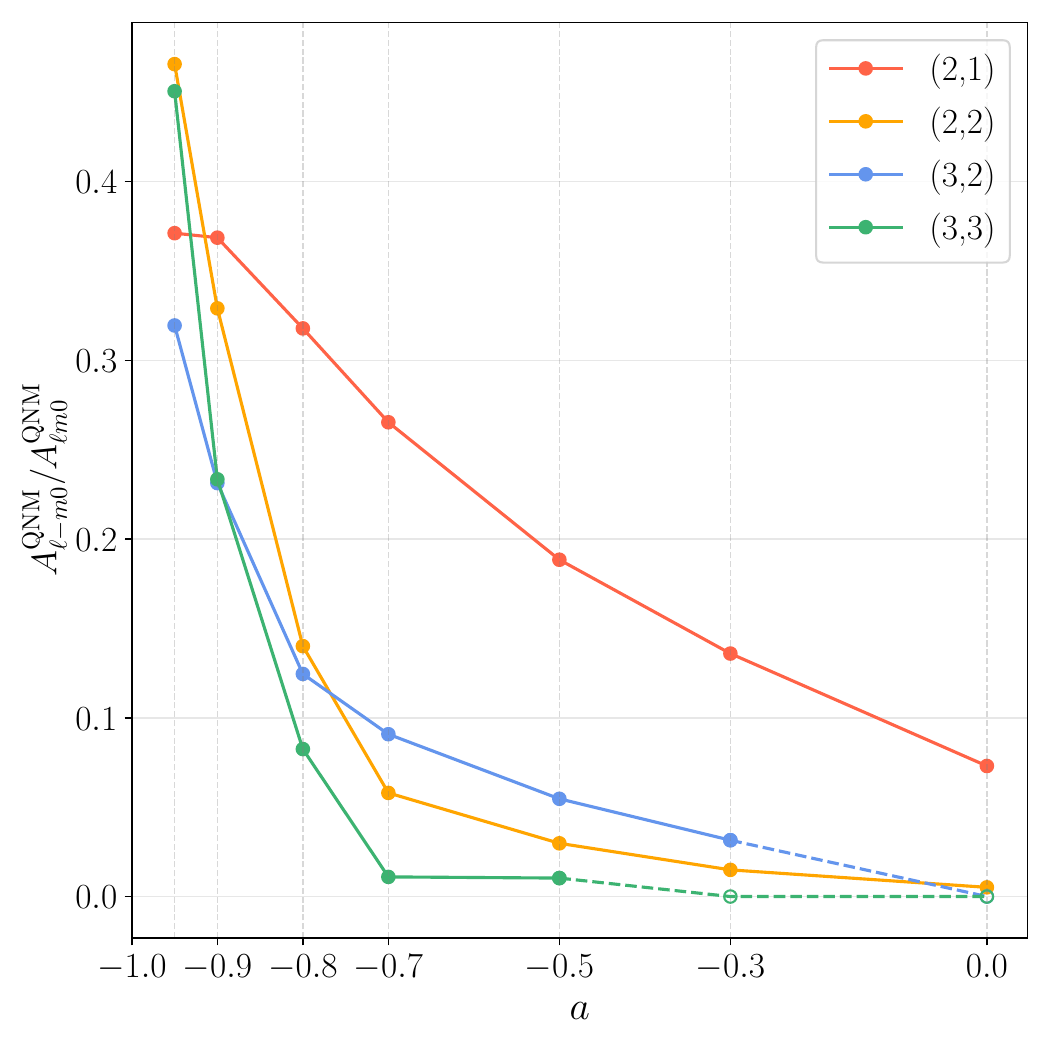}
\caption{Ratio of QNM amplitude $|A_{\ell -m 0}/ A_{\ell m 0}|$ as a function of spins. 
Here, the amplitudes $A_{\ell m n}$ correspond to the $(\ell, m, n)$ spheroidal QNMs mapped onto the $(\ell, m)$ spherical harmonic basis and evolved back to the attachment time.
Data are shown for the four dominant $(\ell, m)$ modes; non-filled circles indicate cases where no stable retrograde mode was identified by \texttt{qnmfinder}.}
\label{fig:Almm0Alm0_ratio}
\end{figure}
First-order BH perturbation theory predicts additional non-QNM content in the ringdown waveform, including the prompt response (the direct propagation of the perturbation to the observer~\cite{PhysRevD.34.384}) and power-law tails from back-scattering off the BH potential~\cite{PhysRevD.5.2419,PhysRevD.5.2439}. 
These features are not simple damped sinusoids and can contaminate QNM fits. 
To guard against this, \texttt{qnmfinder} imposes a stability requirement: over a range of extraction times, the projected QNM amplitudes must consistently track the expected damped sinusoid evolution. 
If stability is violated, the candidate QNM is rejected as spurious. 
The algorithm further improves robustness in two key ways. 
First, the QNM model is constructed over the entire 2-sphere rather than restricted to a single $(\ell,m)$ spherical harmonic mode, which helps constrain mixing and suppress spurious solutions. 
Second, the procedure exploits the fact that QNMs have different decay times: the longest-lived modes are identified from the late portion of the ringdown, while faster-decaying modes are extracted by progressively shifting to earlier time. 

After the prompt response, the emitted GW strain can be expressed as a sum of damped sinusoids 
\begin{equation}
h(t,\theta,\phi)
= \sum_{
    \ell \ge 2,\; |m|\le \ell, \; n \ge 0}
  C_{\ell m n} e^{-i\sigma_{\ell m n} t}\; _{-2} S_{\ell m n}(\theta,\phi).
\label{eq:QNMqnmfinder}
\end{equation}
where $C_{\ell m n}$ are the complex amplitudes and ${}_{-2}S_{\ell m n}$ are the spin-weighted $-2$ spheroidal harmonics.
Since numerical waveforms are typically decomposed in the spherical rather than spheroidal basis, one expands the spheroidal harmonics into spherical harmonics using spherical–spheroidal mixing coefficients~\cite{Stein:2019mop}. 
As a result, within a given $(\ell,m)$ spherical harmonic mode, \texttt{qnmfinder} can also identify contributions from spheroidal QNMs.
These QNM frequencies are fully determined by the remnant BH's mass $M_f$ and spin $a$, in accordance with the no-hair theorem.
In contrast, the complex amplitude $C_{\ell m n p}^{\rm QNM}$ depends on the progenitor binary's parameters and must be extracted from numerical waveforms.
To mitigate errors from residual supertranslations in numerical data, \texttt{qnmfinder} performs the fits in terms of the \emph{news}---the first time derivative of the strain---rather than the strain itself, since the news always decays to zero in any gauge~\cite{Boyle:2015nqa}.
The corresponding strain-domain amplitudes are then obtained directly from the news-domain fits via
\begin{equation}
\begin{aligned}
C_{\text{QNM}}^{\text{strain}} &= \int C_{\text{QNM}}^{\text{news}}  e^{-i \sigma_{\text{QNM}} t}  dt / e^{-i \sigma_{\text{QNM}} t} ,\\
&= C_{\text{QNM}}^{\text{news}} / (-i \sigma_{\text{QNM}})
\end{aligned}
\label{eq:CQNM}
\end{equation}

To illustrate the extraction of QNM amplitudes using \texttt{qnmfinder}, we consider the TD Teukolsky waveform for the negative-spin cases, which, as noted earlier, exhibits enhanced mode mixing in the TML. 
Figure~\ref{fig:stableAqnm} displays the resulting fits for the QNM strain amplitudes, which are defined as the magnitude of the complex coefficients $C_{\ell m n}^{\rm QNM}$ in Eq.~\eqref{eq:QNMqnmfinder}.
In the figure, thin lines represent the extracted amplitudes across all candidate start times, while the thick segments highlight the stability interval.
The vertical dashed line marks the peak of the $(2,2)$ spherical mode.
We restrict our analysis to the fundamental modes and neglect overtones ($n>0$). This is due to their rapid decay and the fact that the early merger–RD is already handled by our phenomenological modeling; our focus here is on the longer-lived modes that dominate the late-time mixing.
To maintain clarity and avoid overlap, the figure is split into two panels: the left for $\ell = m$ and the right for $\ell \neq m$. Within these panels, dark colors indicate the prograde modes $(\ell, m, 0)$, while lighter colors represent the corresponding retrograde modes $(\ell, -m, 0)$.

This visualization confirms that for negative-spin cases, the retrograde QNM contribution is most significant for the $(2,1)$ mode. This is reflected by the high ratio of $|C^{\rm QNM}_{2-10}/C^{\rm QNM}_{210}|$, indicating that retrograde-mode mixing is strongest in this specific harmonic. These findings connect directly to the behavior observed in Figure~\ref{fig:h21attach}, where the $(2,1)$ mode exhibits the most prominent mode-mixing effects in the negative-spin regime.
Figure~\ref{fig:stableAqnm} also shows that the retrograde mode contribution increases as the spin becomes more negative, reflecting the longer portion of the orbit with $\Omega < 0$.
This trend is quantified in Figure~\ref{fig:Almm0Alm0_ratio}, which shows the amplitude ratio $A^{\rm QNM}_{\ell -m 0}/A^{\rm QNM}_{\ell m 0}$ as a function of the spin. 
Here, $A^{\rm QNM}_{\ell m n}$ represents the amplitude of the $(\ell, m, n)$ spheroidal QNM coefficients after they have been mapped onto the $(\ell, m)$ spherical harmonic basis.
This mapping ensures that the QNM contributions are expressed in a form that is easily incorporated into our EOB framework, where the multipolar waveforms are constructed in the spherical basis.
While stable QNMs are identified within their specific stability windows, we evolve the extracted QNM amplitudes back to the attachment time using their respective complex frequencies. We specifically choose this reference point because the merger-RD model starts at the attachment time, where this QNM information is required for the construction (see Sec.~\ref{sec:ModelingMixingMode} for more details).
In this figure, two observations are noteworthy. First, stable retrograde modes could not be identified for $(3,2)$ at $a = 0$ and for $(3,3)$ at $a > -0.5$, which are indicated by non-filled circles at zero ratio. Inspection of the corresponding Teukolsky waveforms confirms that mode-mixing behavior was negligible in these cases, so excluding them does not highly affect the accuracy of our waveform model.
Second, even in the non-spinning case, the retrograde contribution for the $(2,1)$ mode is non-negligible; in fact, a stable $(2,-1,0)$ QNM was found for $a = 0.3$, but not for $a = 0.5$, showing that retrograde-mode excitation can persist even for small positive spins. 
When incorporating this extracted QNM information into our model, we must fit not only the parameters $c_{1,f}, c_{2,f}, d_{1,f}$ and $d_{2,f}$ from Eqs.~\eqref{eq:MRansatz_amp} and \eqref{eq:MRansatz_phase}, but also the ratio of $A_{\ell -m 0}/ A_{\ell m 0}$ across the range of spin values.
As shown in Figure~\ref{fig:Almm0Alm0_ratio}, the retrograde-mode contribution generally increases as the spin decreases.
Consequently, we interpolate these ratios across the spin parameter space by fitting them with rational functions.

\section{Modeling Mixing Modes}
\label{sec:ModelingMixingMode}
Having established the phenomenological ansatz for the EOB framework, identified the necessity of including mixing modes in the TML, and demonstrated the extraction of QNM coefficients, we now proceed to construct the merger–RD waveforms. This construction explicitly incorporates the contributions from mixing modes.
\subsection{Retrograde spin ($\ell = m$ modes)}

\begin{figure*}[t]
\centering
\includegraphics[width=\textwidth]{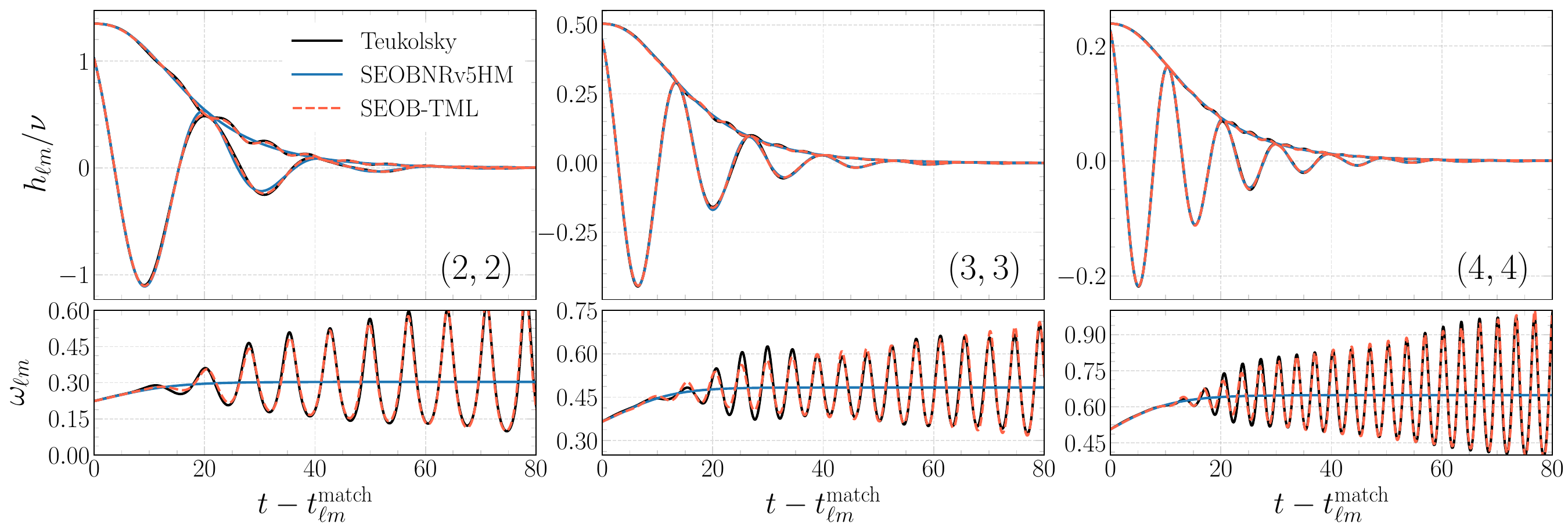}
\caption{Merger–RD modeling for $\ell=m$ modes with retrograde spin $a = -0.8$, compared against TD Teukolsky waveforms (black).
Top panels show the waveform amplitude and the real part of the strain, and bottom panels show the gravitational frequency for the $(2,2)$, $(3,3)$, and $(4,4)$ modes.
Blue curves represent the \texttt{SEOBNRv5HM} merger–RD model, which includes only the fundamental $(\ell,m,0)$ QNM. Red curves represent the \texttt{SEOB-TML} model, which incorporates the retrograde $(\ell,-m,0)$ mode via Eq.~\eqref{eq:RD_with_MixMode}.
The \texttt{SEOB-TML} ansatz successfully captures the modulations in both amplitude and frequency induced by mode mixing.
The time coordinate is shifted such that $t=0$ corresponds to the merger–RD attachment time.
}
\label{fig:am080Mixellm.png}
\end{figure*}

\begin{figure*}[t]
\centering
\includegraphics[width=\textwidth]{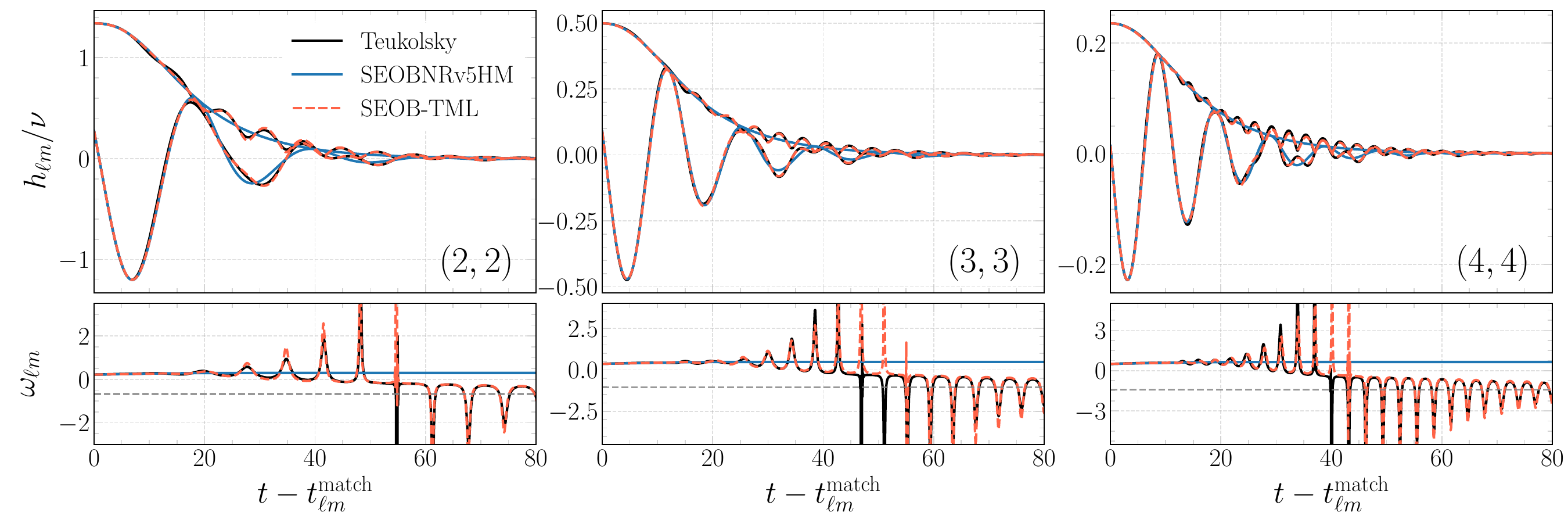}
\caption{Merger–RD modeling for $\ell=m$ mode with retrograde spin $a = -0.9$.
The layout and color scheme follow Figure~\ref{fig:am080Mixellm.png}.
The horizontal dashed line marks the retrograde QNM frequency $\sigma_{\ell -m 0}^{R}$, illustrating how the Teukolsky frequency drifts from the prograde value $\sigma_{\ell m 0}^{R}$ toward the retrograde limit as the latter becomes the dominant contribution at late times.}
\label{fig:am090Mixellm.png}
\end{figure*}

We first focus on $\ell = m$ modes in the negative-spin case. 
As discussed in Sec.~\ref{sec:ModeMixing}, the reversal of the orbital frequency during the plunge excites retrograde QNMs. 
To capture this retrograde-mode mixing, we include the $(\ell,-m,0)$ QNM contribution within the MR ansatz in Eq.~\eqref{eq:hlm_MR}, i.e.
\begin{align}
h_{\ell m}^{\text{MR}} (t) 
&= \tilde{A}_{\ell m 0}(t) \; e^{i \tilde{\phi}_{\ell m}(t)} 
\Big[ e^{-i \sigma_{\ell m 0} (t- t_{\text{match}})} \nonumber\\
&\quad + \mathscr{S}(t) 
\frac{A_{\ell -m 0}^{\text{QNM}}}
     {A_{\ell m 0}^{\text{QNM}}}
 e^{i \sigma_{\ell -m 0}^{*}(t - t_{\text{match}})} \Big]
\label{eq:RD_with_MixMode}
\end{align}
Here, $A_{\ell m n}^{\text{QNM}}$ is the extracted complex amplitude of the QNM coefficient $(\ell,m,n)$ evaluated at the attachment time and translated into the spherical-harmonic basis.
The phenomenological ansatz for the QNM-rescaled amplitude and phase, $\tilde{A}_{\ell m 0}(t)$ and $\tilde{\phi}_{\ell m 0}(t)$, is the same as in Eqs.~\eqref{eq:MRansatz_amp}–\eqref{eq:MRansatz_phase}.  
Four free parameters, $c_{1,f}, c_{2,f}, d_{1,f}, d_{2,f}$, are determined by a least-squares fits to Teukolsky waveforms while ensuring consistency with the $(\ell,m,0)$ QNM amplitude at late times.   
The function $\mathscr{S}(t)$ governs the activation of the retrograde mode contribution. 
In Ref.~\cite{Taracchini:2014zpa}, this role was played by a sigmoid function $\big[1+\tanh[(t-t_s)/\tau_s]\big]/2$.
Here we adopt a refined activation function,
\begingroup
\small
\begin{align}
\mathscr{S}(\tau) 
&= e^{i \gamma \, \text{sech}\!\left( \tfrac{\tau}{\tau_{p}} \right)} \nonumber\\
&\quad \times \Bigg[
\frac{
    \tau_{s} \tanh\!\left( \tfrac{t_{s}}{\tau_{s}} \right)
    - \text{sech}^2\!\left( \tfrac{t_s}{\tau_s} \right) \tanh(\tau)
    + \tau_{s} \tanh\!\left( \tfrac{\tau - t_{s}}{\tau_{s}} \right)
}{
    \tau_{s} - \text{sech}^2\!\left( \tfrac{t_s}{\tau_s} \right)
    + \tau_{s} \tanh\!\left( \tfrac{t_s}{\tau_{s}} \right)
}\Bigg]
\label{eq:ActivationFunction}
\end{align}
\endgroup
with $\tau = t - t_{\text{match}}$. 
This construction retains the overall sigmoid-like behavior, approaching unity as $t \to \infty$. 
Consequently, in this asymptotic limit, Eq.~\eqref{eq:RD_with_MixMode} reduces to a superposition of two fundamental QNMs,
\begin{equation}
h_{\ell m}^{\text{MR}} \;\to\; A_{\ell m 0}^{\text{QNM}} e^{-i \sigma_{\ell m 0} (t - t_{\text{match}})} 
+ A_{\ell -m 0}^{\text{QNM}} e^{i \sigma_{\ell -m 0}^{*} (t - t_{\text{match}})}.
\end{equation}
At the same time, the refined form $\mathscr{S}(t)$ differs from Ref.~\cite{Taracchini:2014zpa} in two important respects.
First, it satisfies $\mathscr{S}(0)=0$ and $\dot{\mathscr{S}}(0)=0$, so that at the attachment point, the waveform reduces smoothly to the pure $(\ell,m,0)$ QNM-rescaled mode of Eq.~\eqref{eq:hlm_MR}, thus ensuring the continuity and differentiability of the merger–RD ansatz.
Second, the prefactor $e^{i \gamma \; \text{sech}(\tau/\tau_p)}$ provides additional phase flexibility, which becomes especially important for large negative spins.
In this regime ($a \leq -0.9$), the early ringdown contains not only $(\ell,\pm m,0)$ modes but also other QNMs that are absent from our explicit model.
The extra phase term therefore plays a phenomenological role, compensating for this missing physics.
This treatment is purely phenomenological; a physically motivated overtone structure would be a more rigorous way to capture these contributions. However, we rely on the added flexibility of $\mathscr{S}(t)$ to handle these early-time transients in order to maintain a manageable level of complexity.
Finally, the parameters $(t_s,\tau_s,\tau_p)$, introduced in the activation function defined in Eq.~\eqref{eq:ActivationFunction} are tuned for each mode to control the onset and steepness of the activation.
For the $(2,2)$ mode, for example, we use $(t_s,\tau_s,\tau_p)=(20,7.5,7.5)$ for $a > -0.9$, while for $a \leq -0.9$ we increase $t_s$ to $25$ and $t_p$ to $10$ in order to account for the delayed excitation of the $(2,-2,0)$ contribution, 
while allowing a wider window for the phenomenological activation function to capture the effects of the missing modes. 
As already discussed in Sec.~\ref{sec:ModeMixing} and illustrated in Figure~\ref{fig:Almm0Alm0_ratio}, the ratio of QNM coefficients $\big|A_{\ell -m 0}^{\text{QNM}} / A_{\ell m 0}^{\text{QNM}}\big|$ grows steadily as the spin decreases, underscoring the increasing importance of retrograde-mode mixing in the highly retrograde regime.

\begin{figure*}[t]
\centering
\includegraphics[width=\textwidth]{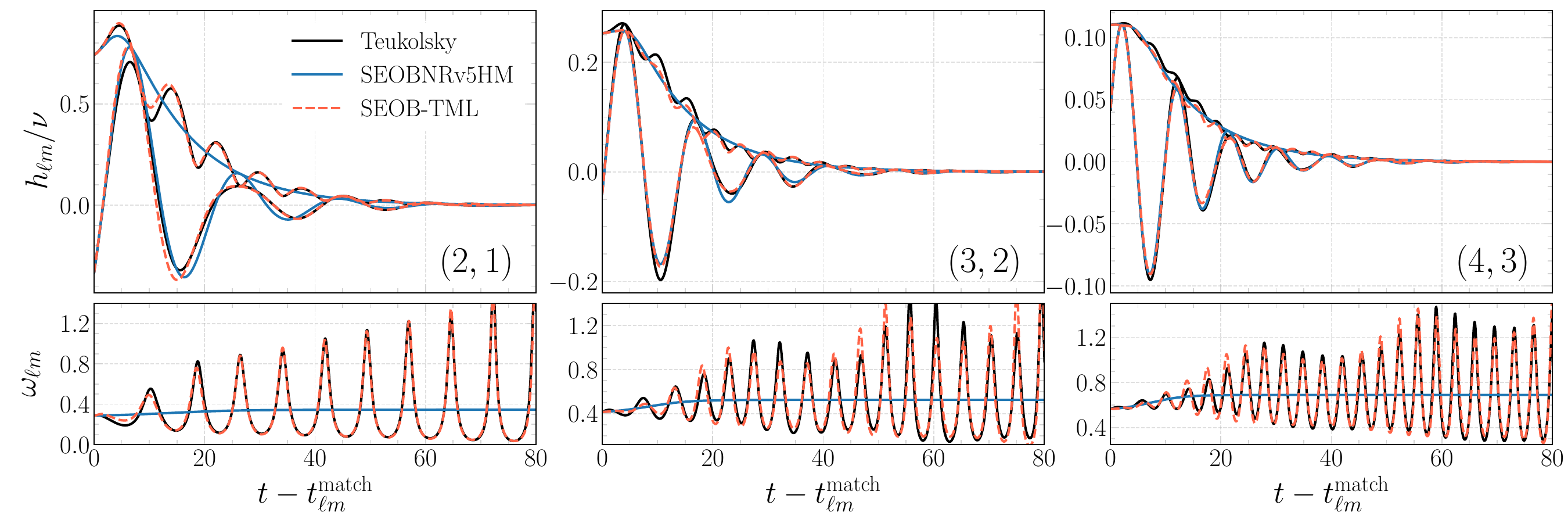}
\caption{Merger–RD modeling for $\ell \neq m$ mode with retrograde spin $a = -0.8$.
Blue curves represent the \texttt{SEOBNRv5HM} merger–RD model, which includes only the fundamental $(\ell,m,0)$ QNM. 
Red curves denote the \texttt{SEOB-TML} model, which incorporates the retrograde $(\ell,-m,0)$ mode; for the $(3,2)$ and $(4,3)$ modes, this model further includes the $(\ell-1,-m,0)$ contribution via Eq.~\eqref{eq:RDMix320retro}. 
Compared with the $\ell = m$ case, the amplitude and frequency modulations are less accurately reproduced, particularly at early times,
reflecting the limitations of the phenomenological ansatz in capturing multiple interfering contributions immediately after attachment. }
\label{fig:am080Mixellnotm.png}
\end{figure*}

The effectiveness of the merger–RD ansatz for $\ell = m$ modes with negative spins is illustrated in Figs.~\ref{fig:am080Mixellm.png} and \ref{fig:am090Mixellm.png}. 
Figure~\ref{fig:am080Mixellm.png} shows the case with spin $a=-0.8$. The upper panels display the real part of the waveform together with its amplitude, while the lower panels show the GW frequency for the $(2,2)$, $(3,3)$, and $(4,4)$ modes.
The Teukolsky waveforms are shown in black. The merger–RD model following the \texttt{SEOBNRv5HM} approach—constructed with only the $(\ell,m,0)$ QNM contribution (Eqs.~\eqref{eq:MRansatz_amp}–\eqref{eq:MRansatz_phase})—is shown in blue. Finally, the \texttt{SEOB-TML} model, which includes the retrograde $(\ell,-m,0)$ mode (Eq.~\eqref{eq:RD_with_MixMode}), is shown in red.
In the frequency panels, the \texttt{SEOBNRv5HM} ansatz asymptotes to the fundamental QNM frequency $\sigma_{\ell m 0}^{R}$, around which the Teukolsky frequency oscillates.
These oscillations in frequency in turn induce modulations in the waveform amplitude. By contrast, the model with retrograde-mode mixing reproduces both the frequency oscillations and the associated amplitude modulation, achieving much better agreement.

For more negative spins the limitations of the model become evident. 
Figure~\ref{fig:am090Mixellm.png} shows the $a=-0.9$ case. 
In the early merger–RD, the red curve does not align well with the Teukolsky waveform, both in amplitude and in frequency. 
Introducing the additional phase flexibility through the prefactor $e^{i \gamma \,\text{sech}(\tau/\tau_p)}$ in the activation function improves the agreement, but does not fully resolve the discrepancy, 
particularly at early times. 
This discrepancy becomes even more pronounced for $a=-0.95$, suggesting that a future refinement of the activation function $\mathscr{S}(t)$ (Eq.~\eqref{eq:ActivationFunction}) may be needed to accurately model the large negative spin regime. 
An additional noteworthy feature emerges for $a \leq -0.9$: the average gravitational frequency of the $(\ell,m)$ mode transitions from $\sigma_{\ell m 0}^{R}$ toward the retrograde value $\sigma_{\ell-m0}^{R}$. 
For example, in the $(2,2)$ mode, the bottom panel of Figure~\ref{fig:am090Mixellm.png} shows that the Teukolsky frequency initially oscillates around $\sigma_{220}^{R}$, but at later times it drifts and begins to oscillate about $\sigma_{2-20}^{R}$, indicated by the gray dashed line, leading to $2 \Omega_H \sim \sigma_{2-20}^{R}$. 
This transition to retrograde behavior is also visible in the higher $\ell=m$ modes, reflecting the growing dominance of retrograde-mode contributions in the highly retrograde regime.  

\begin{figure}[t]
\centering
\includegraphics[width=\columnwidth]{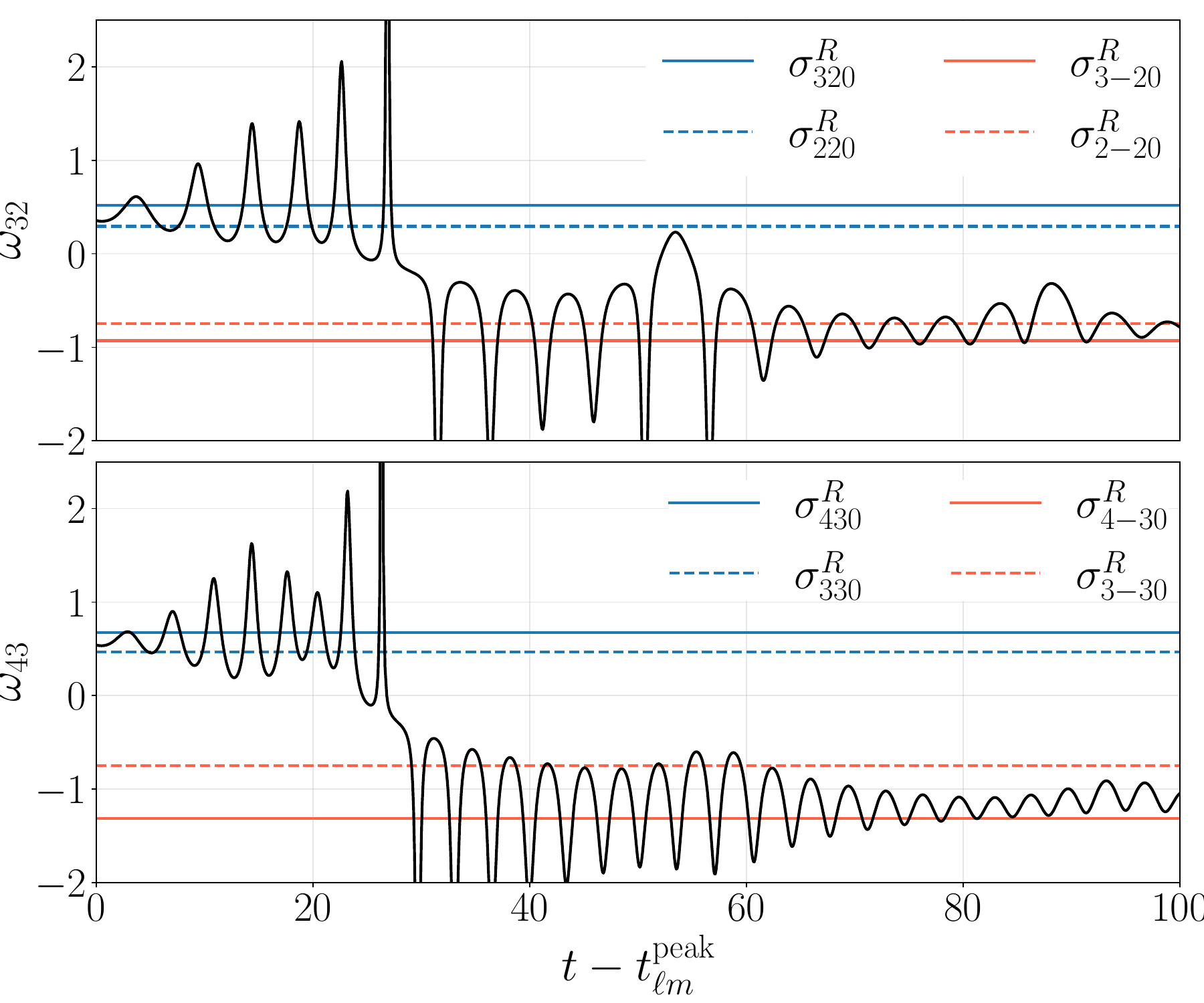}
\caption{Teukolsky GW frequency for the $(3,2)$ mode (top) and $(4,3)$ mode (bottom) for $a=-0.95$, shown as a function of time relative to the mode peak.
Horizontal lines indicate the real parts of selected QNM frequencies: the prograde $(\ell,m,0)$ mode (solid blue), the neighboring prograde $(\ell-1,m,0)$ mode (dashed blue), the retrograde $(\ell,-m,0)$ mode (solid red), and the neighboring retrograde $(\ell-1,-m,0)$ mode (dashed red).
At late times, those frequencies oscillate between the negatives of the two retrograde QNM frequencies.}
\label{fig:am095_32_43}
\end{figure}

\begin{figure*}[t]
\centering
\includegraphics[width=\textwidth]{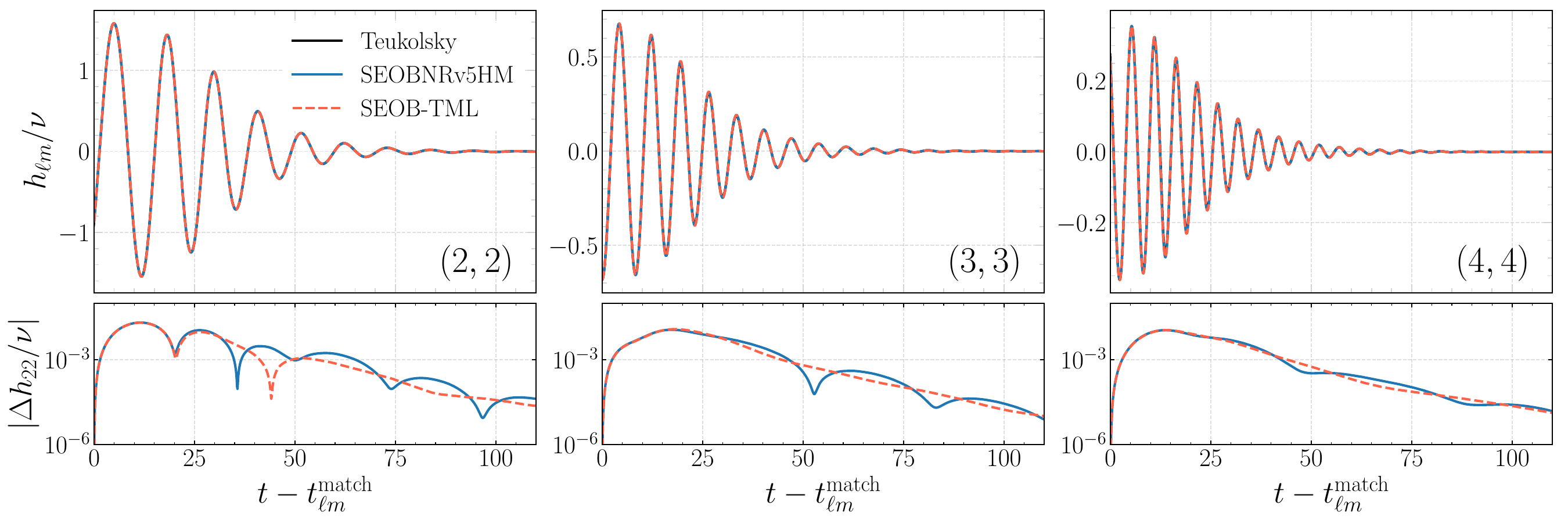}
\caption{Merger–RD modeling for $\ell=m$ modes with prograde spin $a =0.8$, compared against TD Teukolsky waveforms (black).
Top panels show the real part of the strain, while bottom panels display the absolute residual for the corresponding modes.
Blue curves represent the \texttt{SEOBNRv5HM} merger–RD model which includes only the fundamental $(\ell,m,0)$ QNM, whereas red curves additionally include the $(\ell+1,m,0)$ mixing contribution via Eq.~\eqref{eq:RD_with_MixModeProg}.
The two models are visually indistinguishable at the level of the strain.
A modest improvement from the additional mode is visible only in the residuals, where small late-time modulations are partially flattened. }
\label{fig:a080Mixellm.png}
\end{figure*}

\subsection{Retrograde spin ($\ell \neq m$ modes)}
We now turn to the case $\ell \neq m$, still with negative spin.  
In contrast to $\ell = m$ modes, the merger–RD attachment does not occur at the peak of each mode but rather at the times prescribed in Eq.~\eqref{eq:AttachmentTime}.
For the $(2,1)$ mode, the attachment is set when the orbital frequency crosses zero, consistent with the onset of retrograde excitation. For the $(3,2)$ and $(4,3)$ modes, the attachment is further delayed by $2.5$ and $3.5$, respectively.  
The merger–RD modeling of the $(2,1)$ mode follows the same form as Eq.~\eqref{eq:RD_with_MixMode}, incorporating the $(\ell,-m,0)$ QNM. 
By contrast, the $(3,2)$ and $(4,3)$ modes also receive contributions due to spheroidal-spherical translation, which introduces contributions from spheroidal modes with the same $m$ but different $\ell$.
Therefore, the merger–RD ansatz for the $(3,2)$ and $(4,3)$ modes becomes 
\begingroup
\footnotesize
\begin{align}
h_{\ell m}^{\text{MR}}
&(t) = \tilde{A}_{\ell m0}(t) e^{i \tilde{\phi}_{\ell m0}(t)} 
\Big[ e^{-i \sigma_{\ell m0} (t- t_{\scriptscriptstyle \mathrm{match}})} \nonumber\\
&\hspace{-1.3em} + \mathscr{S}(t)\!\left( 
   \frac{A_{\ell\!-\!m0}^{\text{QNM}}}{A_{\ell m0}^{\text{QNM}}} e^{i \sigma_{\ell\!-\!m0}^{*}(t - t_{\scriptscriptstyle \mathrm{match}})} 
 + \frac{A_{\ell' m0}^{\text{QNM}}}{A_{\ell m0}^{\text{QNM}}} e^{-i \sigma_{\ell' m0}(t - t_{\scriptscriptstyle \mathrm{match}})} 
   \right) \Big] .
\label{eq:RDMix320retro}
\end{align}
\endgroup
where $\ell'=2$ for the $(3,2)$ mode and $\ell'=3$ for the $(4,3)$ mode. 
Higher-order spheroidal contributions with $\ell' > \ell$ are neglected, as their effect is subdominant and would overcomplicate the ansatz. 
The comparison with the Teukolsky waveforms is shown in Figure~\ref{fig:am080Mixellnotm.png} for spin $a=-0.8$.
The $(2,1)$ mode exhibits rapidly growing oscillations about $\sigma_{210}^{R}$, well captured by the mixing ansatz at late times. 
For the $(3,2)$ and $(4,3)$ modes, the GW frequency oscillates around $\sigma_{\ell m0}^{R}$ while developing a slower envelope modulation. 
These two timescales arise from the interference of the $(\ell,-m,0)$ and $(\ell', m, 0)$ QNMs with the dominant $(\ell, m, 0)$ mode, producing high-frequency oscillations and low-frequency amplitude beating, respectively. 
The overall mixing behavior is reproduced, though the early ringdown again exposes the limitations of a phenomenological activation.
Strictly speaking, the activation function $\mathscr{S}(t)$ governing the $(\ell,-m,0)$ and $(\ell',m,0)$ contributions does not need to be identical, since their decay times differ. 
A more complete model would assign distinct activations to each term, but for simplicity we employ a single common $\mathscr{S}(t)$. 
At larger negative spins ($a \leq -0.95$), the modeling becomes more challenging.
For example, for the $(3,2)$ mode, the frequency no longer cleanly asymptotes to $\sigma_{320}^{R}$ but instead drifts between $\sigma_{2-20}^{R}$ and $\sigma_{3-20}^{R}$, showing irregular patterns, illustrated in Figure~\ref{fig:am095_32_43}.
This behavior underscores the need for refined activation functions and possibly additional QNM contributions in order to extend the model reliably into the highly retrograde regime.

\subsection{Prograde spin ($\ell = m$ mode)}
So far, we have focused on modeling the merger–RD waveform for negative-spin cases. We now turn to positive-spin configurations.
In principle, similar to the retrograde case, one could include an additional QNM contribution to capture mode mixing in the prograde regime.  
For $\ell = m$ modes with positive spin, the $(\ell+1, m, 0)$ mode is the dominant secondary contributor.
Applying the same formalism, the merger–RD ansatz becomes
\begin{align}
h_{\ell m}^{\text{MR}} (t) 
&= \tilde{A}_{\ell m0}(t) \; e^{i \tilde{\phi}_{\ell m0}(t)} 
\Big[ e^{-i \sigma_{\ell m 0} (t- t_{\text{match}})} \nonumber\\
&\quad + \mathscr{S}(t) 
\frac{A_{\ell' m 0}^{\text{QNM}}}
     {A_{\ell m 0}^{\text{QNM}}}
 e^{-i \sigma_{\ell' m 0}(t - t_{\text{match}})} \Big]
\label{eq:RD_with_MixModeProg}
\end{align}
In this configuration, $\ell'$ corresponds to $\ell + 1$.
The results for spin $a = 0.8$ are shown in Figure~\ref{fig:a080Mixellm.png}. 
The top panel compares the merger-RD model with and without the $(\ell+1, m, 0)$ QNM contribution against the TD Teukolsky waveform. 
Visually, the two waveforms are nearly identical, completely overlapping and indistinguishable. 
However, the bottom panel shows the absolute residual, where the modulation is noticeably flattened for the model including the additional QNM. This indicates that the late-time mixing behavior is slightly better captured when $(\ell+1, m, 0)$ is included. 
Although this confirms the validity of the extended ansatz, the improvement is minor. Therefore, for the sake of simplicity, we do not include this additional QNM information in the model, at least for the quasi-circular spin-aligned case. 
Consequently, for $\ell = m$ modes with positive spin, the merger–RD waveform reduces to the original \texttt{SEOBNRv5HM} ansatz given in Eq.~\eqref{eq:hlm_MR}.
\begin{figure}[t]
\centering
\includegraphics[width=\columnwidth]{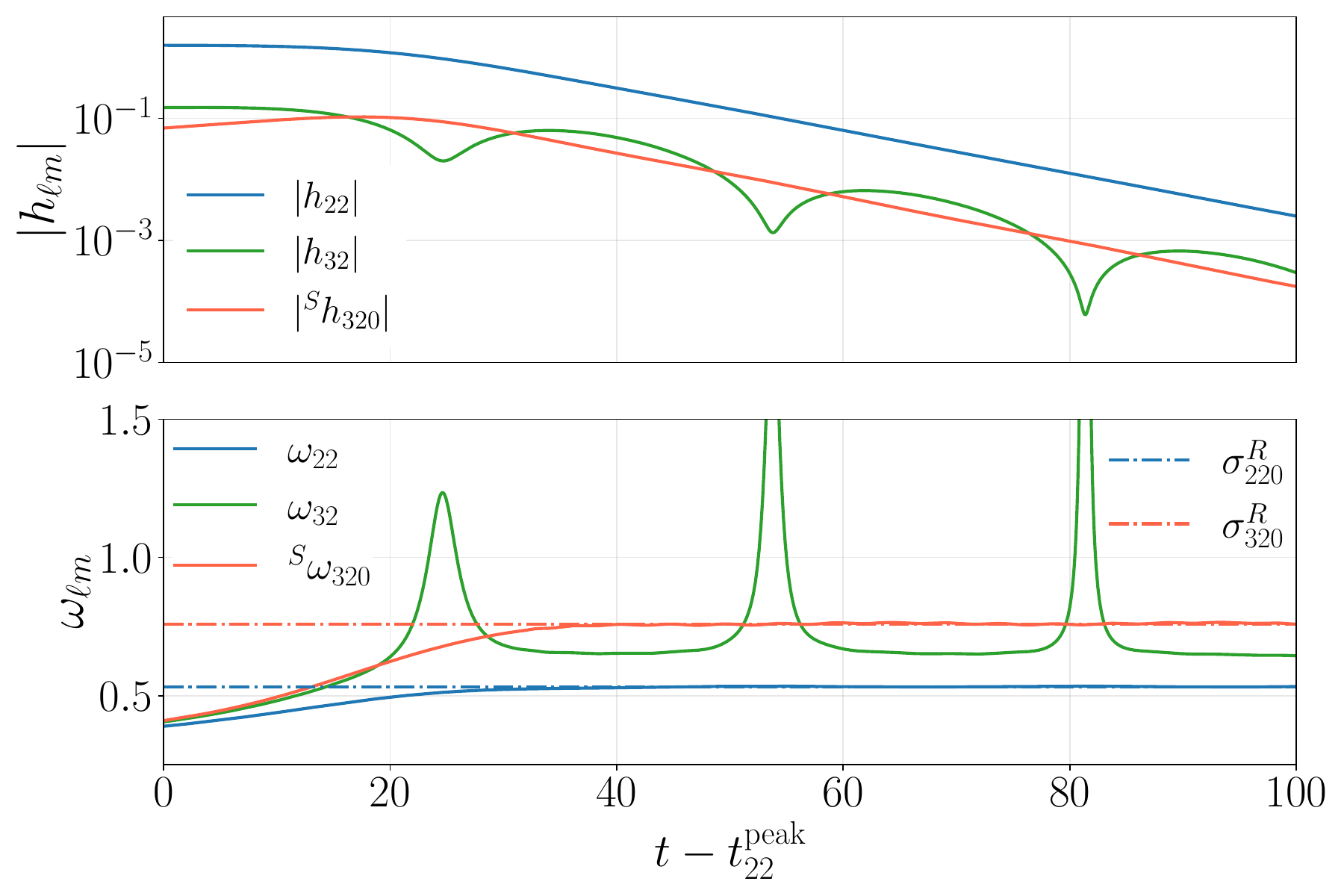} 
\centering
\includegraphics[width=\columnwidth]{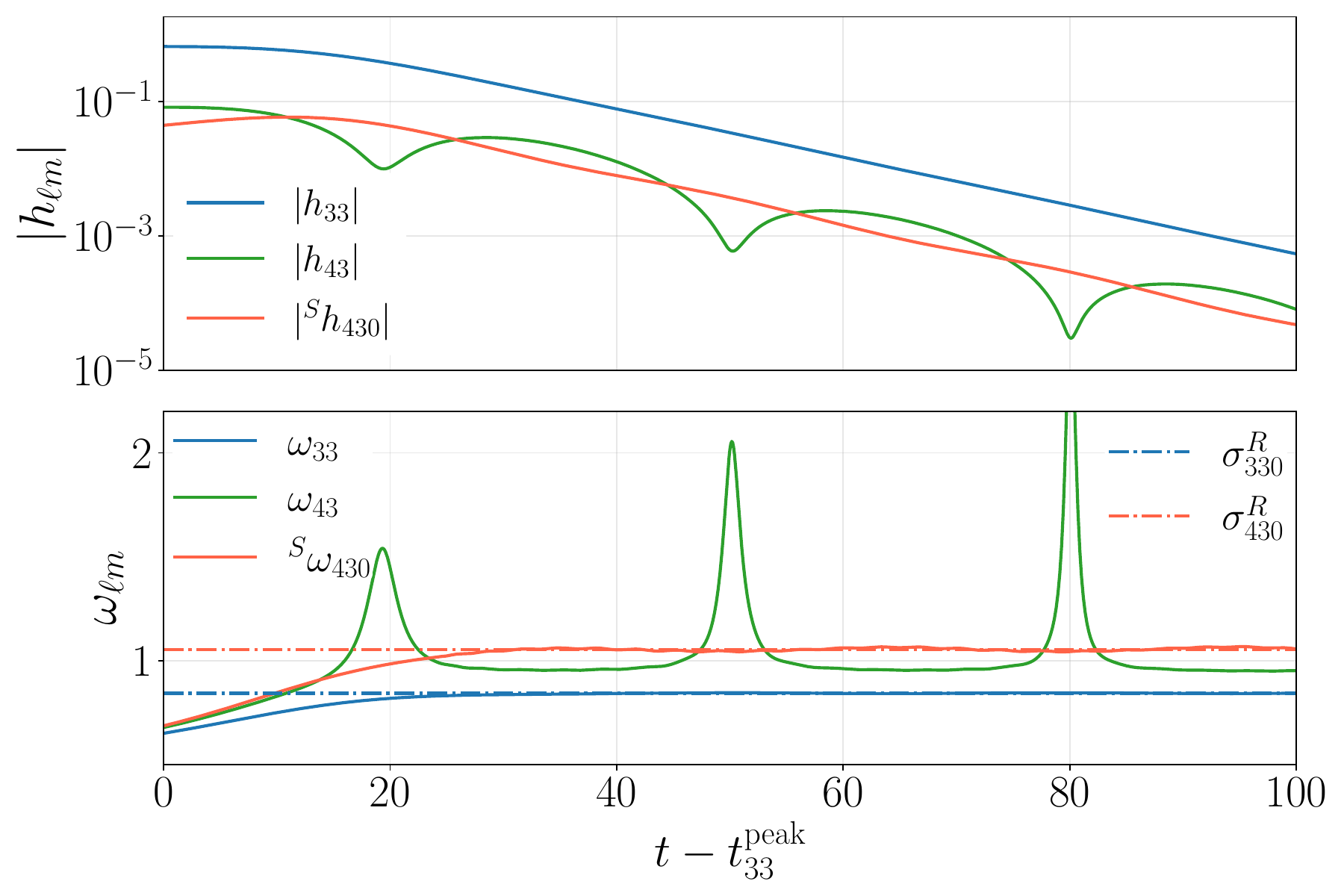} 
\caption{
For each mode, the upper subpanel displays the amplitude of the Teukolsky waveform $h_{\ell m}$ together with the reconstructed spheroidal mode ${}^S h_{\ell m 0}$ defined in Eq.~\eqref{eq:h320}, while the lower subpanel shows the corresponding GW frequencies.
At late times, the frequencies of ${}^S h_{320}$ and ${}^S h_{430}$ approach the fundamental QNM frequencies $(3,2,0)$ and $(4,3,0)$, respectively, indicated by dashed horizontal lines.
The configuration shown corresponds to a prograde spin $a = 0.8$.}
\label{fig:h32Mix}
\end{figure}

\subsection{Prograde spin ($\ell \neq m$ modes)}
\label{sec:ProgradeMix3243}
We first note that the $(2,1)$ mode is modeled exactly as the $\ell = m$ positive-spin modes discussed in the previous section, since it does not exhibit noticeable mode mixing.
In principle, one could extend Eq.~\eqref{eq:RD_with_MixModeProg} to the $(3,2)$ and $(4,3)$ modes by including the additional QNMs $A_{220}$ and $A_{330}$, respectively. 
While this ansatz can capture the mixing behavior using extracted QNM information, it relies heavily on a phenomenological activation function, which—as we have seen—may not fully capture the early merger–RD behavior in some cases. 
To avoid this complication, we instead adopt the strategy used in \texttt{SEOBNRv5HM}, which effectively removes mode mixing by modeling the waveform in the spheroidal basis. 
This approach approximates the spherical-spheroidal transformation while neglecting overtones ($n>0$) and higher-order spheroidal modes ($\ell' > \ell$). 
Despite these approximations, this method accurately reproduces the main features of $(3,2)$ and $(4,3)$ modes.

As an example, consider the merger–RD of the $(3,2)$ mode for spin $a=0.8$ (Figure~\ref{fig:h32Mix}). 
The top panel shows the amplitudes of the $(2,2)$ and $(3,2)$ spherical modes, while the bottom panel shows their instantaneous frequencies. 
The $(3,2)$ mode exhibits pronounced oscillations in both amplitude and frequency, whereas the $(2,2)$ mode is smooth. 
At late times, $\omega_{22}$ asymptotes to the $(2,2,0)$ QNM, while $\omega_{32}$ oscillates around the $(3,2,0)$ QNM. 
Similar behavior is observed for the $(4,3)$ mode.
\begin{figure*}[t]
\centering
\includegraphics[width=\textwidth]{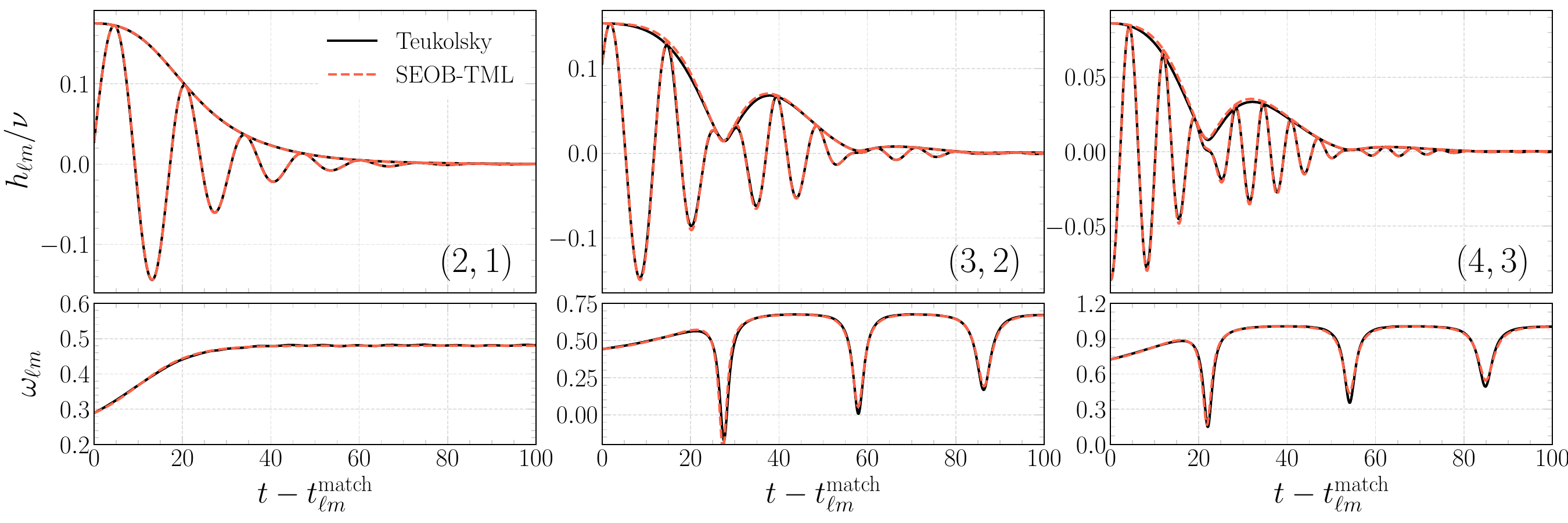}
\caption{Merger–RD modeling of $\ell \neq m$ modes for prograde spin $a = 0.8$, compared against TD Teukolsky waveforms (black).
For each mode, the upper panel shows the amplitude and the lower panel shows the GW frequency.
The merger–RD model (red) is constructed by modeling the reconstructed spheroidal modes ${}^S h_{320}$ and ${}^S h_{430}$ and transforming back to the spherical basis, following the same procedure used in \texttt{SEOBNRv5HM}.
This approach accurately reproduces the late-time amplitude and frequency modulations of the $(3,2)$ and $(4,3)$ modes.}
\label{fig:a080Mixellnotm}
\end{figure*}
In principle, the spherical-spheroidal transformation in Eq.~\eqref{eq:SphericalSpheroidalMode} requires summing over all overtones $n$ and all $\ell'$. 
In practice, contributions from $n>0$ decay rapidly and spheroidal modes with $\ell' > \ell$ have suppressed amplitudes, so we retain only $n=0$ and $\ell' \leq \ell$.
Under these approximations, the spherical modes reduce to 
\begin{subequations}\label{eq:mixing_coeffs_defs}
\begin{align}
h_{22}(t) &\simeq \mu^{*}_{2220}\, {}^S h_{220}(t), \\
h_{33}(t) &\simeq \mu^{*}_{3330}\, {}^S h_{330}(t), \\
h_{32}(t) &\simeq \mu^{*}_{2320}\, {}^S h_{220}(t) + \mu^{*}_{2330}\, {}^S h_{320}(t), \\
h_{43}(t) &\simeq \mu^{*}_{3430}\, {}^S h_{330}(t) + \mu^{*}_{3440}\, {}^S h_{430}(t),
\end{align}
\end{subequations}
From these relations, the spheroidal $(3,2,0)$ and $(4,3,0)$ modes can be reconstructed as
\begin{subequations}\label{eq:h320}
\begin{align}
{}^S h_{320}(t) &\simeq \frac{h_{32}(t)\mu^{*}_{2220} - h_{22}(t)\mu^{*}_{2320}}{\mu^{*}_{2330}\, \mu^{*}_{2220}}, \\
{}^S h_{430}(t) &\simeq \frac{h_{43}(t)\mu^{*}_{3440} - h_{33}(t)\mu^{*}_{3430}}{\mu^{*}_{3330}\, \mu^{*}_{3440}}
\end{align}
\end{subequations}
As illustrated in Figure~\ref{fig:h32Mix}, these spheroidal modes exhibit much less modulation in amplitude and frequency than their spherical counterparts, despite the truncation to $n=0$ and $\ell' \leq \ell$. 
We exploit this monotonic behavior by applying the phenomenological merger–RD ansatz (Eqs.~\eqref{eq:MRansatz_amp} - \eqref{eq:MRansatz_phase}) directly to ${}^S h_{320}$ and ${}^S h_{430}$. 
Minor adjustments are made: in Eq.~\eqref{eq:MRansatz_phase}, $\phi_{\ell m}^{\text{match}}$ is replaced with ${}^{S}\phi_{\ell m}^{\text{match}}$, the phase of ${}^{S}h_{\ell m 0}$ at the attachment time, and in Eqs.~\eqref{eq:c1c}–\eqref{eq:c3c}, the amplitude of $h_{\ell m}$ is replaced by the amplitude of ${}^{S}h_{\ell m 0}$.
Once ${}^S h_{320}$ and ${}^S h_{430}$ are modeled, the spherical $(3,2)$ and $(4,3)$ modes are obtained by inverting Eq.~\eqref{eq:h320} using the previously modeled $(2,2)$ and $(3,3)$ modes.
Figure~\ref{fig:a080Mixellnotm} shows the merger–RD model for $a=0.8$ and $\ell \neq m$ modes. The modulations in both amplitude and frequency for $(3,2)$ and $(4,3)$ are well reproduced by modeling the spheroidal modes ${}^S h_{320}$ and ${}^S h_{430}$ phenomenologically and then transforming back to the spherical basis.

\section{Complete IMR waveforms}
\label{sec:TotalIMR}
Having constructed both the inspiral-plunge and the merger–RD components of the waveform, we now assess the performance of the complete \texttt{SEOB-TML} model.
To quantify the accuracy of our proposed formalism, we compare our model against the TD Teukolsky waveforms, while simultaneously comparing its performance to the current \texttt{SEOBNRv5HM} model to demonstrate the specific gains achieved by our updates.
The key structural differences between our framework and the original \texttt{SEOBNRv5HM} model are summarized below:
\begin{itemize}
  \item \textbf{$Q$-factorized flux}: Introduced a new factorization that maps the total flux onto a single $(2,2)$ mode baseline (Sec.~\ref{sec:Qfactorized_inf}). 
  \item \textbf{Integrated horizon absorption}: Incorporated the dissipative effects of the BH horizon by applying a consistent $Q$-factorized prescription to the absorption flux (Sec.~\ref{sec:Q_horizon}). 
  \item \textbf{Mode-dependent attachment time}: Adopted a flexible attachment prescription that anchors to the individual mode peaks for prograde spins and to the zero-crossing of the orbital frequency for retrograde $\ell \neq m$ modes, accounting for the onset of mode-mixing (Sec.~\ref{sec:AttachmentTime}).
  \item \textbf{Late inspiral-plunge}: Replaced NQC corrections with a phenomenological hyperbolic ansatz, providing the flexibility to handle a wider range of attachment times and spin configurations (Sec.~\ref{sec:intermediate}).
  \item \textbf{Refined merger–RD amplitude ansatz}: Enforced second-derivative continuity at the attachment time (Sec.~\ref{sec:MRphenomAnsatz}). 
  \item \textbf{Physically motivated multi-mode ringdown}: Enhanced the phenomenological merger–RD ansatz by utilizing extracted QNM coefficients to better capture mode-mixing effects (Sec.~\ref{sec:ModelingMixingMode}).
\end{itemize}
We present comparisons of the dominant $(2,2)$ mode against TD Teukolsky waveforms for three representative cases: the non-spinning limit ($a=0$), and high-spin configurations with $a=0.9$ and $a=-0.9$. 
These cases were selected to highlight the main improvements of our model relative to \texttt{SEOBNRv5HM} across distinct physical regimes. 
Comparisons for the subdominant modes are provided in Appendix~\ref{app:highmodeIMR}. 

\begin{figure*}[t]
\centering
\includegraphics[width=\textwidth]{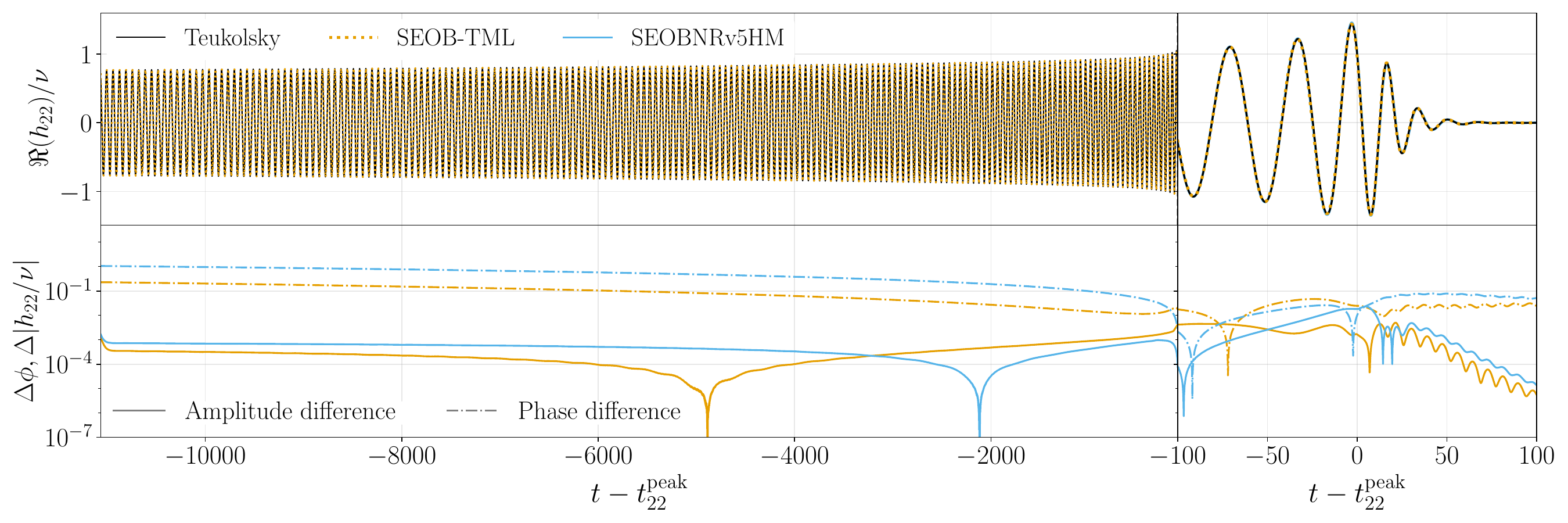} 
\caption{IMR waveform comparison for the non-spinning $(2,2)$ mode.
The top panel displays the real part of the strain, with the TD Teukolsky (black), \texttt{SEOBNRv5HM} (cyan), and \texttt{SEOB-TML} (dashed orange); the time axis is shifted such that $t=0$ corresponds to the amplitude peak.
Alignment is performed by matching peak times and minimizing the phase difference near the merger.
The bottom panel shows the phase difference (dashed) and absolute amplitude residuals (solid) for both models.}
\label{fig:a00IMRwfCompNew_narrow}
\end{figure*}

\begin{figure*}[t]
\centering
\includegraphics[width=\textwidth]{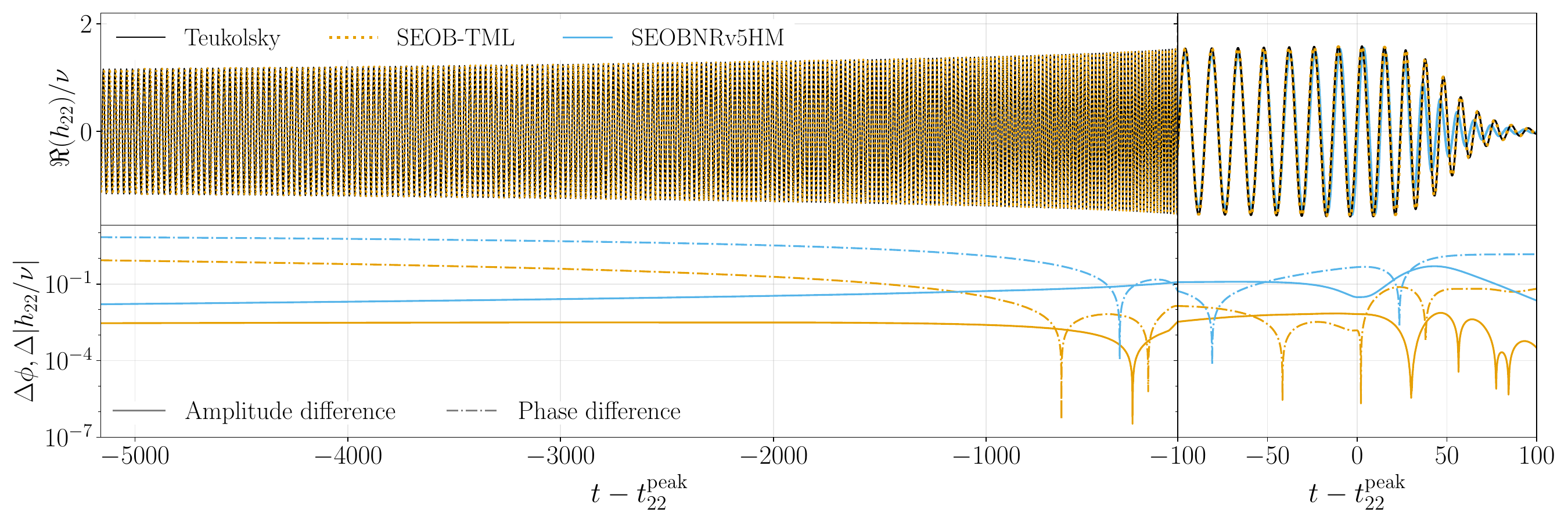}
\caption{IMR waveform comparison for the $(2,2)$ mode with prograde spin $a = 0.9$, following the same format and color scheme as Figure~\ref{fig:a00IMRwfCompNew_narrow}. 
The \texttt{SEOB-TML} model exhibits a substantial reduction in accumulated inspiral dephasing compared to \texttt{SEOBNRv5HM} due to the refined flux formulation.
}
\label{fig:a09IMRwfCompNew_narrow}
\end{figure*}

\begin{figure*}[t]
\centering
\includegraphics[width=\textwidth]{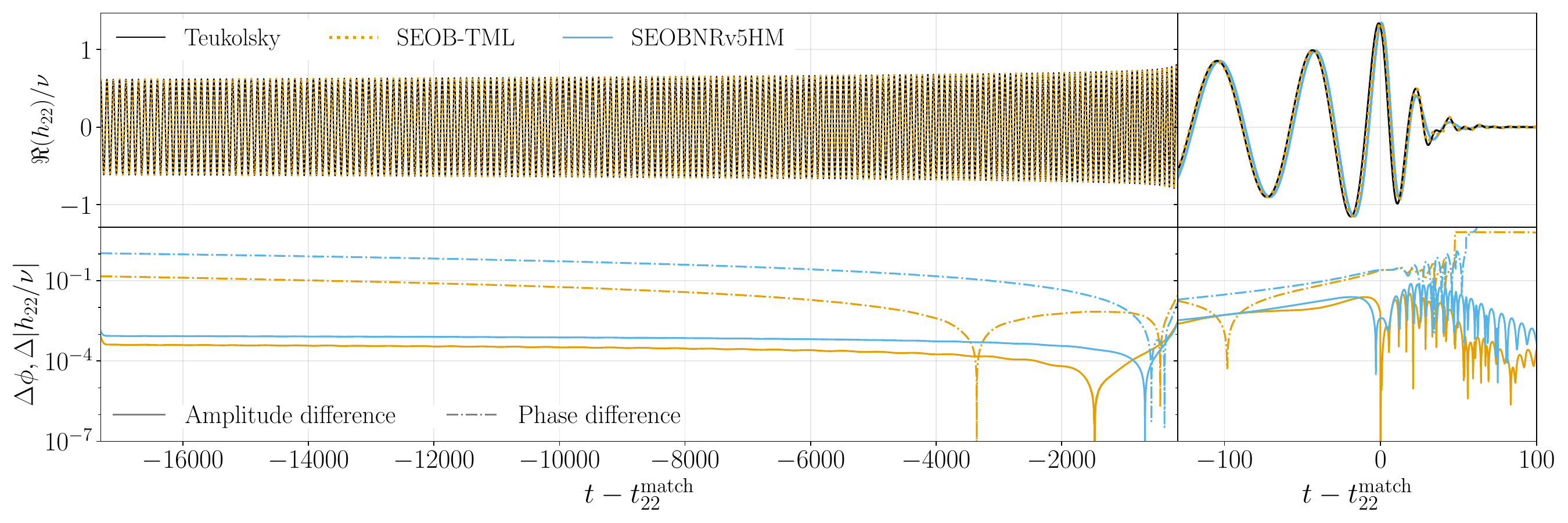}
\caption{IMR waveform comparison for the $(2,2)$ mode with retrograde spin $a = -0.9$, following the same format and color scheme as Figure~\ref{fig:a00IMRwfCompNew_narrow}. 
The refined $Q$-factorized flux treatment reduces the accumulated inspiral dephasing by over an order of magnitude. Significant gains are also observed during the merger–RD stage, where the inclusion of retrograde QNM contributions via our new ansatz captures characteristic oscillatory features in the amplitude and frequency that are absent from \texttt{SEOBNRv5HM}.
}
\label{fig:am09IMRwfCompNew_narrow}
\end{figure*}

Figure~\ref{fig:a00IMRwfCompNew_narrow} presents the $(2,2)$ mode comparison for a non-spinning binary, covering approximately $95$ GW cycles from an initial orbital frequency of $\Omega_0 = 0.048$ through the merger and ringdown. 
The top panel displays the real part of the strain, split into two temporal regions for visual clarity. 
On the left, we focus on the long inspiral phase, comparing the \texttt{SEOB-TML} model (orange dashed) directly against the TD Teukolsky benchmark (black). 
On the right, we zoom into the merger–RD stage and include the \texttt{SEOBNRv5HM} model (cyan). 
We deliberately omit the \texttt{SEOBNRv5HM} data from the dense inspiral panel to avoid visual overcrowding. 
The bottom panel provides a quantitative assessment of the performance by showing the phase difference and the absolute amplitude residuals. 
To facilitate a consistent comparison, the waveforms are aligned by matching the time of the $(2,2)$ mode amplitude peaks. The phase offset $\Delta \phi$ is then fixed by minimizing the phase difference within a time window close to the merger. Because the alignment is anchored near the merger, the accumulated phase difference is reflected at the initial time of the comparison.
For non-spinning configurations, the accumulated dephasing is reduced from $1.06$~rad in \texttt{SEOBNRv5HM} to $0.23$~rad in \texttt{SEOB-TML}, representing a nearly five-fold improvement in accuracy.
This improvement is a direct consequence of the refined energy flux treatment discussed in Sec.~\ref{sec:total_flux_comparison}. 
As shown in Figure~\ref{fig:flux_comparison_contour}, the $Q$-factorized prescription with integrated horizon absorption maintains enhanced agreement with numerical fluxes, with reduced errors persisting from the low-frequency regime up to the ISCO. 
While these gains are relatively modest at $a=0$, the resulting improvements in the orbital dynamics become increasingly significant for spinning configurations.

More substantial improvements are observed for large aligned spins, as illustrated by the $a=0.9$ case in Figure~\ref{fig:a09IMRwfCompNew_narrow}.
This waveform covers approximately $125$ cycles starting from an initial orbital frequency of $\Omega_0=0.115$.
In this regime, the refined flux formulation leads to a notable reduction in dephasing, decreasing from $\Delta\phi \approx 7.02$~rad in \texttt{SEOBNRv5HM} to $0.91$~rad in \texttt{SEOB-TML}—representing a nearly 90$\%$ reduction in accumulated error.
While dephasing naturally grows over longer evolutions—reaching $\Delta\phi=1.79$ after $200$ cycles and $\Delta\phi=2.95$ after $300$ cycles—it is important to note that even after $300$ cycles, the accumulated error in our model remains significantly lower than the $7.02$~rad error found in \texttt{SEOBNRv5HM} over a much shorter evolution.
Although further refinements via additional resummation or calibrated fits could reduce these residuals even further, the current framework provides a much more faithful representation of high-spin dynamics, and we leave such extensions to future work.
Another advancement is visible in the late-plunge through ringdown (right panel of Figure~\ref{fig:a09IMRwfCompNew_narrow}). 
In this region, our refined model achieves a reduction in both amplitude and phase residuals of more than an order of magnitude compared to \texttt{SEOBNRv5HM}.
This localized performance gain is a direct result of replacing traditional NQC corrections with our specialized phenomenological ansatz (Sec.~\ref{sec:intermediate}) and utilizing an updated merger–RD prescription (Sec.~\ref{sec:MRphenomAnsatz}) that eliminates unphysical amplitude peaks in high-spin configurations.

Finally, Figure~\ref{fig:am09IMRwfCompNew_narrow} illustrates the performance for the retrograde case with $a=-0.9$, covering approximately $95$ cycles from an initial orbital frequency of $\Omega_0 = 0.031$. 
Consistent with the trends observed in previous configurations, the refined flux treatment leads to a substantial gain in accuracy; the accumulated dephasing is reduced from $1.02$~rad in \texttt{SEOBNRv5HM} to $0.09$~rad in \texttt{SEOB-TML}, an improvement of over an order of magnitude.
As discussed in Sec.~\ref{sec:total_flux_comparison}, the most pronounced reduction in dephasing is expected for retrograde configurations. In this regime, the total $Q$-factorized flux—which explicitly incorporates horizon absorption—maintains high fidelity from the low-frequency regime all the way to the ISCO.
Another significant gain in this regime occur during the merger–RD stage. 
Both the amplitude and phase behavior are reproduced with substantially higher fidelity than in the \texttt{SEOBNRv5HM} model, which struggles to capture the mode-mixing effects.
As detailed in Sec.~\ref{sec:ModelingMergerRingdown}, this improvement stems from our revised merger–RD ansatz, which explicitly incorporates retrograde QNM contributions. 
By utilizing a phenomenological transition function to smoothly activate these modes, our model successfully captures the characteristic oscillatory features in both frequency and amplitude.

\begin{figure}[t]
\centering
\includegraphics[width=\columnwidth]{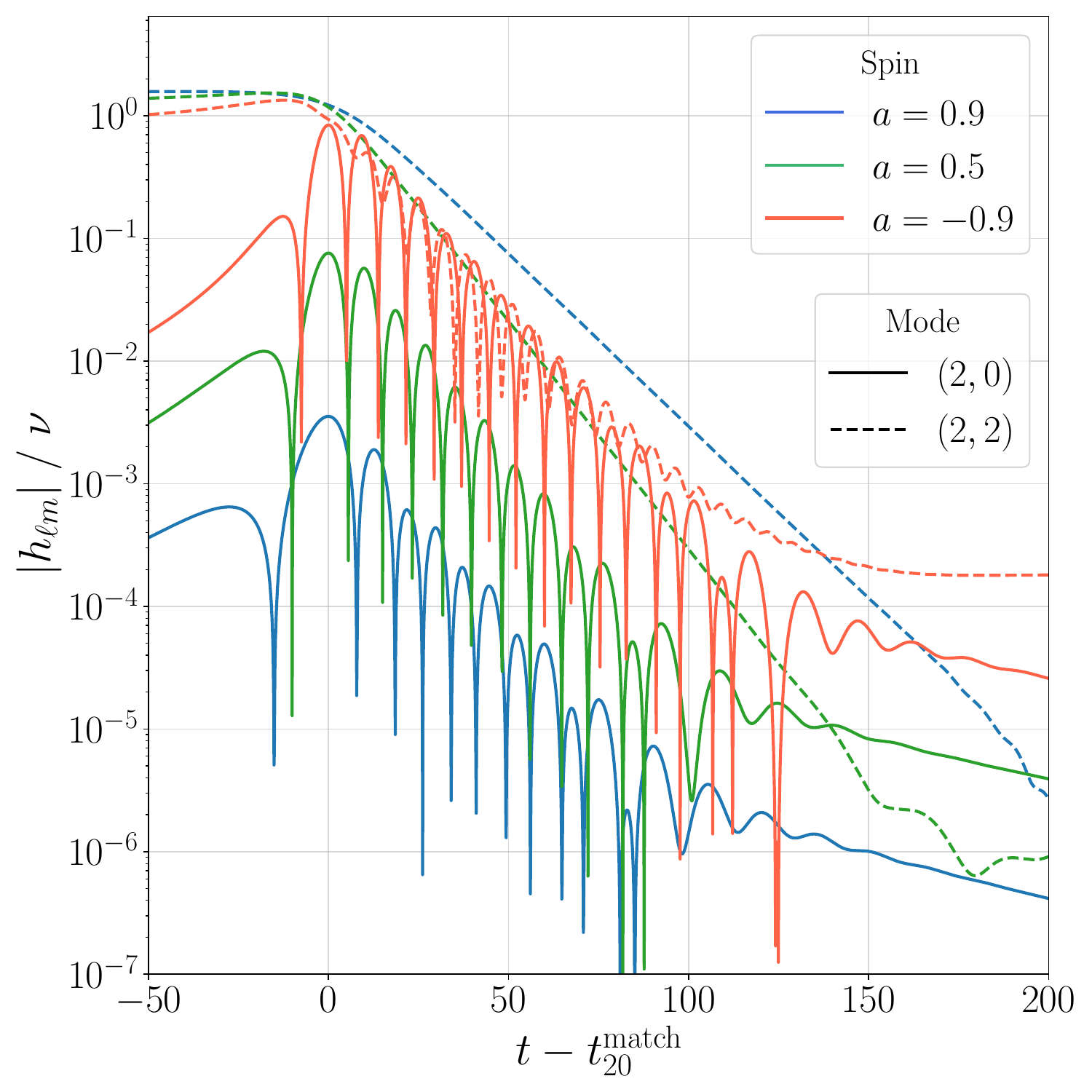} 
\caption{Comparison of the $(2,0)$ (solid) and $(2,2)$ (dashed) mode amplitudes for spin values $a = 0.9$, $0.5$, and $-0.9$.
The time axis is shifted so that $t=0$ corresponds to the peak of the $(2,0)$ mode.
While the $(2,0)$ mode is suppressed for prograde spins, its amplitude becomes comparable to the dominant mode in the retrograde case ($a = -0.9$).}
\label{fig:h20h22comp}
\end{figure}

\section{Modeling (2,0) mode}
\label{sec:h20mode}
\begin{figure*}[t]  
\centering  
\includegraphics[width=\textwidth]{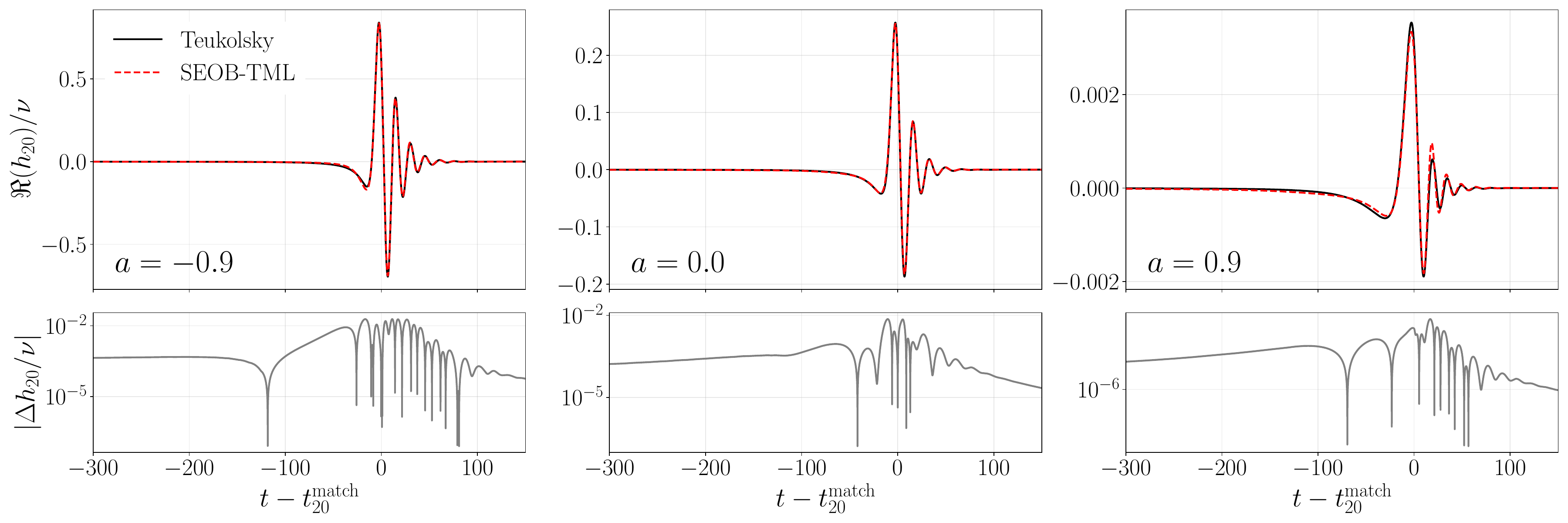}  
\caption{The late-inspiral merger–RD comparison for the $(2,0)$ mode at spins $a = -0.9$, $0.0$, and $0.9$. 
The upper panels show the real part of the strain, comparing TD Teukolsky waveforms (black) with our model (red), while the lower panels display the absolute residuals.
Overall, the waveforms are well-modeled, though some discrepancies persist for high prograde spin.}  
\label{fig:h20IMR}  
\end{figure*} 
While preceding sections focused on enhancing the accuracy of components already present within the \texttt{SEOBNRv5HM} framework, a high-fidelity description of the waveform requires incorporating previously omitted multipoles.
Most notable is the $(2,0)$ mode; as it is intrinsically linked to the radiated energy flux, this mode becomes critical during the late plunge where radial acceleration is large~\cite{PhysRevD.45.520}.
The relevance of the $(2,0)$ mode in the TML strongly depends on the Kerr spin parameter $a$. Because $m=0$ multipoles are primarily driven by radial dynamics, they are significantly more prominent in retrograde configurations. In these cases, the ISCO occurs at a larger radius, resulting in a more prolonged plunge and a post-merger contribution that can become comparable to the dominant $(2,2)$ mode (see Figure~\ref{fig:h20h22comp}). 
Conversely, for prograde spins, the $(2,0)$ mode remains relatively suppressed.

In this section, we describe our approach to modeling the instantaneous, non-linear memory effects of the $(2,0)$ mode from inspiral through ringdown.
Our construction closely follows the implementation in \texttt{TEOBResumS-GIOTTO}~\cite{Albanesi:2024fts}. 
The main difference is that, in the late inspiral, we employ the hyperbolic ansatz described in Sec.~\ref{sec:intermediate} rather than the usual NQC corrections. 
This choice allows us to attach the merger–RD portion of the waveform at the peak of the $(2,0)$ mode itself, instead of at the peak of the dominant $(2,2)$ mode.

\subsubsection{Inspiral modeling}
As discussed in Ref~\cite{Placidi:2021rkh}, the standard factorization approach is ill-suited for the m = 0 mode.
Because the Newtonian factor for $m=0$ is inherently non-circular and vanishes in the circular-orbit limit, factorization can introduce spurious poles and numerical instabilities.
Additionally, it has been demonstrated that keeping all time derivatives explicit in the Newtonian terms—rather than reducing them using PN-expanded equations of motion—generally produces more reliable results, as supported in Ref~\cite{Placidi:2021rkh,Albanesi:2024fts}.
To avoid these issues, we model the inspiral $(2,0)$ mode by retaining the explicit time derivatives in the Newtonian term, rather than reducing them via PN expansions, following the same approach done in Ref~\cite{Grilli:2024lfh}.
The inspiral contribution is given by:
\begin{equation}  
h_{20}^{\text{insp}} = 4 \sqrt{\frac{2 \pi}{15}} \nu (r \ddot{r} + \dot{r}^2)  
\end{equation}  
While higher-order PN corrections have recently been derived in Refs.~\cite{Grilli:2024lfh, Placidi:2023ofj, Cunningham:2024dog}, we restrict our current implementation to the Newtonian term.
Higher-order PN corrections may be included in future extensions, especially this is crucial when considering eccentric orbits.  

\subsubsection{Final stage of inspiral-plunge} 
For the $(2,0)$ mode, particularly in high positive-spin cases, its peak is strongly delayed relative to the $(2,2)$ mode. To improve merger–RD accuracy, we attach the merger–RD waveform at the $(2,0)$ peak. As discussed in Sec.~\ref{sec:ModelingInspPlunge}, we use a phenomenological hyperbolic ansatz instead of the NQC corrections for the last stage of inspiral-plunge. 
Similar to the $(2,1)$, $(3,2)$, and $(4,3)$ modes, we fit the parameter $\gamma$ and $\lambda$ in Eq.~\eqref{eq:hyp_amp} and Eq.~\eqref{eq:hyp_omega} to control the steepness of amplitude and frequency approaching the attachment time. 

\subsubsection{Merger–RD modeling} 
Following the approach in Ref.~\cite{Albanesi:2024fts}, we first complexify the real (2,0) mode in order to apply the standard merger–RD ansatz used for other modes. 
This is accomplished by utilizing the Hilbert transform to construct an analytic signal, which provides a well-defined amplitude and phase for the otherwise real-valued mode.
The Hilbert transform of a function $v(t)$ is defined as:  
\begin{equation}  
\mathcal{H}[v](t) = \frac{2}{\pi} \lim_{\epsilon \to 0} \int_{\epsilon}^{\infty} \frac{v(t-\tau) - v(t+\tau)}{2\tau} d\tau  
\end{equation}  
Its Fourier-domain representation is:  
\begin{equation}  
\mathcal{F}[\mathcal{H}[v]] (\omega) = -i \, \text{sign}(\omega) \, \mathcal{F}[v](\omega)  
\end{equation}  
The Hilbert transform shifts negative frequencies by $\pi/2$. Given a real signal $v_R(t)$, the corresponding analytic signal is:  
\begin{equation}  
v_{\mathcal{H}}(t) = v_R(t) - i \, \mathcal{H}[v](t)  
\end{equation}  
After complexifying the $(2,0)$ mode via the Hilbert transform, we apply the same merger–RD modeling procedure as for other modes, without mixing modes included, as described in Eq.~\eqref{eq:hlm_MR}. 

\subsubsection{Comparison with Teukolsky waveform}  
We now construct the full IMR $(2,0)$ waveforms and compare them with the TD Teukolsky waveforms. 
Figure~\ref{fig:h20IMR} shows the $(2,0)$ mode for spins $a = -0.9, 0.0,$ and $0.9$, together with the residuals in the bottom panels. 
The comparison demonstrates that the late-stage IMR signal is accurately modeled overall.
However, for large positive spins, the merger–RD description becomes less accurate than in the non-spinning case. 
In particular, the complexified amplitude exhibits oscillatory features in the post-merger regime that resemble mode mixing, leading to noticeable amplitude discrepancies. 
We attempted to identify possible additional modes using \texttt{qnmfinder}, but no further stable QNMs were detected. 
As already noted in Ref.~\cite{Albanesi:2024fts}, such behavior may be an artifact of the Hilbert transform: similar oscillations appear when complexifying the real part of the $(2,2)$ mode, though they are less pronounced than for the $(2,0)$ mode.
This suggests that improvements could be achieved by adopting alternative representations for the QNM-rescaled amplitude ansatz. Nevertheless, since the main difficulties arise only for large positive spins—where the $(2,0)$ mode is suppressed and lies low in the mode hierarchy—we defer a more detailed treatment to future work. 

\section{Conclusion}
\label{sec:Conclusion}
In this work, we introduce \texttt{SEOB-TML}, an advanced EOB framework specifically optimized for the TML.
Central to this advancement is our shift away from the traditional mode-sum factorized ($M$-factorized) flux, which relies on an explicit summation over individual multipolar modes.
Instead, we propose a quadrupole-factorized ($Q$-factorized) prescription. 
In this framework, the total energy flux is modeled using only the factorized $(2,2)$ mode as a baseline, multiplied by a polynomial $\beta^4(x)$. 
This polynomial is determined by matching its expansion to high-order PN flux results in the TML, effectively capturing the energy contribution of higher-order multipoles within the dominant mode's structure. 
Horizon absorption, which is non-negligible in the TML, is also incorporated into this formalism following a similar logic.
We validated the $Q$-factorized prescription by comparing its performance against FD Teukolsky numerical fluxes. 
To evaluate the impact of our proposed methodology, we compared it to the traditional $M$-factorized approach using the same 9PN order, summed to $\ell_{\max}=10$.

The most striking improvement occurs in the non-spinning and retrograde-spin regimes. In these regions, the inclusion of horizon absorption in the $Q$-factorized prescription leads to a dramatic reduction in fractional error that persists from the low-frequency inspiral all the way to the ISCO.
While the gains are less pronounced for intermediate prograde spins ($0 \lesssim a  \lesssim 0.8$), the $Q$-factorized flux still provides superior accuracy in the low-frequency regime and maintains a comparable error profile to the M-factorized with $\ell_{\max}=10$ sum in the strong field.
Although the model encounters accuracy degradation in the extremely strong-field prograde regime (e.g. close to the ISCO for $a  \gtrsim 0.95$), it remains highly robust across the vast majority of the inspiral. 
These results establish the $Q$-factorized framework as a highly efficient methodology for modeling the EOB energy flux.
It offers a path to decouple the orbital dynamics from the complexity of multipolar expansions without sacrificing, and often enhancing, physical fidelity.

Building on these dynamical refinements, we then addressed the modeling of the waveform itself. A key improvement in our approach is a more flexible transition to the post-merger stage, achieved by adopting a mode-dependent attachment prescription that moves beyond the reliance on the $(2,2)$ mode peak.
For prograde spins, we allow the merger–RD to begin at the amplitude peak of each individual mode, while for retrograde spins, we specifically account for the onset of mode-mixing in $\ell \neq m$ modes by anchoring the attachment to the zero-crossing of the orbital frequency.
To enable this wider range of attachment times, we replaced the NQC corrections with a phenomenological hyperbolic ansatz. 
By removing the dependence on the dynamical orbital frequency, this framework provides the necessary flexibility to accommodate the increased variability in attachment times.
We introduced additional parameters to the ansatz that specifically improved the modeling of subdominant modes characterized by steep growth near merger, such as the $(2,1)$ mode at large negative spins.

Furthermore, we refined the modeling of the merger–RD stage to achieve better consistency in the strong-field regime of the TML. 
While maintaining the standard phenomenological ansatz based on factorizing the fundamental QNM, we improved the amplitude construction by enforcing continuity up to the second derivative at the attachment time. 
This modification effectively eliminates the unphysical amplitude peaks often observed in high-spin configurations.
Beyond these structural improvements, the most significant advancement is the systematic inclusion of mode-mixing effects. 
By employing the QNM coefficients extracted via \texttt{qnmfinder}, we transitioned from purely phenomenological fitting to a physically motivated framework informed by the underlying QNM content.

For the negative spin configurations, our model explicitly captures the excitation of retrograde $(\ell,-m,0)$ modes triggered by the reversal of the orbital frequency during the plunge. 
By introducing a refined activation function that ensures a smooth attachment while providing the necessary phase flexibility, we successfully reproduce the characteristic modulations in both the waveform amplitude and gravitational frequency. 
This multi-mode approach was further extended for the $(3,2)$ and $(4,3)$ modes to account for the simultaneous presence of retrograde excitations and spheroidal–spherical basis mismatch. 
While the highly retrograde regime ($a \lesssim -0.9$) presents an ongoing challenge due to late-time frequency drifts, our framework provides a substantial accuracy improvement across the majority of the retrograde parameter space.
Furthermore, while the current work focuses on quasi-circular inspirals, this framework for modeling mixing modes is being extended to more general equatorial spin–eccentric mergers, a development that will be presented in a forthcoming study~\cite{Faggioli:prep}.

Regarding prograde configurations, our implementation of $(\ell+1, m, 0)$ mixing for the $\ell=m$ modes led to relatively insignificant changes in the overall waveform strain. 
While we demonstrated that including these extra QNM contributions effectively flattens the late-time residuals, the visual impact remains negligible; consequently, we retain the original simplified ansatz restricted to the fundamental $(\ell,m,0)$ QNM. 
For prograde $\ell \neq m$ modes, such as $(3,2)$ and $(4,3)$, where mode-mixing is more physically prominent, we maintain consistency with the \texttt{SEOBNRv5HM} strategy by modeling the signal in the reconstructed spheroidal basis and transforming back to the spherical basis.
\begin{figure*}[t]
\centering
\includegraphics[width=\textwidth]{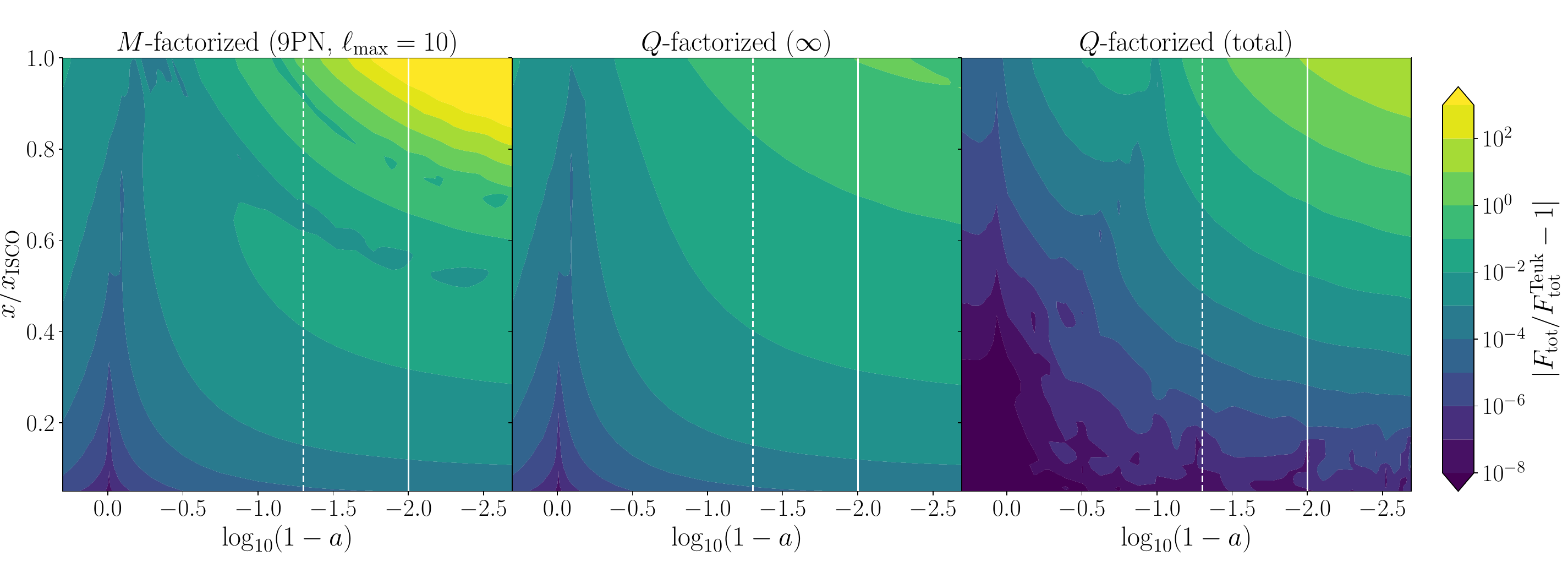}
\caption{Absolute fractional difference $|F_{\text{tot}} / F_{\text{tot}}^{\text{Teuk}} - 1|$ across the spin-frequency parameter space, with an emphasis on the near-extremal regime. 
The $x$-axis is scaled as $\log_{10}(1 - a )$ to resolve the strong-field behavior, and the $y$-axis is normalized to the ISCO frequency.
Vertical dashed and solid white lines indicate $a=0.95$ and $a=0.99$, respectively.
The panels compare the 9PN $M$-factorized flux with $\ell_{\max}=10$ (left), the $Q$-factorized flux at infinity (center), and the total $Q$-factorized flux including horizon absorption (right).
Results show that while $Q$-factorization at infinity outperforms the $M$-factorized approach near extremal spins, the current implementation of horizon absorption introduces inaccuracies that degrade the total flux performance in this limit.}
\label{fig:flux_at_extremal_spin}
\end{figure*}

Finally, we addressed the inclusion of the $(2,0)$ mode, which is currently absent from the \texttt{SEOBNRv5HM} model. 
Following the methodology introduced in Ref.~\cite{Albanesi:2024fts}, we constructed the inspiral signal using the leading-order Newtonian expression. 
The late-inspiral and plunge stages were modeled using our phenomenological hyperbolic ansatz, ensuring a smooth transition to the merger. 
To maintain consistency with our treatment of other multipoles, the real-valued $(2,0)$ mode was complexified via a Hilbert transform, enabling the application of our standard merger–RD ansatz. 
Comparisons with TD Teukolsky waveforms show consistent agreement across all three representative spin configurations.

In summary, while several technical challenges persist—particularly in the near-extremal and highly retrograde regimes—the results presented here establish a robust foundation for future EOB development. 
This work extends the reach of the EOB framework by providing a streamlined architecture tailored for high-accuracy modeling in the TML.
Ultimately, this formulation serves as a promising framework upon which next-generation waveform models can be built and further refined.

\begin{figure}[t]
\centering
\includegraphics[width=\columnwidth]{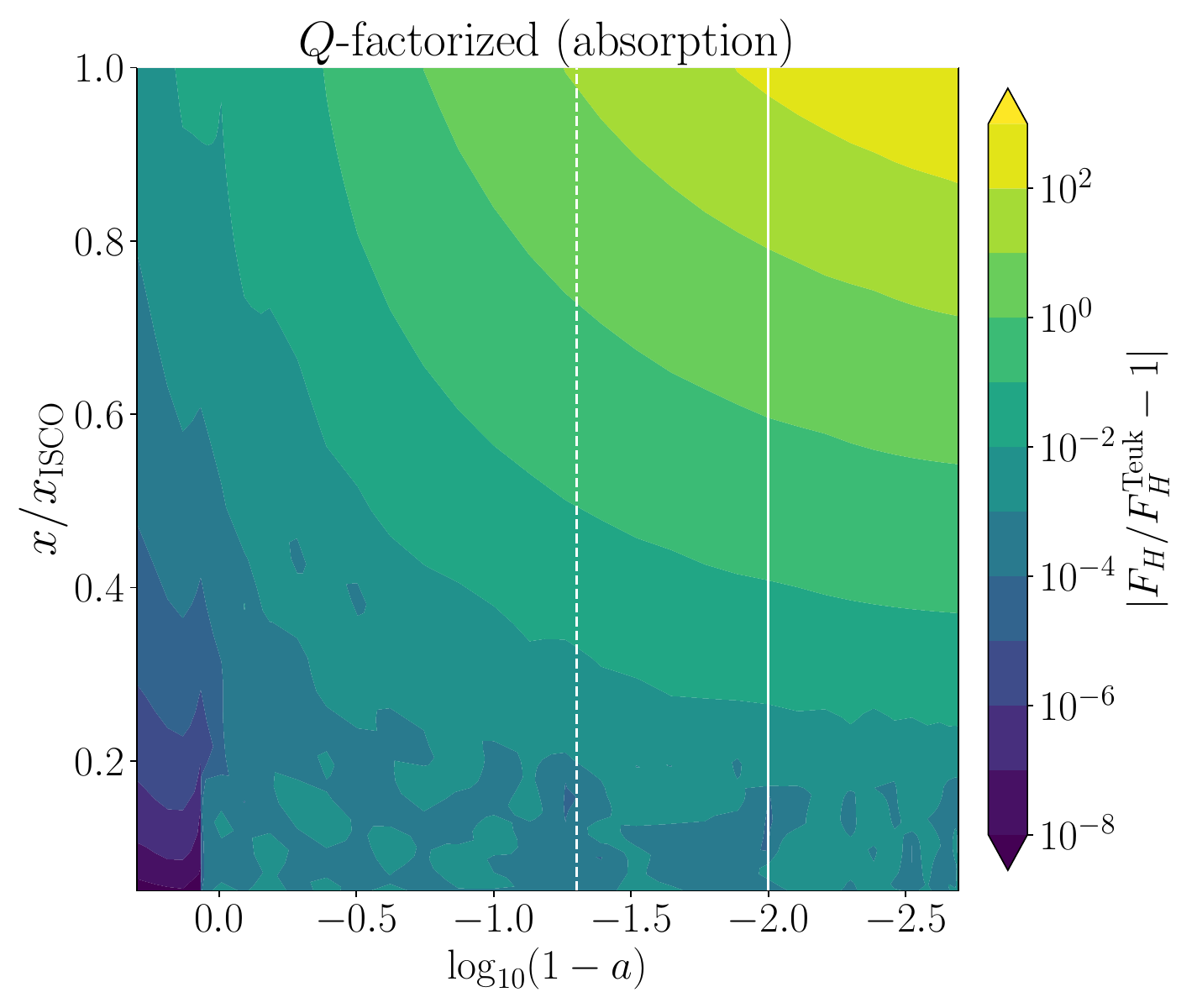}
\caption{Absolute fractional difference $|F_{H} / F_{H}^{\text{Teuk}} - 1|$ between the $Q$-factorized horizon absorption prescription (Eq.~\eqref{eq:SteffH}) and the FD Teukolsky horizon flux. 
The axes and vertical lines are consistent with Fig.~\ref{fig:flux_at_extremal_spin}.
While the prescription remains accurate at low frequencies, the fractional error grows significantly as the system approaches the near-extremal regime ($a \gtrsim 0.95$).}
\label{fig:FH_comparison}
\end{figure}

\begin{figure*}[t]
\centering
\includegraphics[width=\textwidth]{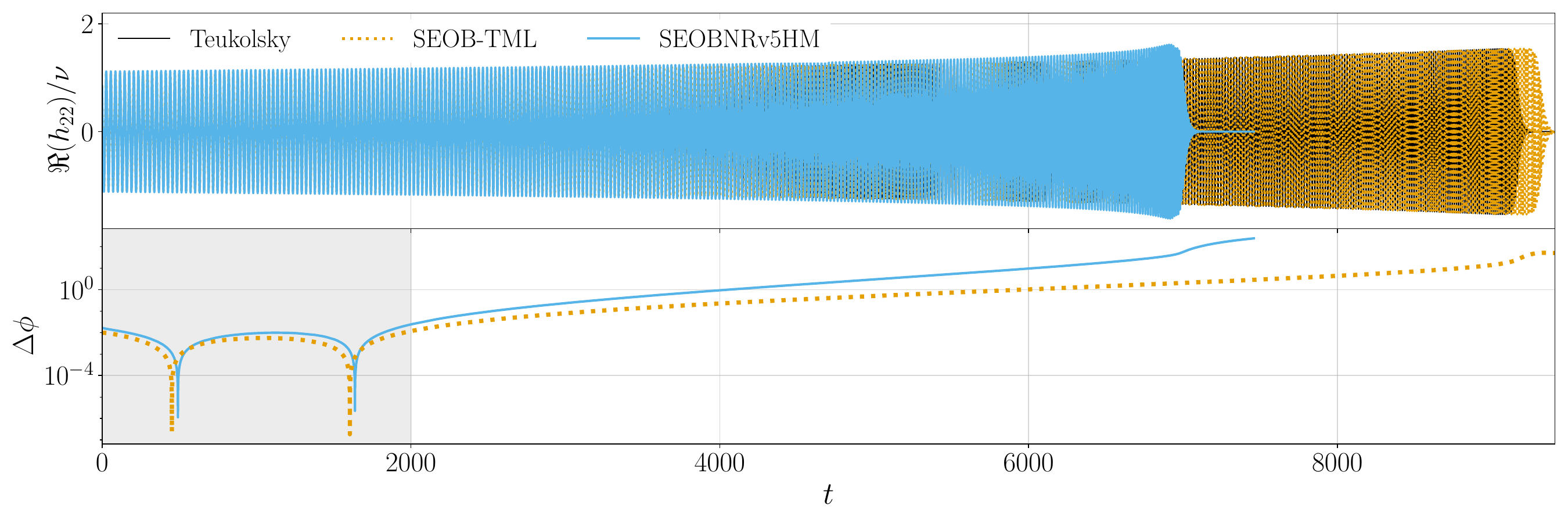} 
\caption{Comparison of the $(2,2)$ mode for a near-extremal spin $a  = 0.95$. 
The top panel shows the real part of the strain for the TD Teukolsky data (black), the \texttt{SEOBNRv5HM} model (cyan), and the \texttt{SEOB-TML} model (orange). 
The bottom panel displays the phase difference relative to the Teukolsky waveform, with the gray shaded region indicating the low-frequency window used for alignment.
Despite the inaccuracies in the horizon flux, our model significantly reduces the accumulated dephasing and provides a more accurate time-to-merger compared to the \texttt{SEOBNRv5HM} implementation.}
\label{fig:h22a095}
\end{figure*}

\section*{Acknowledgements}
The authors are grateful to Héctor Estellés, Cheng Foo, Aldo Gamboa, Marcus Haberland, Peter James Nee, and Lorenzo Pompili for useful comments and discussions.
Some computations were performed on the UMass-URI UNITY HPC/AI cluster at the Massachusetts Green High-Performance Computing Center (MGHPCC).
MvdM acknowledges financial support by 
the VILLUM Foundation (grant no. VIL37766),
the DNRF Chair program (grant no. DNRF162) by the Danish National Research Foundation,
and the EU’s Horizon ERC Synergy Grant “Making Sense of the Unexpected in the Gravitational-Wave Sky” grant GWSky–101167314.  
The Center of Gravity is a Center of Excellence funded by the Danish National Research Foundation under grant No. 184.
This work makes use of the Black Hole Perturbation Toolkit~\cite{BHPToolkit}. 

\appendix

\section{$Q$-factorized flux at near-extremal spin}
\label{app:extremal}
In Section~\ref{sec:total_flux_comparison}, we demonstrated that the $Q$-factorized prescription provides high-fidelity agreement with the FD Teukolsky flux for retrograde and moderate prograde spins. 
However, as the spin approaches the extremal limit ($a  \to 1$), the total $Q$-factorized flux exhibits a notable increase in fractional error near the ISCO.
We investigate this behavior here, focusing specifically on spins where $a  \gtrsim 0.95$.
Figure~\ref{fig:flux_at_extremal_spin} presents the same absolute fractional differences shown previously in Figure~\ref{fig:flux_comparison_contour}, but expressed as a function of $\log_{10}(1 - a )$.
The vertical white dashed and solid lines mark $a  = 0.95$ and $a  = 0.99$, respectively; as noted in Section~\ref{sec:total_flux_comparison}, the former represents the threshold beyond which the performance of the total $Q$-factorized flux begins to degrade significantly.
A key takeaway from these results is that the $Q$-factorized prescription for the flux at infinity (center panel) provides a significantly more stable description in the strong field than the traditional $M$-factorized flux at the same PN order (left panel). This demonstrates that our proposed factorization remains remarkably robust even as the system approaches the extremal limit.
However, the performance of the total $Q$-factorized flux (right panel) is noticeably worse than the infinity-only version, indicating that the current implementation of the horizon absorption flux introduces substantial inaccuracies that propagate to the total flux. 

This breakdown is isolated in Figure~\ref{fig:FH_comparison}, which displays the absolute fractional difference for the absorption flux alone.
While this prescription maintains relatively low errors at low frequencies, its accuracy diminishes rapidly as the system approaches the strong-field, high-spin limit.
These inaccuracies become critical in the near-extremal regime where the horizon contribution is no longer a negligible correction. As shown in Figure~\ref{fig:FhorFinf_LR}, this component can reach nearly $10\%$ of the total flux at the ISCO for $a=0.99$, directly impacting the overall fidelity of the model.
Consequently, the modeling errors in the absorption term—which exceed $10^2$ in the high-frequency corner—become the dominant factor in the degradation of the total flux model.
These observations indicate that future refinement of the horizon flux modeling is required for the near-extremal regime.
A potential solution involves introducing a calibrated fitting parameter at a higher PN order, such as $\alpha_{9/2}$ in Eq.~\eqref{eq:PNalpha}.
Crucially, rather than determining this coefficient by matching to higher-order analytical PN expansions—which may not converge well near the horizon—this term would be calibrated directly against numerical Teukolsky results.
Such an adjustment could effectively tune the total $Q$-factorized prescription, ensuring the model remains robust even in the extreme-spin limit.
We leave this optimization for future work.

Despite the identified inaccuracies in the absorption flux, it is instructive to examine the resulting impact on the full IMR waveform at these high spins. 
In Fig.~\ref{fig:h22a095}, we compare the $(2,2)$ mode for $a  = 0.95$ across the TD Teukolsky data, \texttt{SEOBNRv5HM}, and \texttt{SEOB-TML}. 
To ensure a fair comparison, the analytical waveforms are aligned to the Teukolsky data by minimizing the squared phase difference over a low-frequency window $[t_1, t_2]$:
\begin{align}
\int_{t_1}^{t_2} \left[ \phi_{22}^{\text{Teuk}}(t) - \phi_{22}^{\text{Model}}(t + \Delta t) + \Delta \phi \right]^2 \mathrm{d}t
\end{align}
We opt for this early-time alignment rather than peak-matching because the broad, flat nature of the amplitude at high prograde spins makes the physical merger peak difficult to resolve. 
As shown in the top panel, our model's time-to-merger is significantly closer to the Teukolsky benchmark than the original \texttt{SEOBNRv5HM} implementation. 
The phase difference $\Delta \phi_{22}$ in the bottom panel confirms that, while the agreement is not yet perfect, our new flux formulation provides a substantial reduction in dephasing relative to the baseline model, even as the model approaches its limit of validity.

\section{Coefficients of $Q$-factorized flux}
In this appendix, we provide the explicit expressions for the multiplicative correction factors $\beta(x)$ and $\alpha(x)$ of the $Q$-factorized flux, defined in Eqs.~\eqref{eq:PNbeta} and \eqref{eq:PNalpha}, respectively.
As discussed in Sec.~\ref{sec:Qfactorized_inf}, both factors are constructed using a similar matching procedure: we evaluate the $Q$-factorized prescriptions for the flux at infinity (Eq.~\eqref{eq:Steff}) and the horizon (Eq.~\eqref{eq:SteffH}) in the circular-orbit limit, and then match them order-by-order to the corresponding PN-expanded fluxes in the TML.
\begin{figure*}[t]
    \centering
    \includegraphics[width=\textwidth]{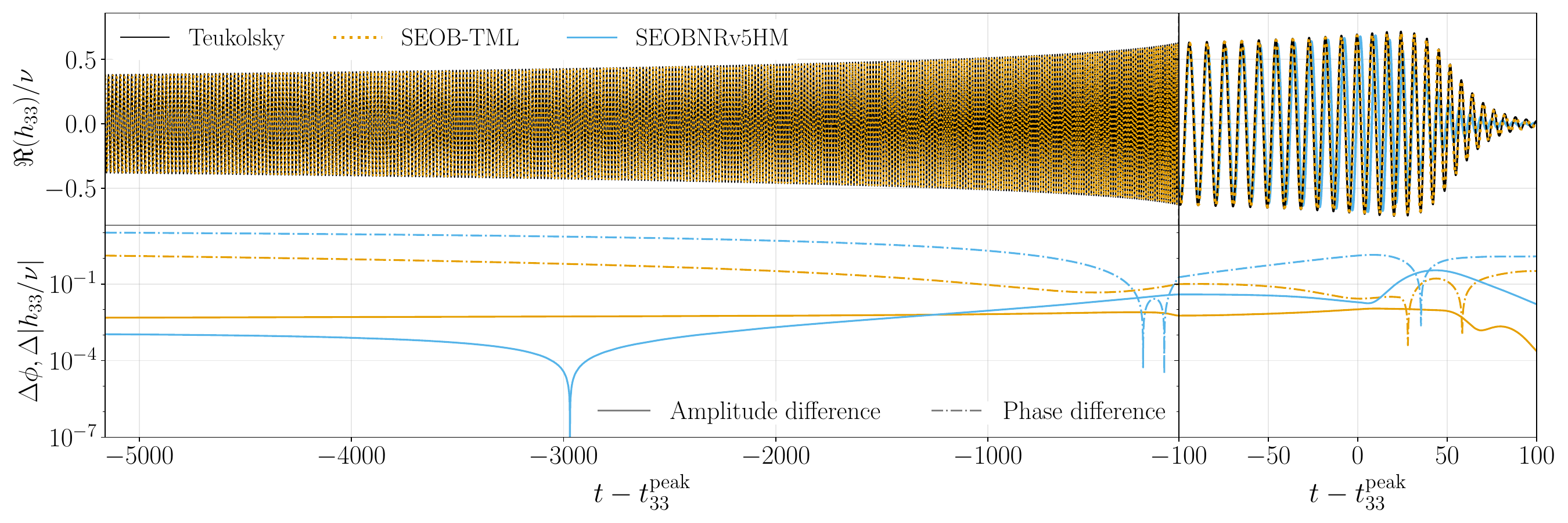}
    \vspace{0.1cm}
    \includegraphics[width=\textwidth]{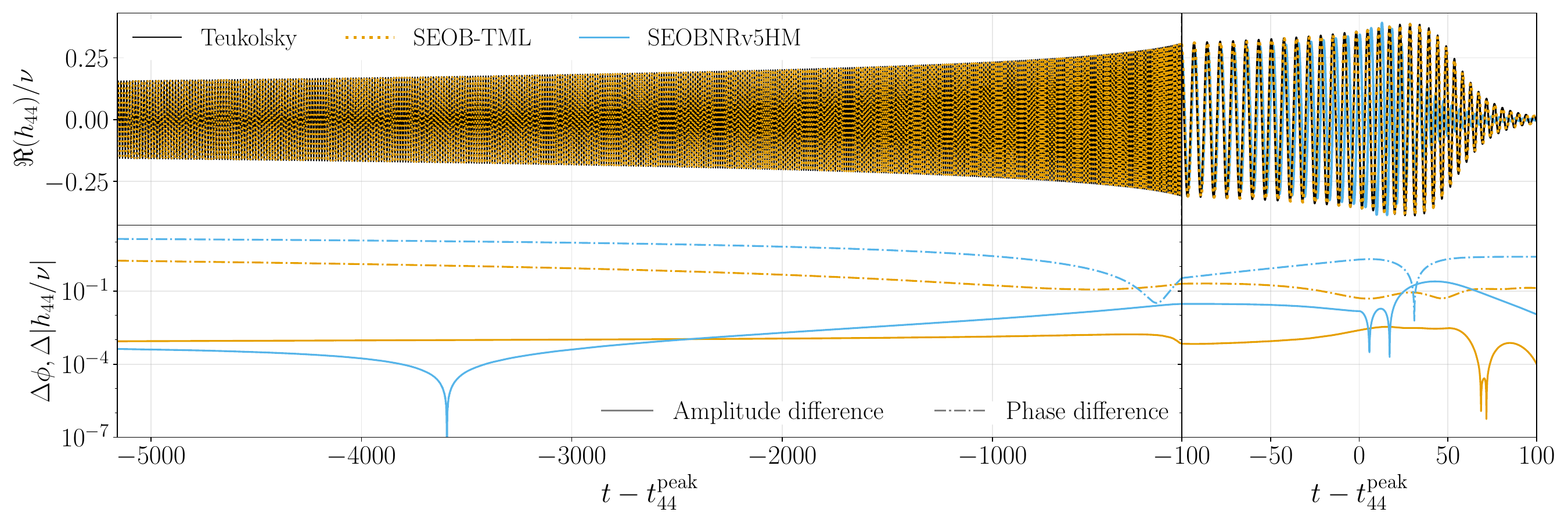} 
    \caption{IMR waveform comparison for the next-to-dominant $(3,3)$ and $(4,4)$ modes at $a = 0.9$. 
    The top panel displays the real part of the strain, with the TD Teukolsky (black), \texttt{SEOBNRv5HM} (cyan), and \texttt{SEOB-TML} (dashed orange); the time axis is shifted such that $t=0$ corresponds to the amplitude peak.
    These waveforms are aligned using the time and phase shifts derived from the $(2,2)$ mode in Fig.~\ref{fig:a09IMRwfCompNew_narrow}. 
    The bottom panels show the phase difference (dashed) and absolute amplitude residuals (solid), demonstrating an 80$\%$ reduction in dephasing and an order-of-magnitude improvement in merger–RD accuracy relative to \texttt{SEOBNRv5HM}.
    }
    \label{fig:subdominanta09}
\end{figure*}
For the construction of both $\beta(x)$ and $\alpha(x)$, we utilize the $(2,2)$ mode amplitude corrections, $\rho_{22}$ and $\rho_{22}^H$, extended to 5PN order.
This choice is primarily motivated by the PN order employed in the \texttt{SEOBNRv5HM} model, as our analysis indicates that maintaining this consistency provides the most robust performance across the explored parameter space. 
The specific expressions for $\rho_{22}$ and $\rho_{22}^H$ used in this work are provided in Ref.~\cite{Fujita:2014eta}.
While our matching procedure for the flux at infinity extends $\beta(v)$ to 9PN order, and the horizon flux implementation includes coefficients up to $\alpha_4$, we omit the full expressions for brevity. 
Instead, we list the expansion for $\beta(x)$ up to 5PN and $\alpha(x)$ up to 2.5PN; the complete results are available in the supplemental \textit{Mathematica} notebook.

\begin{widetext}
\small
\begin{flalign}
& \begin{aligned}
\beta(x) = & 1+ \frac{155 x}{448} - \frac{a x^{3/2}}{96} + \left(\frac{a^2}{256} - \frac{17974267}{32514048}\right) x^2 + \left(\frac{85 \pi }{128} - \frac{33053 a}{129024}\right) x^{5/2} \\
& + \left(\frac{11015 a^2}{114688} + \frac{\pi a}{48} - \frac{753384900379}{480687685632}\right) x^3 + \left(\frac{35 a^3}{24576}-\frac{\pi  a^2}{128}-\frac{1902514045 a}{3121348608}-\frac{2113973 \pi }{4644864}\right) x^{7/2} \\
& + x^4 \Bigg[ \frac{7615303934227510193}{359938939001241600} + \frac{793685485 a^2}{8323596288} + \frac{32509 a^4}{131072} - \frac{34261 \gamma_E}{17640} - \frac{95713 \pi a}{258048} - \frac{95 \pi^2}{224} \\
& \quad + \frac{65249 \log(2)}{17640} - \frac{47385 \log(3)}{6272} - \frac{34261 \log(x)}{35280} \Bigg] \\
& + x^{9/2} \Bigg[ \frac{1876153 a^3}{12386304} + \frac{95969 \pi a^2}{172032} - \frac{107}{840} a \log(x) + a \left(\frac{750542177372677}{192275074252800} - \frac{107 \gamma_E}{420} - \frac{\pi^2}{12} - \frac{107 \log(2)}{180}\right) - \frac{788388765823 \pi }{137339338752} \Bigg] \\
& + x^{5} \Bigg[ \frac{330134256032549418589081}{37249360919360490700800} - \frac{526795 a^4}{11010048} - \frac{19 \pi a^3}{768} - \frac{5127189563 \pi a}{1560674304} + \frac{7428593 \pi^2}{6193152} + \frac{1197441953 \gamma_E}{1005903360} + \frac{11956815 \log(3)}{401408} \\
& \quad - \frac{43960422973 \log(2)}{1005903360} + \left(\frac{107 a^2}{1120} + \frac{1197441953}{2011806720}\right) \log(x) + a^2 \left(-\frac{6821161090442773}{2307300891033600} + \frac{107 \gamma_E}{560} + \frac{\pi^2}{16} + \frac{107 \log(2)}{240}\right) \Bigg] + \mathcal{O}(x^{11/2}).
\end{aligned} \\[5ex]
& \begin{aligned}
\small
\alpha(x) = & 1 + \frac{(4 - 3 a^2) x}{12 a^2 + 4} + \frac{a (3 a^2 - 4) x^{3/2}}{9 a^2 + 3} + \frac{(99 a^4 + 48 a^2 + 110) x^2}{126 a^2 + 42} \\
& + \frac{x^{5/2} \left[ a (117 a^4 - 567 a^2 + 158) + 9 \pi ( -9 a^4 + 9 a^2 + 4 ) \tanh \left( \dfrac{\pi a}{\sqrt{1 - a^2}} \right) \right]}{18 (3 a^2 + 1)^2} + \mathcal{O}(x^3)
\end{aligned}
\end{flalign}
\end{widetext} 

\begin{figure*}[t]
    \centering
    \includegraphics[width = \textwidth]{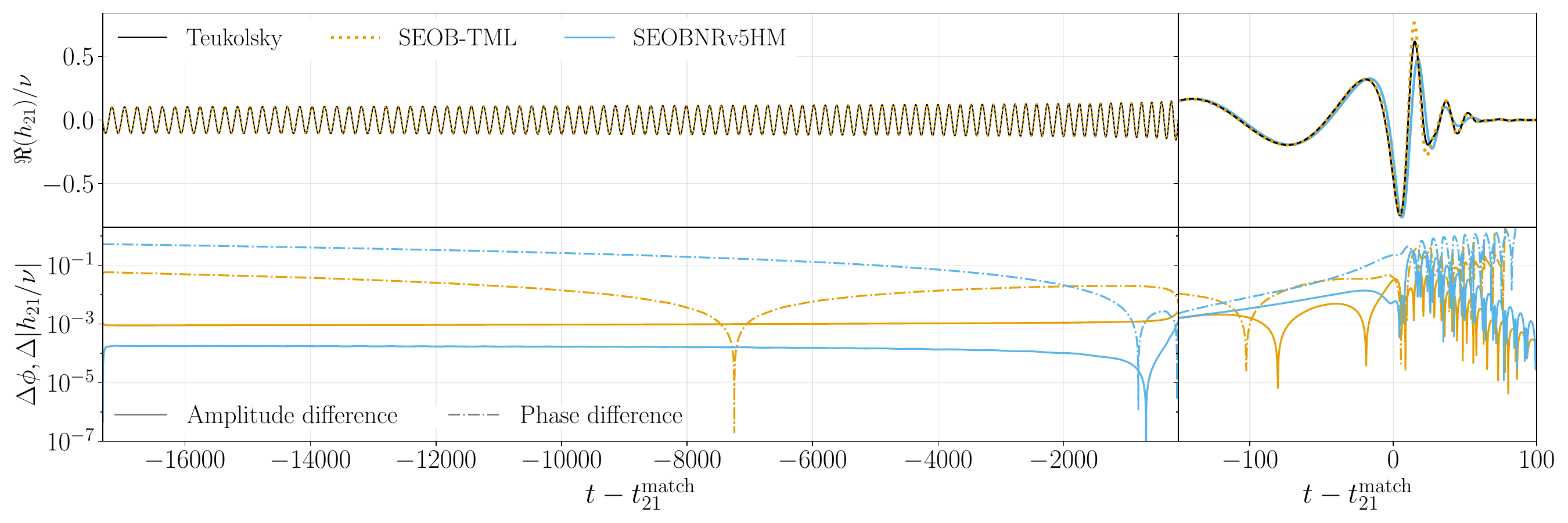} \\
    \vspace{0.1cm}
    \includegraphics[width = \textwidth]{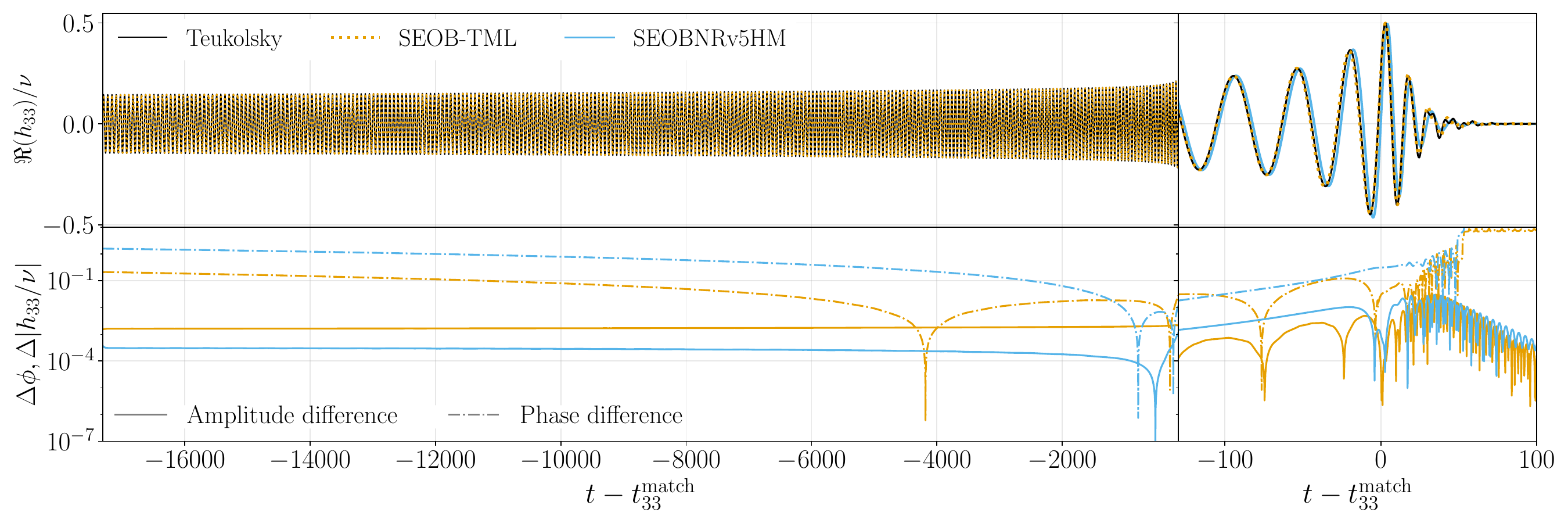}
    \caption{IMR waveform comparisons for subdominant modes $(2,1)$ and $(3,3)$ at $a =-0.9$, following the same format and color scheme as Figure~\ref{fig:subdominanta09}. 
    The refined flux results in a significant reduction in accumulated dephasing for both modes.
    Furthermore, the \texttt{SEOB-TML} ansatz accurately captures the characteristic mode-mixing features during the merger–RD stage, which are absent in the \texttt{SEOBNRv5HM} model.}
    \label{fig:subdominantam09}
\end{figure*}

\section{Complete IMR waveforms: Higher modes}
\label{app:highmodeIMR}
Following the analysis of the dominant $(2,2)$ mode in Sec.~\ref{sec:TotalIMR}, we here extend the IMR comparison to the subdominant sector for high-spin configurations ($a = \pm 0.9$). 
To maintain physical consistency, the time shift $\Delta t$ and phase offset $\Delta \phi$ for all higher-order multipoles are held fixed at the values determined by the $(2,2)$ mode alignment.
Figure~\ref{fig:subdominanta09} illustrates this comparison for the $(3,3)$ and $(4,4)$ modes, which constitute the next-to-dominant contributions in the multipolar hierarchy for $a = 0.9$.
Adopting the format of Figure~\ref{fig:a00IMRwfCompNew_narrow}, the plot displays the real part of the strain across the inspiral and merger–RD stages, alongside the resulting phase and amplitude residuals.
Consistent with the results observed for the dominant mode, the most striking improvement is an approximately 80$\%$ reduction in accumulated dephasing.
For instance, the dephasing in the $(4,4)$ mode is reduced from approximately $7.57$~rad to $1.60$~rad, while the $(3,3)$ mode shows a similar improvement, dropping from $6.24$~rad to $1.22$~rad.
Furthermore, the accuracy of the late-plunge through ringdown is significantly enhanced. This local improvement stems from two specific refinements to the waveform modeling: first, the merger–RD signal is now attached at the peak of each individual mode rather than the $(2,2)$ peak; second, the coefficients $d_{1,c}^{\ell m}$ and $c_{i,c}^{\ell m}$ ($i = 1,2,3$) in the merger–RD ansatz [Eqs.~\eqref{eq:MRansatz_amp} and \eqref{eq:MRansatz_phase}] have been recalibrated specifically for the TML. 
Collectively, these updates reduce residuals in both amplitude and phase by more than an order of magnitude.

Figure~\ref{fig:subdominantam09} presents the $(2,1)$ and $(3,3)$ modes for the retrograde case $a =-0.9$. 
Mirroring the performance gains seen in the dominant sector, the refined flux treatment significantly mitigates the accumulated dephasing in these higher-order modes. Specifically, for the $(2,1)$ mode, the dephasing drops from $0.52$~rad to $0.054$~rad, while for the $(3,3)$ mode, it decreases from $1.58$~rad to $0.21$~rad.
In addition to the improved phase agreement during the inspiral, these panels highlight the model’s ability to capture the complex merger–RD dynamics. Specifically, the inclusion of physically motivated QNM contributions in our revised ansatz allows the model to resolve mode-mixing features that are absent in \texttt{SEOBNRv5HM}.
These comparisons demonstrate that our framework maintains a high degree of fidelity across the most relevant modes.

\clearpage %
\bibliography{references}
\twocolumngrid

\end{document}